\documentclass[usenatbib]{mn2e}
\usepackage{natbibmnfix}
\usepackage{graphicx}
\usepackage{times}
\usepackage{hyperref}

\pdfoutput=1
\usepackage{epstopdf}
\usepackage{float}

\newcommand{\arepo}{{\footnotesize AREPO}}

\newcommand{\msun}{M$_{\odot}$}

\newcommand{\kms}{km s$^{-1}$}
\newcommand{\inv}{$^{-1}$}
\newcommand{\vk}{$v_{\rm k}$}
\newcommand{\vesc}{$v_{\rm esc}$}

\newcommand{\hbeta}{H{$\beta$}}

\newcommand{\hcos}{{\em HST}-COSMOS}
\newcommand{\hst}{{\em HST}}
\newcommand{\jwst}{{\em JWST}}
\newcommand{\chandra}{{\em Chandra}}

\newcommand{\randhigh}{\texttt{random-high}}
\newcommand{\randdry}{\texttt{random-dry}}
\newcommand{\fgasa}{\texttt{fgas-a}}
\newcommand{\fgasb}{\texttt{fgas-b}}
\newcommand{\cold}{\texttt{cold}}
\newcommand{\hot}{\texttt{hot}}
\newcommand{\fivedeg}{\texttt{5deg}}

\title[Detection of recoiling BHs]{Recoiling black holes: prospects for detection and implications of spin alignment}

\author[L.~Blecha et al.]
{\parbox{20cm}{
Laura Blecha$^1$\thanks{Einstein Fellow; E-mail: lblecha@astro.umd.edu}, Debora Sijacki $^3$, Luke Zoltan Kelley $^2$, Paul Torrey $^4$, \\Mark Vogelsberger $^4$, Dylan Nelson $^2$, Volker Springel $^{6,7}$, Gregory Snyder $^{8}$,\\ and Lars Hernquist $^2$}\vspace{0.3cm} \\
$^1$University of Maryland Dept. of Astronomy, 1113 PSC, Bldg. 415, College Park, MD 20742, USA \\
$^2$Harvard-Smithsonian Center for Astrophysics, 60 Garden Street, Cambridge, MA 02138, USA\\
$^3$Institute of Astronomy and Kavli Institute for Cosmology, Cambridge University, Madingley Road, Cambridge CB3 0HA, UK\\
$^4$Department of Physics, Kavli Institute for Astrophysics and Space Research, Massachusetts Institute of Technology, Cambridge, MA 02139, USA\\
$^{6}$Heidelberg Institute for Theoretical Studies, Schloss-Wolfsbrunnenweg 35, 69118 Heidelberg, Germany\\
$^{7}$Zentrum f{\"u}r Astronomie der Universit{\"a}t Heidelberg, ARI, M{\"o}nchhofstr. 12-14, 69120 Heidelberg, Germany\\
$^{8}$Space Telescope Science Institute, 3700 San Martin Drive, Baltimore, MD 21218, USA}

\begin{document}
\maketitle

\begin{abstract}
Supermassive black hole (BH) mergers produce powerful gravitational wave (GW) emission. Asymmetry in this emission imparts a recoil kick to the merged BH, which can eject the BH from its host galaxy altogether. Recoiling BHs could be observed as offset active galactic nuclei (AGN). Several candidates have been identified, but systematic searches have been hampered by large uncertainties regarding their observability.  By extracting merging BHs and host galaxy properties from the Illustris cosmological simulations, we have developed a comprehensive model for recoiling AGN. Here, for the first time, we model the effects of BH spin alignment and recoil dynamics based on the gas-richness of host galaxies. We predict that if BH spins are not highly aligned, seeing-limited observations could resolve offset AGN, making them promising targets for all-sky surveys. For randomly-oriented spins, $\la$10 spatially-offset AGN may be detectable in \hcos, and $> 10^3$ could be found with Pan-STARRS, LSST,  Euclid, and WFIRST. Nearly a thousand velocity-offset AGN are predicted within the SDSS footprint; the rarity of large broad-line offsets among SDSS quasars is likely due in part to selection effects but suggests that spin alignment plays a role in suppressing recoils. Nonetheless, in our most physically motivated model where alignment occurs only in gas-rich mergers, hundreds of offset AGN should be found in all-sky surveys. Our findings strongly motivate a dedicated search for recoiling AGN.
\end{abstract}

\begin{keywords}
accretion, accretion discs -- black hole physics -- gravitational waves -- hydrodynamics --  galaxies: active -- galaxies: interactions 
\end{keywords}

\section{Introduction}
\label{sec:intro}
Mergers between galaxies are important drivers of evolution. They grow stellar bulges, evolve galaxies along the Hubble sequence, and can trigger both star formation and accretion onto central supermassive black holes (BHs), which in turn influence the host via feedback processes \citep[e.g.,][]{sander88a,barher91,barher96,mihher96,wyiloe03,dimatt05,hopkin06a,hopkin08c}. While the majority of active galactic nuclei (AGN) in the local Universe do not appear to be triggered by mergers \citep{cister11,schawi11,kocevs12}, a clear correlation between merging activity and AGN fueling has been observed \citep[e.g.,][]{koss10,elliso11,satyap14}. There is also evidence that the majority of the highest-luminosity quasars are triggered by major mergers \citep[e.g.,][]{urruti08,treist12,hopkin14}. 

In addition to fueling rapid growth, a major merger between comparable-mass galaxies, each containing a central BH, will inevitably lead to the formation of a BH binary. This binary may eventually merge, releasing vast amounts of energy as gravitational waves (GWs). Such events could be detected directly in the coming years via pulsar timing arrays \citep[PTAs, e.g.,][]{sesana08,sesana13}, or with a future space-based GW interferometer such as {\em eLISA} \citep[e.g.,][]{amaros12}. Additionally, any asymmetry in the merging BH system -- unequal masses or spins -- will produce asymmetric GW emission and a net momentum flux, causing the merged BH to ``recoil" in the opposite direction at the moment of merger \citep{peres62,bekens73}. Numerical relativity simulations in recent years have demonstrated that these kicks may be very large, up to 5000 \kms, implying that some BHs could be ejected from galaxies entirely \citep{campan07b,lousto11}. 

These ``superkicks" of more than a thousand \kms\ should be rare, but smaller recoils may still displace BHs from galactic nuclei for substantial periods of time. Recoil events may have been more frequent in the early Universe, when galactic escape speeds were lower and merger rates were higher. This places extra constraints on models of BH seed formation that must produce massive quasars by $z\sim7$ \citep[e.g.,][]{volree06,li07,sijack09}. At lower redshift, BH ejections must be relatively rare in massive ellipticals, at least, owing to the observed ubiquity of central BHs in such galaxies for which mass measurements have been obtained. The sample of galaxies with direct mass measurements is still modest, however, and the lower-mass range is largely unexplored. A large sample of dynamical mass measurements in massive galaxies, such as brightest cluster galaxies (BCGs), could yield strong evidence for a past recoil event if a galaxy with a ``missing" or very undermassive BH were found \citep{gerses15}. Even if the BH is not ejected entirely, a marginally-bound BH may oscillate in the galactic potential for many Gyr before returning to the center \citep[e.g.,][]{madqua04,guamer08,kommer08b,bleloe08}. 

These GW recoil events have implications for BH--galaxy co-evolution. They increase the intrinsic scatter to the observed BH-bulge relations \citep{volont07}, even when the BH is not ejected entirely from the host \citep{blecha11,sijack11}. The displacement of the BH from the galactic center also reduces the amount of AGN feedback imparted to the nuclear region, allowing a denser stellar cusp to form and delaying the galaxy's transition from the blue to the red sequence \citep{blecha11}.

A major uncertainty in predicting the effects of GW recoil is that the kick velocity distribution depends sensitively on the pre-merger BH spin vectors, which are not known. In particular, if the BH binary is embedded in a circumbinary gas disk that drives the binary evolution, torques from the disk may efficiently align the BH spins prior to a merger \citep[e.g.,][]{bogdan07,dotti10,milkro13}. The maximum recoil kick for perfectly aligned spins is $< 200$ \kms, even for maximally spinning BHs, and kicks $\ga 600$ \kms\ do not occur if the spins are aligned to within a few degrees of the orbital angular momentum. However, even in gas-rich systems where a circumnuclear disk forms, spin alignment may not always be efficient \citep[][]{lodger13}. Also, if the accretion flow is chaotic, characterized by low-mass gas streams ($\la$ 1\% $M_{\rm BH}$) infalling from random directions, the BH spins may not be efficiently aligned prior to merger. In certain cases, circumbinary disks could even produce stable {\em counter}-alignment if the disk angular momentum is low and the initial BH misalignment is large \citep{nixon12}, and retrograde disks may drive the BHs to coalescence before alignment can occur \citep{schkro15}. Such configurations may occur if outer regions of the disk become unstable to star formation, or if the nuclear gas inflow is intermittent \citep[e.g.,][]{kinpri06,lodato09,kinnix13}. Furthermore, stellar dynamics may be the dominant driver of BH binary inspiral in some systems, especially in triaxial, gas-poor mergers \citep[e.g.,][]{berczi06,khan11,holkha15}. These stellar interactions typically will not cause any preferential alignment of BH spins \citep{mervas12}. In all cases, even in the final stages of BH inspiral after the BHs have dynamically decoupled from their surroundings, their spins will undergo further evolution via relativistic precession. This precession can substantially alter the final BH spins prior to merger, driving them toward alignment in some cases and increasing misalignment in others, thereby decreasing or increasing the resultant recoil velocity depending on the mass ratio and initial spins of the BH binary \citep{berti12,kesden15,gerosa15}.

In addition to its possible role in aligning BH spins, gas in merging galaxies may strongly influence the recoil trajectories themselves. The inflow of cold gas to central regions during a gas-rich major merger can dramatically increase the central escape speed of the galaxy for a period of time, and gas drag can also help retain BHs in galactic nuclei \citep{blecha11,sijack11}. Thus, even if spin alignment via gas dynamics is less efficient than numerical results suggest, large BH displacements in gas-rich systems may be rare. 

 In contrast, recoil events in gas-poor mergers can be much longer lived, particularly if the inner stellar profile has a shallow density core. Cuspy stellar density profiles may be transformed into cores (and cores may be enlarged) by BH binary inspiral \citep[e.g.,][]{quinla96,yu02,milmer01} and by the impulsive removal of the BH and its bound cusp of stars \citep{merrit04}.

A kicked BH that is actively accreting will carry along its inner accretion disk and broad emission line (BL) region, and may also sweep up gas from its surroundings, such that it could be observed as an AGN that is spatially- or kinematically-offset from its host galaxy\footnote{Following convention, we refer to any actively accreting BH as an ``AGN" but note that in the event of a GW recoil, active BHs will generally {\em not} be coincident with the galactic nucleus.} \citep[e.g.,][]{madqua04,loeb07,bleloe08,volmad08, fujita09}. In the case of kinematic offsets, the line-of-sight (LOS) velocity of the recoiling BH could be detected in the object's spectrum as an offset between the BLs and the narrow emission lines (NLs), which will remain bound to the host galaxy \citep[e.g.,][]{mersto06}. An offset could also be detectable between the AGN BLs and the rest-frame redshift of the host galaxy itself. Observable offset AGN lifetimes can be up to tens of Myr for both spatial and kinematic offsets, for a wide range in kick speeds \citep{blecha11,guedes11}. Prior to the direct detection of GWs, recoiling AGN could offer the best evidence of BH mergers. 

In the last few years, a handful of candidate (spatially- or velocity-) offset AGN have been identified \citep[][]{komoss08, shield09b, comerf09b, robins10, batche10, civano10, steinh12, koss14, lena14}. None have yet been confirmed \citep[see][for a review]{komoss12}, and the recoil scenario is disfavored for some \citep{shield09b, decarl14}, but two candidates in particular appear promising. CID-42 is a highly disturbed galaxy with an AGN that appears both spatially {\em and} kinematically offset from the host nucleus, though the possibility that the galaxy is in a pre-merger phase, with a second, quiescent BH in the nucleus, is difficult to rule out \citep{comerf09b,civano10, civano12b, blecha13a, novak15}. SDSS1133 is a bright point source offset from a nearby dwarf galaxy; it has no evidence for an extended emission component down to a scale of 12 pc \citep{koss14} and is observed over a 65-year baseline. An alternative scenario in which the object experienced an extreme, 50-year luminous blue variable type outburst followed by an unusually long-lived Type IIn supernova cannot be excluded, but further monitoring of the source should be conclusive in the near future. 

These largely serendipitous discoveries motivate a systematic search for offset AGN. To date, a few searches for velocity-offset quasars have been carried out \citep{bonshi07,tsalma11,eracle12}. \citet{bonshi07} searched for consistent, symmetric velocity shifts in broad \hbeta\ and Mg II lines in SDSS DR5 quasar spectra and found a null result for $\Delta v > 800$ \kms. \citet{tsalma11} searched for offset BLs among SDSS DR7 quasars, focusing on binary BH candidates, and identified 32 quasars with significant BL shifts. \citet{eracle12} similarly examined the SDSS DR7 quasars for broad \hbeta\ lines offset by $\Delta v > 1000$ \kms\ and found 88 such objects. In these samples, some objects have BL shifts $> 5000$ \kms, the theoretical maximum recoil velocity, and some are best explained as double-peaked emitters. Offset BLs may also be produced by high-velocity outflows or the orbital motion of a single active BH in a sub-parsec binary pair. These results indicate that recoiling AGN with observable velocity offsets are rare among SDSS quasars, if present at all.

Spatially-offset AGN remain largely unexplored. \citet{lena14} have conducted the only systematic search for spatial offsets thus far, focusing on very small-scale offsets ($< 10$ pc) in nearby giant core ellipticals. They find displacements between the AGN and the galactic photocenter in 6 of 14 ellipticals, including M87 where a displacement was previously noted by \citet{batche10}. However, the alignment of the displacement axis with a radio jet in four of the AGN argues against the recoil scenario for these objects. No searches have yet been carried out for AGN with larger spatial offsets.

Designing a targeted search for offset AGN requires better theoretical understanding of when and where significant displacements are most likely to occur. Using semi-analytic models, and assuming randomly-oriented BH spins, \citet[][hereafter VM08]{volmad08} predict rates of observable spatially-offset AGN; their results are of particular relevance to this work and are compared with our findings  in Section \ref{ssec:prevwork}. The effect of pre-merger spin alignment remains a major open question, and the influence of gas on recoil dynamics has been studied only for isolated galaxy mergers, rather than in a cosmological framework. The host galaxy properties of offset AGN are also unexplored, with the exception of a recent study \citep{gerses15}, which concludes that a larger sample of BH mass measurements in brightest cluster galaxies (BCGs) could find evidence for past superkicks. 

Here, we address these open questions by constructing a model for the observability of spatially- and velocity-offset AGN, utilizing data from the state-of-the-art cosmological hydrodynamic simulation Illustris \citep[e.g.,][]{vogels14a,vogels14b,nelson15b,genel14}. The Illustris simulation has been shown to successfully reproduce many observed properties of galaxies and their BHs, including the stellar and BH mass functions, cosmic star formation rate density, galaxy merger rate, baryonic Tully-Fisher relation, and quasar luminosity function \citep{vogels14b, genel14, sijack15}. We examine the offset lifetimes and space density of observable offset AGN that could be detected in various surveys, and we make the first predictions regarding the host galaxy properties of offset AGN. We consider a range of possible pre-merger BH spin distributions in order to study the effect of spin alignment on offset AGN observability, and we discuss how detections of recoils might be used to constrain the degree of alignment. These predictions will be testable in existing and future surveys with the {\em Hubble Space Telescope} (\hst), \chandra, the Panoramic Survey Telescope \& Rapid Response System (Pan-STARRS), the {\em James Webb Space Telescope} (\jwst), the Large Synoptic Survey Telescope (LSST), Euclid, and the Wide-Field Infrared Survey (WFIRST).

The outline of this paper is as follows. The cosmological simulations are described in Section \ref{ssec:sims}, BH spin models and recoil velocity distributions are detailed in Section \ref{ssec:kicks}, and the model for recoiling AGN is outlined in Sections \ref{ssec:model} - \ref{ssec:agnmodel}. Section \ref{sec:results} describes our results concerning the BH merger rate (Section \ref{ssec:mrgrate}), recoil trajectories (Section \ref{ssec:traj}), recoiling AGN properties (Section \ref{ssec:recoilingagn}), spatial and velocity offsets (Section \ref{ssec:offsets}), offset AGN observability (Section \ref{ssec:observe}), and host galaxy properties (Section \ref{ssec:hosts}). The dependence of our results on model assumptions is discussed in Section \ref{ssec:paramstudy}. We discuss the prospects for identification and follow-up of promising recoil candidates in Section \ref{ssec:confirmation} and compare with previous work in Section \ref{ssec:prevwork}. Our conclusions are summarized in Section \ref{sec:summary}. 

\section{Methodology}
\label{sec:methods}

Our basic procedure for constructing a recoiling AGN model is as follows: 1) extract the properties of merging BH pairs and their host galaxies from the Illustris cosmological simulations, 2) assign pre-merger spins to the BHs and calculate the resulting recoil kick velocity, according to an assumed spin distribution, 3) construct an analytic potential model for the galaxy merger remnant at the time of BH merger, based on the properties of the progenitor galaxies, 4) integrate the trajectory of the recoiling BH in this potential, including stellar dynamical friction, and 5) calculate the AGN luminosity of the recoiling BH at each timestep and determine its observability as an offset AGN. Here we describe this procedure in detail. 

\subsection{Cosmological simulations}
\label{ssec:sims}
We use the data from the Illustris cosmological simulation project \citep[e.g.,][]{vogels14b,genel14} as the basis for our models. The Illustris simulations were conducted with the moving-mesh hydrodynamics code \arepo\ \citep{spring10}. The highest-resolution simulation (``Illustris-1") spans a comoving volume of (106.5 Mpc)$^3$ and has $2\times1820^3$ resolution elements, with a dark matter (DM) mass resolution of $6.26\times 10^6$ \msun\ and a typical gas cell mass of $1.26\times 10^6$ \msun. The ``Illustris-2" and ``Illustris-3" simulations have the same volume with $2\times910^3$ and $2\times455^3$ resolution elements, respectively. The following cosmological parameters, which are consistent with the Wilkinson Microwave Anisotropy Probe (WMAP)-9 measurements, are used in the simulation and assumed throughout this paper: $\Omega_{\rm m} = 0.2726$, $\Omega_\Lambda = 0.7274$, $\Omega_{\rm b} = 0.0456$, $\sigma_8 = 0.809$, and $H_0 = 100 h$ \kms\ Mpc\inv\ with $h=0.704$. Sub-resolution physical models are included for primordial and metal-line cooling, star formation and stellar feedback \citep{sprher03}, gas recycling and chemical enrichment, and BH seeding, accretion, and feedback. These prescriptions are described in detail in \citet{vogels13} and \citet{torrey14}. Here, we outline the relevant aspects of the prescriptions for BH formation, growth, and feedback in the Illustris simulation. These models successfully reproduce BH and AGN populations in good agreement with observational constraints, including the BH mass function at $z=0$ and the quasar luminosity function \citep{sijack15}.

Throughout the simulation, an on-the-fly Friends-of-Friends (FOF) algorithm is used to identify bound subhalos, and a BH particle with a seed mass of $10^5 h$\inv\ \msun\ is placed in each subhalo more massive than $7.1\times10^{10}$ \msun\ that does not already contain a BH. Gas accretion onto BHs is parametrized in terms of a Bondi-like formula, capped at the Eddington limit. A repositioning scheme is implemented to center BHs on the gravitational potential minimum of their hosts, which prevents spurious BH wandering from numerical two-body scattering. As a result, BH velocities are ill-defined, and the velocity of the BH relative to the surrounding gas is not taken into account in calculating the accretion rate. Additionally, a pressure criterion is introduced to reduce the BH accretion rate if the gas pressure near the BH is insufficient to compress the gas against thermal feedback, which prevents the formation of unphysically large, hot gas bubbles around accreting BHs. 

In addition to gas accretion, BHs are allowed to grow via mergers. Two BHs are assumed to have merged when their separation falls below the gravitational softening length. Owing to the effect of the repositioning scheme on BH velocities, no criterion is imposed on the relative BH velocity in order for them to merge. Furthermore, in some instances, when a satellite halo containing a BH is determined to have merged with the central halo, the repositioning scheme may cause the low-mass satellite BH to merge with the central BH on an artificially short timescale. We find that this effect is largely confined to BHs with mass $< 10^6$ \msun, and we accordingly exclude from our analysis all mergers in which either BH mass is below this value. 

Three distinct modes of BH feedback are included in the simulation. Thermal or ``quasar mode" feedback is introduced by coupling 5\% of the AGN luminosity to surrounding gas particles as thermal energy \citep{spring05b}. When the accretion rate falls below 0.05 of the Eddington rate, AGN feedback is assumed to be dominated by a radiatively inefficient ``radio mode," in which hot, buoyant bubbles with a radius of 100 kpc are injected into the ambient gas \citep{sijack07}. Additionally, ``radiative" feedback is included by assuming a fixed AGN spectral energy distribution (SED) and modifying the net gas cooling rate in the presence of strong ionizing radiation. This radiative feedback is most relevant for BHs accreting near the Eddington limit.  

In addition to this constraint on the BH mass, we also require that each host halo have a minimum of 300 DM and 80 stellar particles, which in Illustris-1 corresponds to masses of $M_{\rm DM} = 2\times 10^9$ \msun\ and $M_* = 10^8$ \msun. In practice, the BH mass constraint removes most of the lowest-mass dwarf galaxies from our sample, and those remaining are largely satellites that undergo minor mergers with larger galaxies. We find that less than 0.1\% of mergers have a total halo mass $< 10^{10}$ \msun, and $\sim 0.1\%$ have a total stellar mass $< 10^9$ \msun. 

The BH merger timescales for more massive BHs can also be underestimated in the simulation to a lesser degree, owing to the repositioning scheme and the resolution limit, which impose the condition that BH binary inspiral is always rapid on sub-resolution scales. This may be a reasonable assumption in the case of gas-rich major mergers, where circumnuclear gas disks and highly asymmetric stellar potentials can greatly reduce the BH inspiral timescales \citep[e.g.,][]{escala05,berczi06,dotti07,mayer07}, but BH binary lifetimes in gas-poor or minor mergers are uncertain and may be substantially longer. 

To account for this uncertainty, we impose a delay for {\em all} BH mergers between the merger time in the simulation and the merger time assumed in our analysis. The time delay is assumed to scale with the BH mass ratio and the host galaxy gas fraction. Specifically, we add a time delay of 0.1 Gyr / $q$ to each merger, where $q$ is the BH mass ratio, motivated by the 1/$M_{\rm BH}$ scaling of the \citet{chandr43} dynamical friction timescale for a given primary galaxy mass. Because gas-driven BH inspiral should be efficient only in the regime where the gas disk is massive compared to the BHs, we add an additional 0.5 Gyr delay if the merged host galaxy has a cold (star-forming) gas fraction of $f_{\rm gas,sf} < 0.1$, where $f_{\rm gas,sf} \equiv M_{\rm gas,sf}/(M_* + M_{\rm gas,sf})$. Our results are not sensitive to the exact choice of these parameters; in practice, the effect of the delay imposed on gas-poor mergers is subdominant to the $q$-dependent delay. With this prescription, 26\% of BH ``mergers" in Illustris are still unmerged at $z=0$, but only 2.5\% of binaries with log $q > -1.5$ have not coalesced by $z=0$. This delayed merger time is used as the actual merger time in all subsequent analysis, and only mergers occurring at $z>0$ are considered.

In addition to the simulation snapshots, data for each BH merger in Illustris, as well as gas accretion data during active BH phases, is written at much higher time resolution.\footnote{These data will be made publicly available in the coming months at the permanent site \href{http://www.illustris-project.org/data}{\texttt{www.illustris-project.org/data}}, where the other simulation data are already available \citep{nelson15b}.} We note that a small fraction of these data files for Illustris-1 in the redshift range 0.15-0.38 were corrupted. This has only a modest effect on the statistics of our results, well within the uncertainties from model assumptions and resolution convergence. Except where otherwise noted, the results below refer to the Illustris-1 simulation.  We discuss the convergence of the Illustris-1, 2, and 3 results in Appendix \ref{ssec:converge}.

We also note that in some snapshots, a small number of black holes are not associated with any subhalo. Detailed analysis indicates that this can occur if there is a mismatch between the time when a subhalo is determined to have merged with a larger halo or FOF group and the time when its central BH is repositioned on the new potential minimum. Such phases are short-lived and affect $< 0.7\%$ of BHs at any point in the simulation. We therefore ignore these BHs in our analysis, with negligible impact on our results.

\subsection{BH spin models and kick velocities}
\label{ssec:kicks}

The recoil velocity of a merged BH depends on the mass ratio and spin vectors of the progenitor BHs. We obtain the BH mass ratio just prior to merger directly from the simulation data. For the spins, we must assume a physically motivated distribution. As discussed above, BH spins will be preferentially aligned with the orbital angular momentum and with each other prior to merger if their inspiral is driven by torques from a circumbinary gas disk. If the BH inspiral is instead driven predominantly by stellar interactions, or if the accretion flow is chaotic, the spin orientations may be closer to random. Coherent gas accretion will also spin up the BHs on average, while repeated mergers at random orientations cause BH spin magnitudes to tend toward an approximately thermal distribution. The efficiency of each of these processes in various merger environments is not known, but by considering a range of models for BH spins, we can explore their relative effects on the resulting recoil kicks and on the observability of offset AGN.

\begin{figure}
\centering
\includegraphics[width=0.49\textwidth,trim=8 12 0 0]{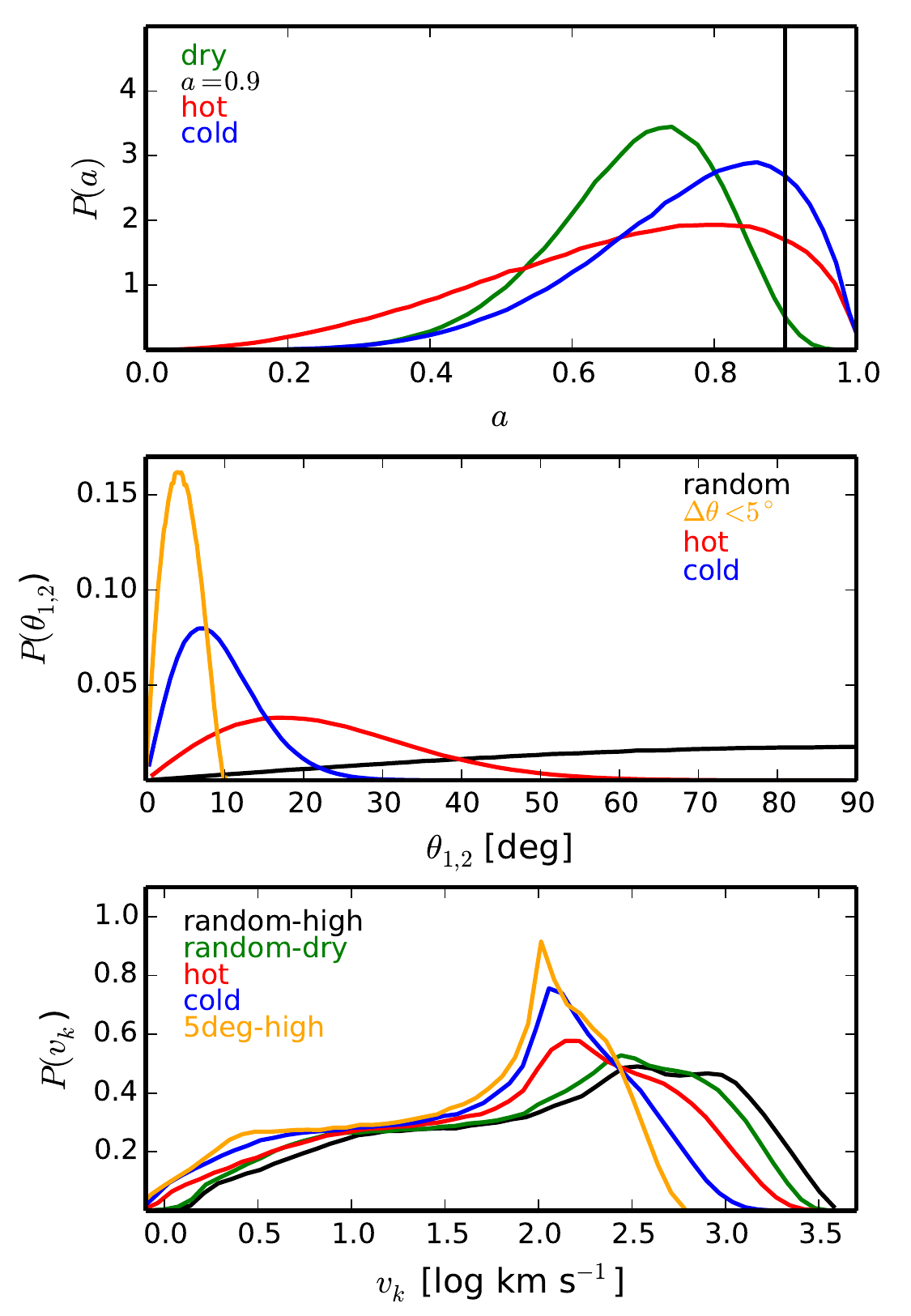}
\caption{Normalized distributions of spin magnitude, misalignment angle, and recoil kick velocity. {\em Top panel:} spin magnitude distributions for the \hot\ (red), \cold\ (blue), and ``dry merger" (green) models, as defined in the text; for comparison, the ``high" spin value of $a=0.9$ (black) is also shown. {\em Middle panel:} Distributions of the angle of misalignment between the BH spin vectors prior to merger, $\theta_{1,2}$, for the models used. The black curve is a distribution uniform in cos($\theta_{1,2}$); for clarity, we map $\theta \rightarrow 180^{\circ} - \theta$ for $\theta_{1,2} > 90^{\circ}$. The red and blue curves again denote the \hot\ and \cold\ distributions, respectively. For the orange curve, the misalignment between each BH spin and the orbital angular momentum ($\Delta\theta)$ is constrained to be $< 5^{\circ}$; the distribution is uniform in cos($\theta$) over this range. {\em Bottom panel:} recoil kick distributions resulting from Equation (\ref{eqn:kick}) for the spin models used, assuming the mass ratio $q$ has a log-uniform distribution in the range $-2 < $ log($q$) $< 0$. The \randhigh\ model uses $a=0.9$ and random orientations, the \randdry\ model uses the ``dry" spin magnitude distribution and random orientations, the \hot\ and \cold\ models correspond to partially- and nearly-aligned spins, and the \fivedeg\ model assumes $a=0.9$ and $\Delta\theta < 5^{\circ}$. The high-velocity tail of kicks is strongly suppressed when spin alignment is efficient. \label{fig:spinkick}}
\end{figure}

The most optimistic assumption for BH spins -- yielding the largest kicks -- is that the BHs are always rapidly spinning and that their spins are randomly oriented prior to merger. We refer to this (using a dimensionless spin parameter $a = 0.9$) as the \randhigh\ model. We also consider a spin distribution resulting from repeated randomly-oriented mergers, which could arise from a succession of dry mergers. \citet{lousto12} fit this spin magnitude distribution with a beta function that peaks at $a \sim 0.7$ and has a large tail toward low spins (Figure \ref{fig:spinkick}). This is denoted as the \randdry\ model. 

At the other extreme, we may assume that BH spins always undergo some amount of alignment prior to merger. We use the \hot\ and \cold\ spin evolution models as described in \citet{dotti10} and \citet{lousto12}, which result in partial alignment (within $\sim$30$^{\circ}$) or near alignment ($\la 10^{\circ}$) of BH spins, respectively. We adopt the analytic approximations to the spin angle and magnitude distributions given in \citet{lousto12}, which are based on the simulations of \citet{dotti10}. Finally, we consider the case in which spin alignment is always very efficient, such that spins are aligned to within 5$^{\circ}$ of the binary orbital angular momentum. The misalignment angle of each BH is chosen at random in cos$(\theta)$ over the interval (0$^{\circ}$,5$^{\circ}$), and the spin magnitude is assumed to be high ($a=0.9$). We refer to this as the \fivedeg\ model.

Of course, it is likely that reality lies somewhere between these extremes, with efficient spin alignment occurring in some BH mergers and not in others. Recent studies of BH spin evolution using semi-analytic models  have considered a mixture of coherent and chaotic accretion \citep{baraus12,sesana14}. \citet{sesana14} find that empirical constraints on BH spins are better fit by this type of intermediate scenario than if accretion is assumed to be always coherent or always chaotic. As binary BH spin alignment may be driven by circumbinary gas disks, we construct hybrid spin models by utilizing information about the cold gas content of the progenitor host galaxies in Illustris. In general, circumnuclear disks with $M_{\rm disk} \ga M_{\rm BH}$ can significantly influence the inspiraling BHs. We do not attempt to estimate the fraction of gas in the progenitor galaxies that will end up in a circumnuclear disk, but rather assume that gas-rich galaxy mergers are likely to trigger the formation of massive circumnuclear disks. Thus, we define galaxies as ``gas-rich" if they have a cold (star-forming) gas fraction of $> 10\%$, where again the gas fraction $f_{\rm gas,sf}$ is defined relative to the total stellar mass of the progenitor subhalos. With this definition, the large majority of galaxy mergers are classified as gas-rich -- more than 85\% of those at $z<1$, and nearly all of those at higher redshift. We note that the gas-rich fraction could be overrepresented in Illustris, as feedback processes do not sufficiently suppress star formation at the high- and low-mass end of the halo mass function \citep[e.g.,][]{vogels14b,genel14,snyder15}. However, the stellar mass function is otherwise in reasonable agreement with observations, as are the molecular gas fractions and the star formation rate density over cosmic time \citep{vogels14b,genel14}. Moreover, we find that a lower gas-rich merger fraction would only {\em increase} the observability of recoiling AGN. Alternate definitions of gas-rich mergers are discussed in detail in Section \ref{ssec:paramstudy}, but our results are not very sensitive to the exact definition used. 

We define one hybrid model in which BHs in gas-rich mergers have nearly-aligned spins (drawn from the \cold\ distribution), and BHs in gas-poor mergers have spins drawn from the \randhigh\ model. We denote this spin model as \fgasa. We also consider a more conservative model in which the same critical $f_{\rm gas,sf}$ determines whether spins are drawn from the \randdry\ model (for gas-poor mergers) or the \fivedeg\ model (for gas-rich mergers); this hybrid model is denoted \fgasb.  

Throughout this paper, we refer to these three classes of spin models as ``random", ``hybrid" and ``aligned". Table \ref{table:models} summarizes the definitions of each spin model. We finally note that the inclusion of relativistic precession effects on BH spins is beyond the scope of this work, but that recently developed efficient techniques for calculating spin precession create a promising avenue for future studies \citep{kesden15,gerosa15}.

Once the pre-merger BH spins have been obtained from the distribution for a given spin model, the recoil kick velocity can be calculated using a fitting formula based on results from numerical relativity simulations. We use the folllowing formula given by \citet{lousto12}:
\begin{equation}
{\mathbf v_{\rm recoil}} = v_{\rm m}  {\mathbf {\hat e}_{\perp,1}}+ v_{\perp} ({\rm cos} \,\xi \, {\mathbf {\hat e}_{\perp,1}} + {\rm sin} \,\xi \, {\mathbf {\hat e}_{\perp,2}}) + v_{\parallel} {\mathbf {\hat e}_{\parallel}},
\label{eqn:kick}
\end{equation}
\begin{equation}
v_{\rm m} = A \eta^2 \sqrt{1 - 4\eta} \,(1 + B \eta), 
\end{equation}
\begin{equation}
v_{\perp} = {H \eta^2 \over (1 + q )} (a_{2\parallel} - q a_{1\parallel}),
\end{equation}
\begin{eqnarray}
v_{\parallel} = {16 \eta^2 \over (1+ q)} \left [ V_{1,1} + V_{\rm A} \tilde S_{\parallel} + V_{\rm B} \tilde S_{\parallel}^2 + V_{\rm C}\tilde S_{\parallel}^3 \right ] \times \\ \nonumber 
|\, {\mathbf a_{2\perp}} - q {\mathbf a_{1\perp}} | \, {\rm cos}(\phi_{\Delta} - \phi_1),
\end{eqnarray}
where $\eta \equiv q/(1+q)^2$ is the symmetric mass ratio, $\perp$ and $\parallel$ refer to vector components perpendicular and parallel to the orbital angular momentum, respectively, and ${\mathbf {\hat e}_{\perp,1}}$ and ${\mathbf {\hat e}_{\perp,2}}$ are orthogonal unit vectors in the orbital plane. The vector ${\mathbf {\tilde S}} \equiv 2({\mathbf a_2} + q^2 {\mathbf a_1})/(1+q)^2$, and $\phi_{\Delta}$ is the angle between the in-plane component ${\mathbf \Delta_{\perp}}$ of the vector ${\mathbf \Delta} \equiv M^2({\mathbf a_2} - q {\mathbf a_1})/(1+q)$ and the infall direction at merger. The phase angle $\phi_1$ depends on the initial conditions of the binary and is assumed to be random. The best-fit values of $A = 1.2\times 10^4$ \kms, $B = -0.93$, $H = 6.9\times10^3$ \kms, and $\xi = 145^{\circ}$ are taken from \citet{gonzal07a} and \citet{louzlo08}, and the coefficients $V_{1,1} = 3677.76$, $V_{\rm A} = 2481.21$, $V_{\rm B} = 1792.45$, and $V_{\rm C} = 1506.52$ (all in \kms) are defined in \citet{lousto12}.

The kick distributions resulting from each spin model are shown in Figure \ref{fig:spinkick} assuming a log-uniform mass ratio distribution in the range $-2 < $ log($q$) $< 0$. It is clear that the high-velocity tail is greatly suppressed when spin alignment is efficient.

\setlength{\tabcolsep}{4pt}

\begin{table}
\begin{center}
\begin{tabular}{l r r c r r r}
& \multicolumn{2}{c}{Gas-rich} & & \multicolumn{2}{c}{Gas-poor} & \\
\cline{2-3} \cline{5-6} \\[-6pt]
Name & $|\mathbf{a_{1,2}}|$ & cos($\theta_{1,2}$) & & $| \mathbf{a_{1,2}} |$ & cos($\theta_{1,2}$) & $f_{\rm gas,sf}$ \\ \hline \\[-6pt]
\multicolumn{7}{c}{\em Random} \\ \hline
\randhigh\ & 0.9 & (-1,1) & & 0.9 & (-1,1) & - \\
\randdry\ & dry &  (-1,1) & & dry &  (-1,1) & - \\[6pt]
\multicolumn{7}{c}{\em Hybrid} \\ \hline
\fgasa\ & cold & cold & & 0.9 &(-1,1)  & 0.1 \\
\fgasb\  & 0.9 & (0.996,1) & & dry &  (-1,1) & 0.1 \\[6pt]
\multicolumn{7}{c}{\em Aligned} \\ \hline
\hot\ & hot &hot &  & hot & hot &- \\ 
\cold\ & cold & cold & & cold & cold &  - \\
\fivedeg\  & 0.9 & (0.996,1) &  & 0.9 & (0.996,1) & - \\
\end{tabular}
\end{center}
\caption{BH spin models used to determine the recoil kick distributions. We consider three classes of spin models: {\em random}, {\em hybrid}, and {\em aligned}. Columns: (1) Name of model, (2) spin magnitude distribution for gas-rich mergers, (3) distribution of spin misalignment angle (relative to the orbital angular momentum) for gas-rich mergers, (4) spin magnitude distribution for gas-poor mergers, (5) spin misalignment angle distribution for gas-poor mergers, (6) critical fraction of star-forming gas used to define gas-rich versus gas-poor mergers ($f_{\rm gas,sf} \equiv M_{\rm gas,sf} / (M_{\rm gas,sf} + M_*)$). The spin angle and magnitude distributions denoted as ``dry", ``cold", and ``hot" are taken from \citet{lousto12} and shown in Figure \ref{fig:spinkick}. Otherwise, the spin angle distributions are randomized in cos($\theta$) over the interval shown. \label{table:models}}
\end{table}

\subsection{Host galaxy model}
\label{ssec:model}

To model the merged host galaxy, we first match the BHs in each merging pair to their progenitor subhalos using data from the closest prior simulation snapshot. While the simulations provide detailed information about the density profiles of the progenitor host galaxies of merging BHs, the structure of the merger remnant at the time of the BH merger (which generally occurs between snapshots) is not known. Major mergers in particular should produce remnant galaxies with drastically different structure than their progenitors, and these systems are most likely to produce observable recoils. Thus, we opt to use the mass and stellar half-mass radii of the progenitors to construct an analytic model for the gravitational potential of the merger remnant. 

In many cases, the BHs already inhabit a common subhalo in the snapshot prior to their merger, and we use the (DM, stellar, and gas) mass of this common subhalo as the total mass of the merged host galaxy. If the BHs inhabit unique subhalos, the sums of the DM, stellar, and gas masses of each progenitor are used for the total mass in the merger remnant. We note that \citet{rodgom15} have found that the mass ratio of two merging subhalos can vary dramatically between the time of first infall and the time of halo merger, owing to the difficulty in assigning particles between two nearby halos. However, the {\em combined} mass of the merging halos is generally robust, and this is the quantity used in our calculations. This provides further motivation for our choice to construct analytic merger-remnant potential models rather than extracting density profiles directly from the simulation. Again, we consider only BH mergers for which each progenitor has at least 300 DM and 80 stellar particles.  
 
The DM component is modeled as a \citet{hernqu90} potential matched to an equivalent NFW profile \citep{navarr97} with concentration parameter $c_{\rm vir} = 10.5 (M_{\rm vir} / 10^{12} h^{-1}$ \msun$)^{-0.11} (1 + z)^{-1}$ \citep[following][]{bulloc01,maccio07}. The stellar component is modeled as a spherical bulge with a softened isothermal density profile, $\rho_* = \sigma_*^2/(2\pi G (r^2 +r_{\rm soft}^2)$, with velocity dispersion $\sigma_*^2 = G M_* / R_{\rm bulge}$ and r$_{\rm soft} = r_{\rm infl}$, where the influence radius $r_{\rm infl}$ of the BH is $G M_{\rm BH} / \sigma_*^2$. The profile is truncated at an inner radius $R_{\rm ej}$ within which matter remains bound to the recoiling BH. This truncated, softened profile accounts for the removal of the cusp of stars that is carried along with the BH and prevents unphysically large central densities.

 We truncate the stellar bulge at an outer radius $R_{\rm bulge}$ as well; this radius is scaled to the stellar half-mass radii of the progenitor galaxies. Specifically, we take the maximum radius of the stellar bulge in the merger remnant to be a simple average of the progenitor half-mass radii; this choice is based on the assumption that the inner density profile is steep during a major merger, and is also meant to account for the fact that the stellar half-light radii in Illustris are up to a factor of two larger than in observed galaxies at $z=0$, at least at the low mass end  \citep{snyder15}. To additionally ensure that the stellar bulges do not have artificially low densities in low-mass galaxies, we compare the calculated stellar velocity dispersion to that inferred from the $M_{\rm BH}$ -- $\sigma_*$ relation of \citet{mccma13}. If the calculated $\sigma_*$ derived from $M_*$ and $R_{\rm bulge}$ falls more than $3\sigma$ below the empirical relation (assuming 0.4 dex scatter in log $M_{\rm BH}$), we set $\sigma_*$ to lie on the relation and calculate $R_{\rm bulge}$ from this instead of using the value inferred from the simulation. In practice, this criterion is imposed on most low-mass bulges ($M_* \la$ 1 - 2 $\times 10^{10}$ \msun), affecting 14\% of all merger remnants. Finally, an absolute maximum bulge radius of 15 kpc is also imposed; this affects about 1\% of all merger hosts. The assumption of compact stellar bulges likely overestimates the inner stellar density in some cases, particularly in dry mergers between elliptical galaxies, when the BH inspiral timescale is very long, or if a disk is able to re-form after the merger. However, we consider this to be a conservative choice, as the steeper inner profile ensures that we do not overestimate the fraction of recoiling BHs that can escape the bulge.  

We also include a gas disk in our galaxy models; hydrodynamic simulations of recoiling BHs in isolated galaxy mergers indicate that the dense, circumnuclear gas disks that form during major gas-rich mergers can strongly suppress recoil trajectories \citep{blecha11, sijack11}. We assume that the cold, star-forming gas in the progenitor halos has condensed into a circumnuclear gas disk by the time of the merger, and this gas is modeled with a $1/r$ Mestel surface density profile. This is a reasonable approximation to the surface density profiles obtained from zoom-in simulations of gas-rich mergers by \citet{hopqua10}, spanning radii of $\sim 0.1$ pc - 1 kpc. Motivated by these results, we assume the inner disk surface density scales with the star-forming gas fraction as log $\Sigma_{0.1}$ $= 2$ log $(f_{\rm gas,sf}/0.1)+12$, where $\Sigma_{0.1}$ is the surface density at 0.1 pc in \msun/kpc$^2$. The disk component is ignored in galaxies with log f$_{\rm gas,sf} < -1.5$. Typical disk masses are log ($M_{\rm disk}$/\msun) $\sim 9.5 - 10.5$ at $z < 1$ and log ($M_{\rm disk}$/\msun) $\sim 10 - 11$ at $z>1$. The disk is truncated at an inner radius $R_{\rm ej}$ within which gas is bound to the recoiling BH, and at an outer radius $R_{\rm max} = G M_{\rm disk} / v_0^2$, where the fiducial circular velocity $v_0$ of the Mestel disk is set by the above scaling of $\Sigma_{0.1}$. An absolute maximum disk radius of 15 kpc is also imposed; this affects about 6\% of merger hosts.

\subsection{Recoil trajectories}
\label{ssec:integration}

Once the merged galaxy model is assigned, the recoiling BH trajectory is integrated in this potential for each recoil event with \vk/\vesc$ > 0.1$. All spatial and velocity offsets are calculated relative to the center of the static host potential. We include stellar dynamical friction via the Chandrasekhar formula \citep{chandr43}, taking $M_{\rm BH} + M_{\rm disk,ej}$ as the total mass of the recoiling object. Gas dynamical friction is neglected; because the disk is assumed to be thin ($h/r \la 0.1$ for gas-rich systems), only a few percent of recoiling BHs will be kicked directly into the disk if their orientations are random. If BH spin alignment is efficient and the angular momenta of the accretion disk and the large-scale disk are also aligned, there may be a preference for kicks into the disk plane, but in this case the kicks will be small and difficult to observe. For large kicks, the $v^{-2}$ scaling of dynamical friction ensures that it will be relatively inefficient as the BH leaves the center of the galaxy. The main uncertainty introduced by neglecting gas dynamical friction is its larger possible effect on recoil events with moderate \vk/\vesc\ that return to the center on short timescales, in which case the damping of their motion may be underestimated in our model. However, we demonstrate below that these objects do not dominate the population of observable recoils.

The integration is stopped when one of the following occurs: a) the velocity and galactocentric distance of the BH are less than critical values $v<0.2\sigma_*$ and $R<R_{\rm infl}$, b) the AGN luminosity falls below the observable limit and \vk$>$\vesc, or c) the integration reaches a Hubble time. 

Because the GW recoil kicks are implemented in post-processing, we must explicitly remove from our analysis mergers involving a BH that either a) was previously ejected from its host galaxy or b) experienced a previous recoil kick and has not yet returned to the galactic center. For the fiducial galaxy potential model, this removes 5\% of BH mergers that meet all other criteria, if the \randhigh\ spin model is assumed. 2\% of mergers are excluded in the \randdry\ model, and $< 1\%$ are excluded in all other spin models, where large kicks are less common. 

Our final sample therefore includes BH mergers that meet the following criteria: (i) each BH has M$_{\rm BH}\ge10^6$ \msun, (ii) each BH is hosted in a progenitor with $\geq$ 300 DM (80 stellar) particles, (iii) the ``delayed" BH merger time (defined above) occurs at $z>0$, and (iv) the merger occurs at a sufficiently high redshift to appear in the past light cone of an observer at $z=0$ situated in the center of the simulation box. (Because we are concerned only with the statistics of observable recoils over the entire sky, this is relevant only at extremely low redshifts; position in the box is a negligible consideration at cosmological redshifts). For Illustris-1, this yields a sample of 8993 BH mergers. Finally, we ensure that (v) neither BH is still displaced from a previous recoil event at the time of merger, which primarily affects the random spin models. The final sample sizes for the \randhigh\ and \randdry\ spin models are 8576 and 8800, respectively.

\subsection{AGN luminosities}
\label{ssec:agnmodel}

\cite{bleloe08,blecha11} have developed an analytic model for the mass and extent of the accretion disk carried along with a recoiling BH that is actively accreting at the time of the kick. The model assumes a thin ``$\alpha$-disk" that becomes self-gravitating beyond the radius where the Toomre $Q$ parameter \citep{toomre64} equals unity. The radius of the disk bound to the ejected BH is that at which the orbital speed $v_{\rm orb} \approx $ \vk. Typical ejected disk masses are a few percent of the BH mass for observable offset AGN (for further details see Figure \ref{fig:fmdisk}).

The BH accretion rate at the time of the kick is taken directly from the simulation. The ejected, isolated accretion disk will then diffuse outward over time, causing the accretion rate to monotonically decline $\propto t^{-19/16}$ \citep[assuming Thomson scattering opacity,][]{canniz90,pringl91}. Following \citet{blecha11}, we use this to calculate the accretion rate at each integration timestep of the recoil trajectory. The accretion rate is converted into a bolometric AGN luminosity $L_{\rm bol} = \epsilon \dot M$c$^2$, where the radiative efficienty $\epsilon$ is assumed to be 0.1 at high Eddington ratios. At low Eddington ratios, the accretion flow is assumed to become radiatively inefficient. Following \citet{narmcc08}, $\epsilon$ is multiplied by an accretion rate dependent factor $\dot M / (f_{\rm riaf} \dot M_{\rm Edd}$) when the Eddington ratio falls below a critical value $f_{\rm riaf}$. We adopt $f_{\rm riaf} = 0.05$, which is the critical Eddington ratio for the transition to ``radio mode" feedback in Illustris.  Also, because very low Eddington ratios are unrealistic in the context of our thin $\alpha$-disk accretion model, a minimum Eddington ratio must be imposed to prevent unfeasibly long offset AGN lifetimes. We conservatively adopt a minimum observable Eddington ratio of $10^{-2}$. While lower Eddington ratios may be detectable for some massive BHs, this choice is motivated by the distribution of Eddington ratios in observed quasars \citep{shekel12}. Section \ref{ssec:paramstudy} examines how our results are affected if a lower minimum value of $f_{\rm Edd}$ is assumed. 

Our procedure allows the observable offset AGN lifetime to be calculated for arbitrary telescope resolution and sensitivity. For each recoil event, a random viewing angle is assigned, and projected separations ($\Delta R_{\rm proj}$) of twice the angular resolution are defined as resolvable spatial offsets. For most of our analysis, the minimum resolvable LOS velocity offset is assumed to be $\Delta v_{\rm LOS} > 600$ \kms. However, in some spectra, especially those with very broad lines (FWHM $\gg 1000$ \kms), offsets this small may not be measurable. Thus, we also consider a more conservative minimum $\Delta v_{\rm LOS} = 1000$ \kms.

 The minimum observable $L_{\rm bol}$ for each event is determined by first defining a desired flux sensitivity and band, and then applying an inverse bolometric correction to the corresponding luminosity at the redshift of the recoil event. We use the luminosity-dependent optical and hard X-ray bolometric corrections from \citet{hopkin07a}. For the near-IR, we assume a constant bolometric correction of 11.0, derived from the mean SED of \citet{richar06} at 2 $\mu$m. AGN bolometric corrections are not well constrained, particularly at optical wavelengths where obscuration and host galaxy contamination have large effects. Even at X-ray wavelengths, the fraction of Compton-thick sources is uncertain, and bolometric corrections have been found to vary with Eddington ratio \citep{vasfab07,lusso12}. We neglect obscuration effects here, as these should be less significant for {\em offset} AGN (though the early post-recoil stage of velocity-offset AGN may be a notable exception, as discussed in Section \ref{ssec:observe}). 

Despite the inherent uncertainties, our results depend only weakly on the choice of bolometric correction. This is primarily because at low to moderate redshift, the minimum $f_{\rm Edd} = 10^{-2}$ criterion is typically a more stringent constraint on the AGN lifetime than the absolute flux limit derived from the (reverse) bolometric correction. The consequences of this are discussed in detail in Section \ref{ssec:recoilingagn}.

\section{Results}
\label{sec:results}

\subsection{Merging BH population}
\label{ssec:mrgrate}

\begin{figure}
\begin{center}
\includegraphics[width=0.495\textwidth,trim= 8 20 0 0]{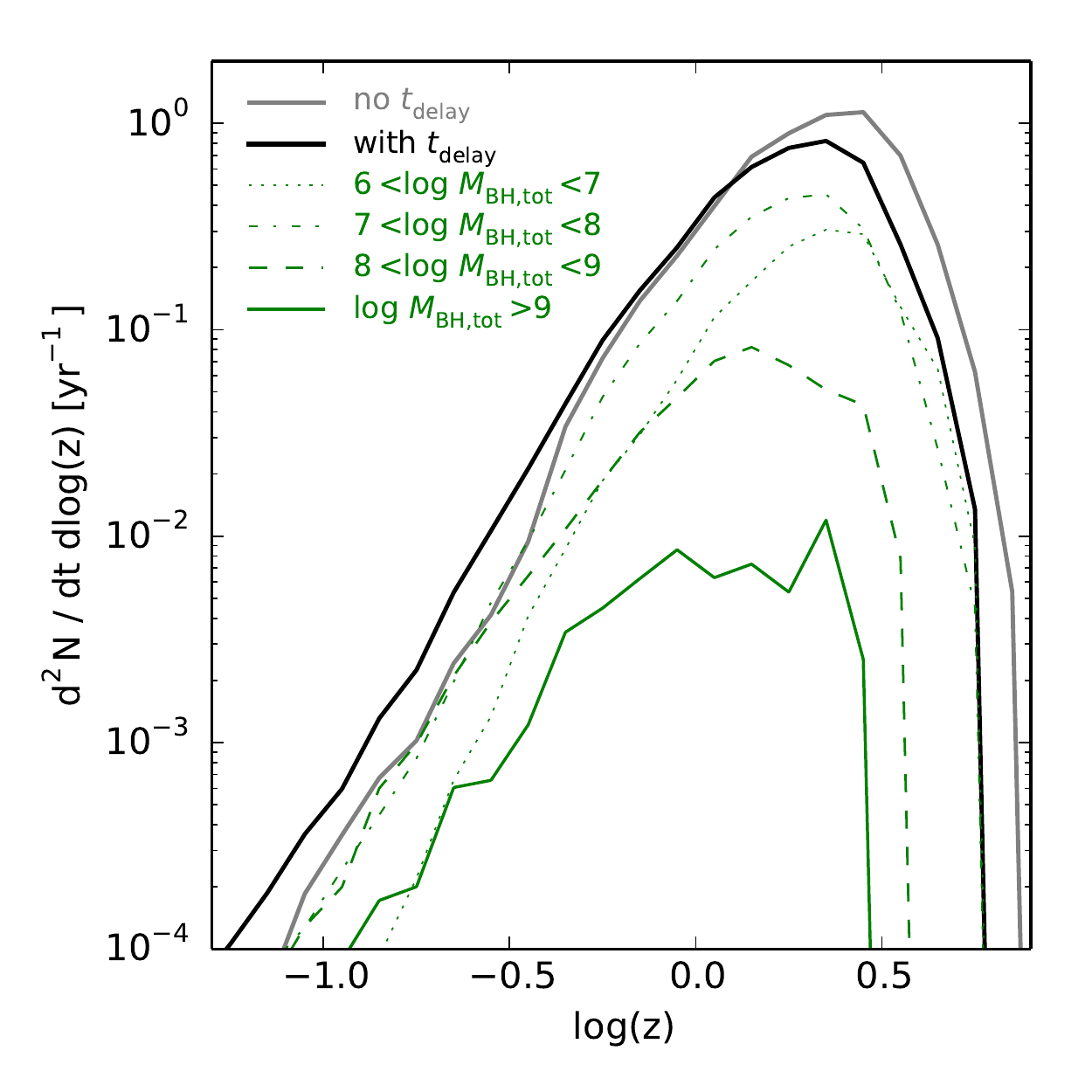}
\caption{The BH merger rate in Illustris at $z=0$, per unit redshift (in logarithmic bins). The thick gray curve denotes the ``raw" merger rate for all BH mergers that meet the selection criteria defined above, except that the actual BH merger redshift from the simulation is assumed instead of the ``delayed" redshift. The thick black curve shows the merger rate with a delay timescale added to each merger that depends on BH mass ratio and host gas fraction, as described in the text. The thin green lines denote the merger rate for different remnant BH mass bins, as indicated in the plot legend. The raw merger rate peaks at $z \sim 2$, in good agreement with semi-analytic models, and the delayed merger rate peaks at $z \sim 1.5$. The cumulative merger rate over cosmic time also agrees well with semi-analytic models, but more low-redshift mergers occur in Illustris. \label{fig:mrgrate}}
\end{center}
\end{figure}

The differential all-sky merger rate of BHs in Illustris at $z=0$, d$^2N$/dlog$(z)$\,d$t$, is shown in Figure \ref{fig:mrgrate}. Two BHs are allowed to merge in the simulation when their separation is less than a smoothing length. In cosmological simulations, this spatial scale ($\sim$ 0.1-1 kpc in Illustris) is far larger than the regime in which BH inspiral is dominated by GW emission. Thus, as described above, we add a delay timescale $t_{\rm delay} = 0.1 {\rm Gyr}/q + t_{\rm gas}$ to the time of each BH merger, where $t_{\rm gas} = 0.5$ Gyr if the fraction of mass in cold (star-forming) gas $f_{\rm gas,sf} < 0.1$ and zero otherwise. More detailed modeling of binary BH inspiral on sub-resolution scales is beyond the scope of this work, and will be examined in an upcoming study (Kelley et al., in preparation).

The green lines in Figure~\ref{fig:mrgrate} show the merger rate in different mass bins, for the remnant BH mass $M_{\rm tot} = M_1 + M_2$. Over a large redshift range $0.5 < z < 3$, the merger rate is dominated by BHs in the mass range $10^7$ - $10^8$ \msun. Lower-mass BHs dominate at higher redshifts, and higher-mass BHs dominate at lower redshifts, consistent with expectations from hierarchical growth. Low-mass BH mergers ($\la 10^6$ \msun) are of interest as GW sources for a future space-based interferometer such as {\em eLISA}, but these are excluded from our analysis owing to resolution limits, as discussed above. 

The cumulative merger rate of this BH sample at $z=0$, integrated over cosmic time, is 0.42 yr\inv, and the merger rate for BHs with total mass $10^7$ - $10^9$ \msun\ is 0.27 yr\inv. This is in reasonable agreement with results from semi-analytic modeling \citep[e.g.,][]{sesana04,baraus12}, especially considering the very different BH seeding and growth prescriptions in the semi-analytic models versus Illustris. The BH merger rate also peaks at similar redshifts ($z \ga 2$) in both Illustris and the semi-analytic models, though after the merger delay timescale is implemented, the peak shifts to slightly lower redshift ($z \sim 1.5$) in our sample. However, there are more high-redshift mergers in the semi-analytic models, owing to the BH seed formation prescriptions, which form seeds at higher redshifts than in Illustris. The first BHs in Illustris form at $z \sim$ 10 - 11, and the first mergers occur at $z < 7$. The merger rate at low redshift ($z<1$) is correspondingly higher in Illustris than in the models of \citet{sesana04} (for BH binaries with total mass $> 10^6$ \msun). In the ``heavy seed" model of \citet[][]{baraus12}, the cumulative merger rate (for $M_{\rm tot} > 10^6$ \msun) is about a factor of three higher than in Illustris, such that the merger rate per unit redshift is slightly higher than the Illustris rate at all redshifts. Further comparison with comparable cuts on host halo masses improves the agreement with both the \citet{baraus12} and \citet{sesana04} models, though some differences in the mass and redshift distribution of the merger rates remain, as does the overall trend toward more high-redshift mergers in the semi-analytic models (E.~Barausse and A.~Sesana, private communication). Empirical constraints cannot yet distinguish between these differences in the details of BH formation, growth, and merger histories. These distinctions between cosmological simulations and semi-analytic models are important to bear in mind, and they contribute to the differences in offset AGN rates predicted here and in \citet{volmad08} (Section \ref{ssec:prevwork}). However, the level of agreement of the overall merger rates in Illustris versus semi-analytic models is encouraging. We finally note that the {\em galaxy} merger rates in Illustris are in reasonable agreement with observations \citep{rodgom15}, though they disagree qualitatively with rates derived from semi-analytic models based on cosmological N-body simulations \citep[][]{guowhi08}.

The most massive BH mergers ($M_{\rm tot} \ga 10^9$ \msun) will create luminous GW signals that could be detected with PTAs; we find a cumulative merger rate of  0.006 yr\inv\ for these massive mergers. This is consistent with predictions from semi-analytic models that individual GW events resolvable with PTAs should be rare \citep[e.g.,][]{sesana09}. However, the stochastic background from unresolved massive BH mergers may be detectable with PTAs in the coming years. The predicted GW signal from BH mergers in Illustris will be explored in detail in Kelley et al. (in prep).

\begin{figure}
\begin{center}
\includegraphics[width=0.5\textwidth,trim= 12 20 6 4]{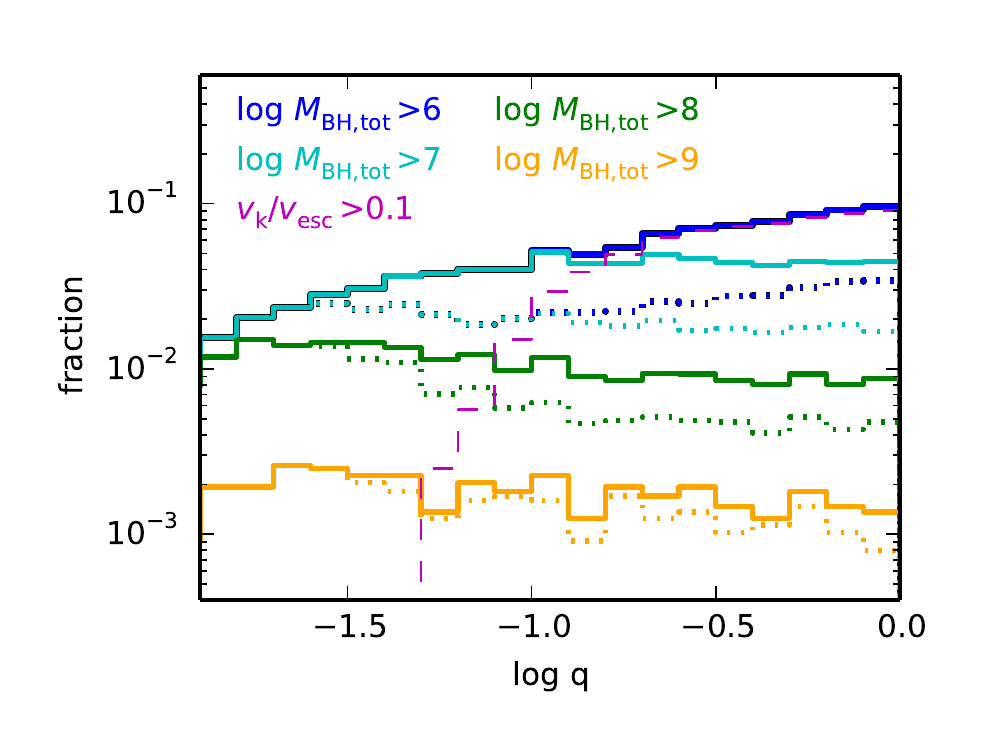}
\caption{The merging BH mass ratio distribution in Illustris, separated into total BH mass bins. As indicated in the plot legend, the blue curve shows the distribution for all BH masses (above the minimum $2 \times 10^6$ \msun, and the cyan, green, and orange curves correspond to remnant BH masses of log $(M_{\rm BH}$/\msun) $>$ 7, 8, and 9, respectively. For most BH masses, the $q$ distribution is nearly flat in log space over the range -2 $<$ log $q$ $<$ 0. The dotted lines show the distributions for only BH mergers occurring at $z < 1$. The thin magenta line denotes mergers that yield a recoil velocity $> 0.1$ \vesc; no mergers with log $q \la -1.3$ produce significant recoils. \label{fig:massratio}}
\end{center}
\end{figure}

\begin{figure*}
\begin{center}
\includegraphics[width=0.33\textwidth,trim=14 18 18 12]{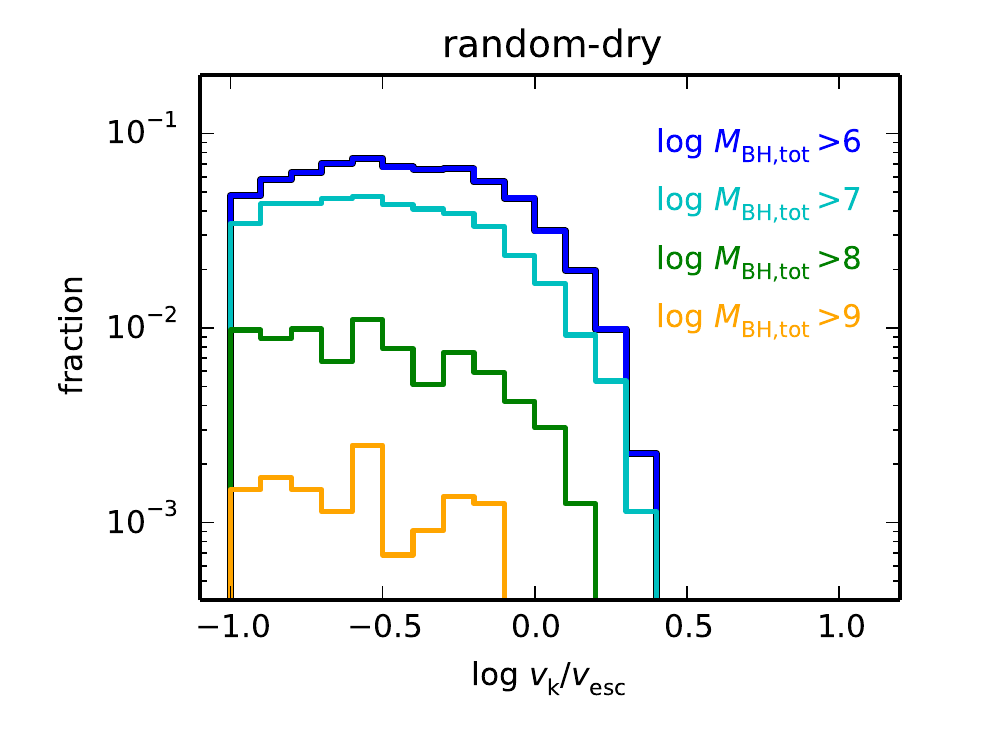}
\includegraphics[width=0.33\textwidth,trim=14 18 18 12]{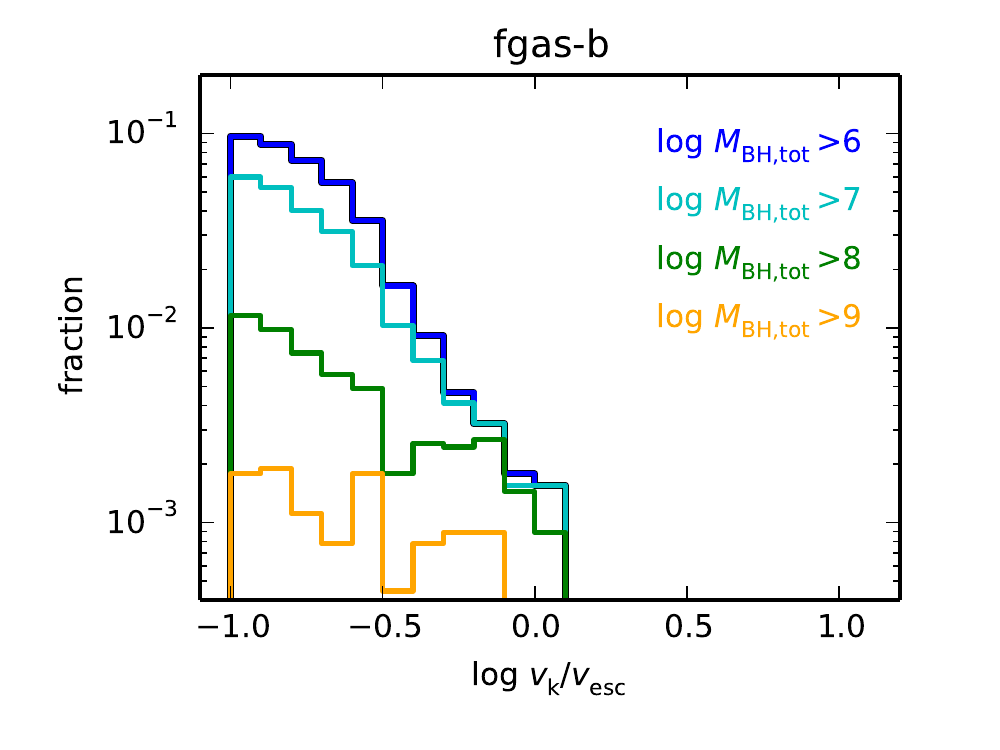}
\includegraphics[width=0.33\textwidth,trim=14 18 18 12]{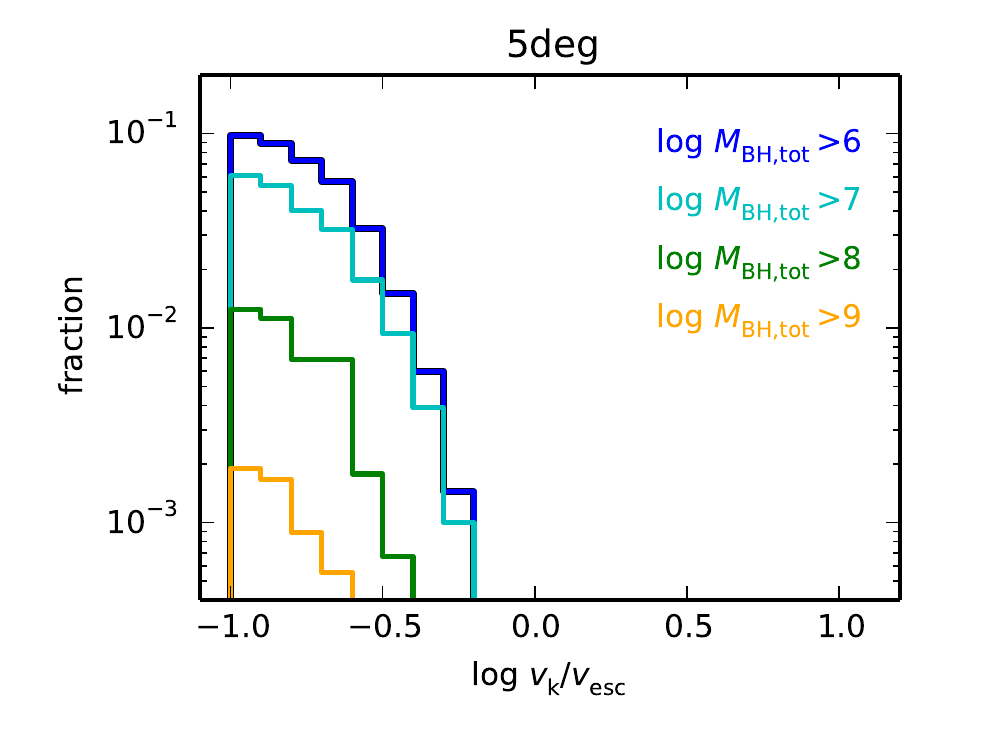}
\caption{The distribution of kick velocities scaled to the host escape speed, \vk/\vesc, is shown for a random spin model (\randdry, left panel), a hybrid model (\fgasb, middle panel), and an aligned model (\fivedeg, right panel). The distributions are separated by total merging BH mass using the same color scheme as in Figure \ref{fig:massratio}, and as indicated in the plot legend. The random model has a large tail of escaping BHs, the hybrid model has only a few, and the aligned model has none.  \label{fig:vkvesc}}
\end{center}
\end{figure*}

Figure \ref{fig:massratio} shows the distribution of merging BH mass ratios $q$, separated by total BH binary mass. For a wide range of BH masses, the log $q$ distribution is nearly flat over the range $-2 <$ log $q < 0$. Low mass ratios are favored slightly for total BH masses $> 10^8$ \msun; these massive BHs are fewer in number, so they more often undergo unequal-mass mergers. The trend reverses when mergers with total mass $M_{\rm BH} = 10^7$ - $10^8$ \msun\ are included, and when all BH masses $> 10^6$ \msun\ are considered, the distribution clearly rises toward log $q=0$. Recall that a lower mass limit of $10^6$ \msun\ has been imposed for individual merging BHs, such that BH binaries with total mass $< 10^7$ \msun\ cannot have log $q < -1$. 

The dotted lines indicate the $q$ distribution for only mergers occurring at $z<1$. We see that major mergers are more common at high redshift. This is partly a result of hierarchical growth; minor mergers become more frequent as larger galaxies (and BHs) form. It is also a product of the long dynamical friction timescales for minor mergers, reflected in the $q$-dependent delay timescales we have imposed on the simulation merger time. The limited box size of Illustris, (106.5 Mpc)$^3$, could further contribute to the limited number of major mergers between low-redshift, massive BHs. For mergers at $z<1$, the distribution is nearly flat or declining with log $q$ for all BH masses. 

Also shown in this figure is the subset of BH mergers that produce a recoil kick with \vk/\vesc\ $> 0.1$ (in the \randdry\ spin model), which is the minimum kick velocity for which recoil trajectories are calculated. Only mergers with log $q \ga -1.5$ produce such recoils, with most requiring log $q > -1$.

\begin{figure*}
\begin{center}
\includegraphics[width=0.32\textwidth]{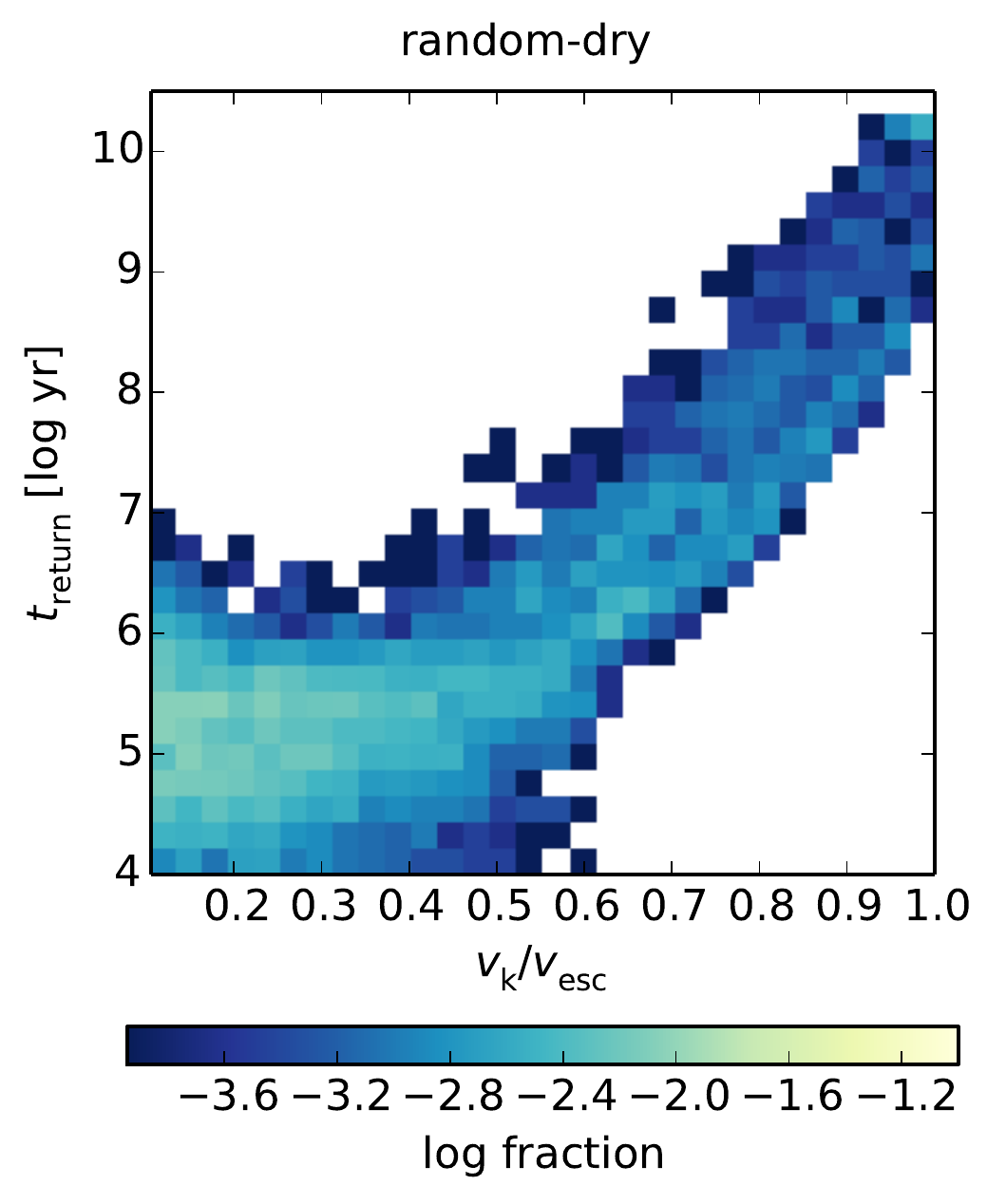}
\includegraphics[width=0.32\textwidth]{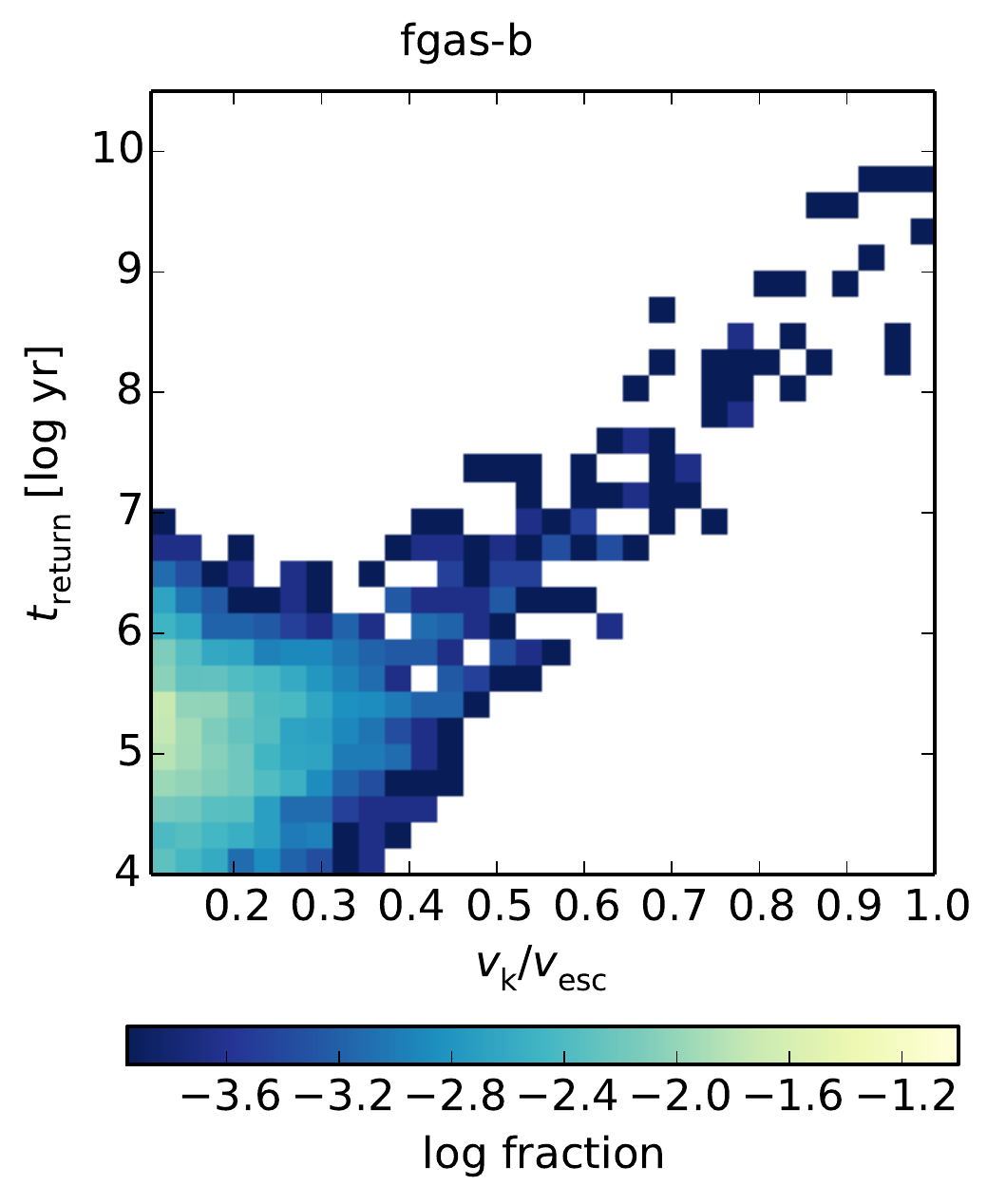}
\includegraphics[width=0.32\textwidth]{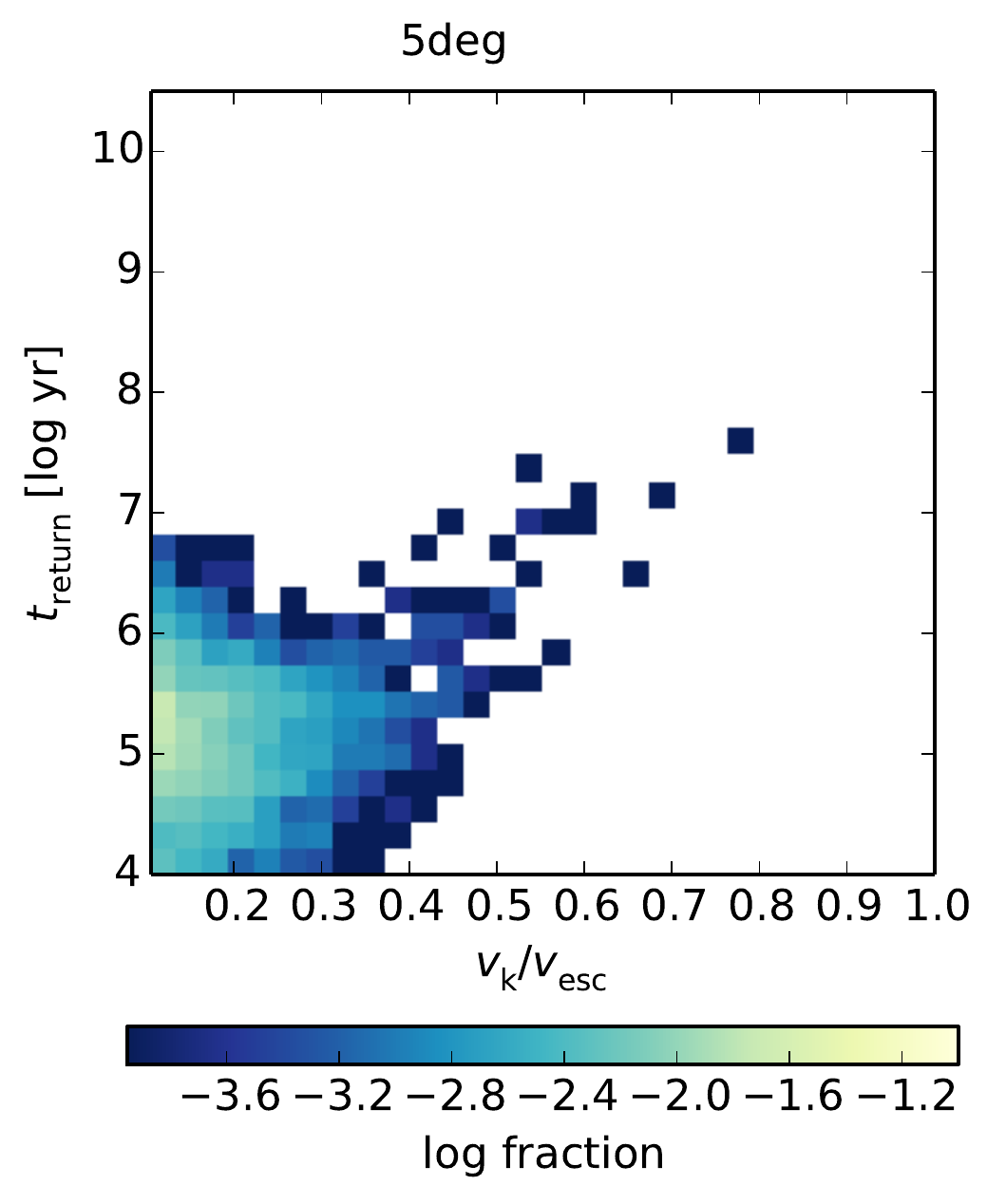}
\includegraphics[width=0.32\textwidth]{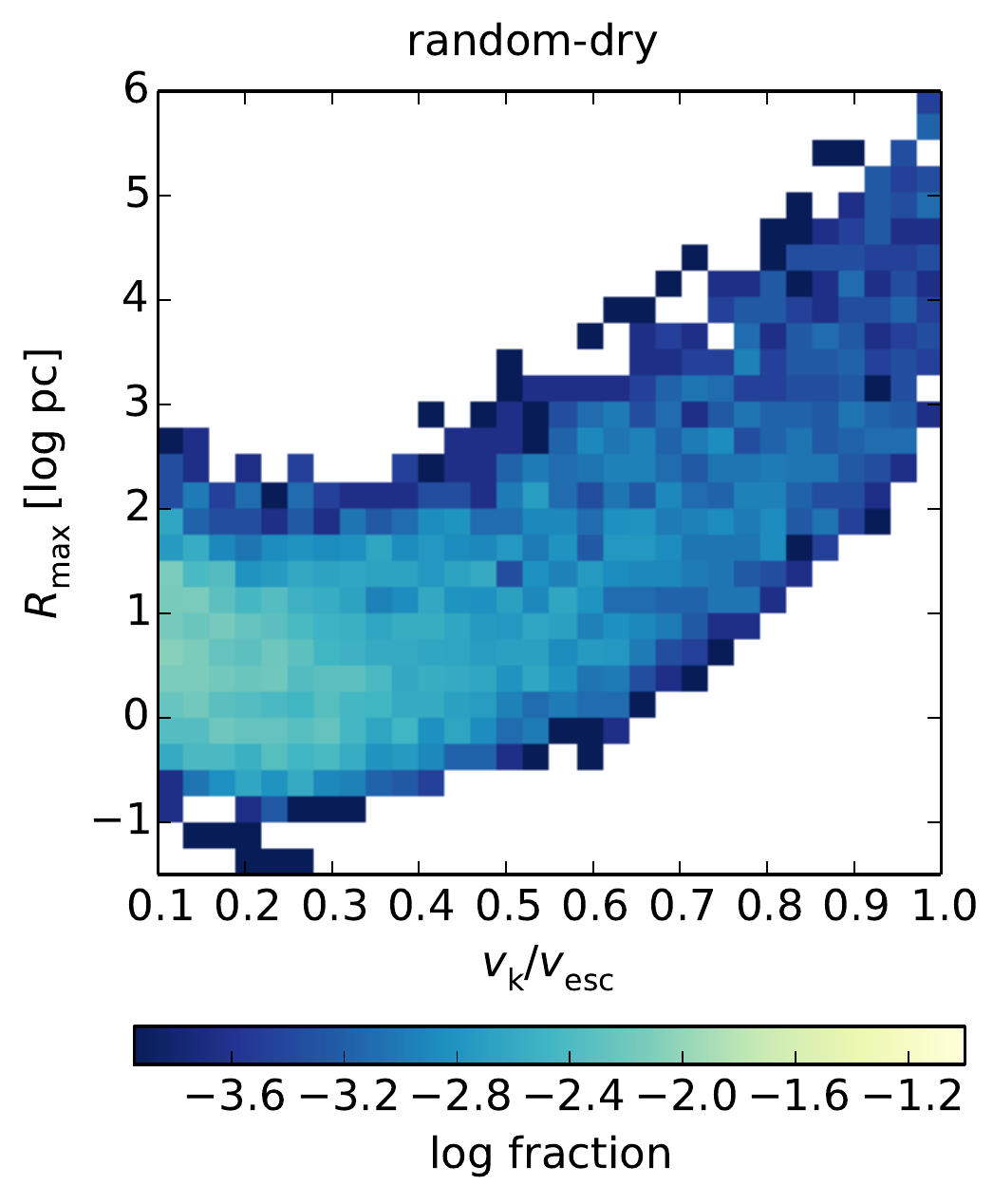}
\includegraphics[width=0.32\textwidth]{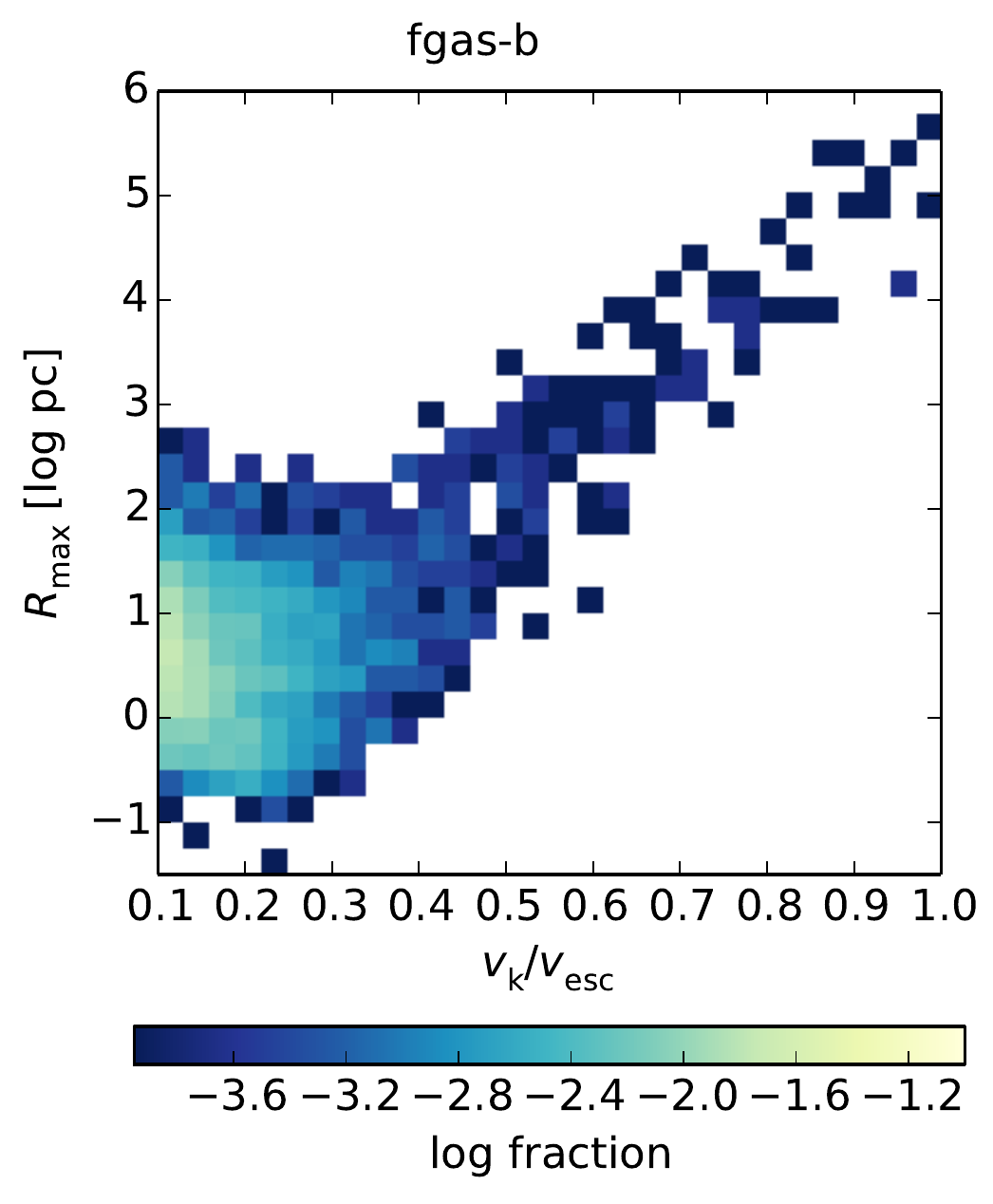}
\includegraphics[width=0.32\textwidth]{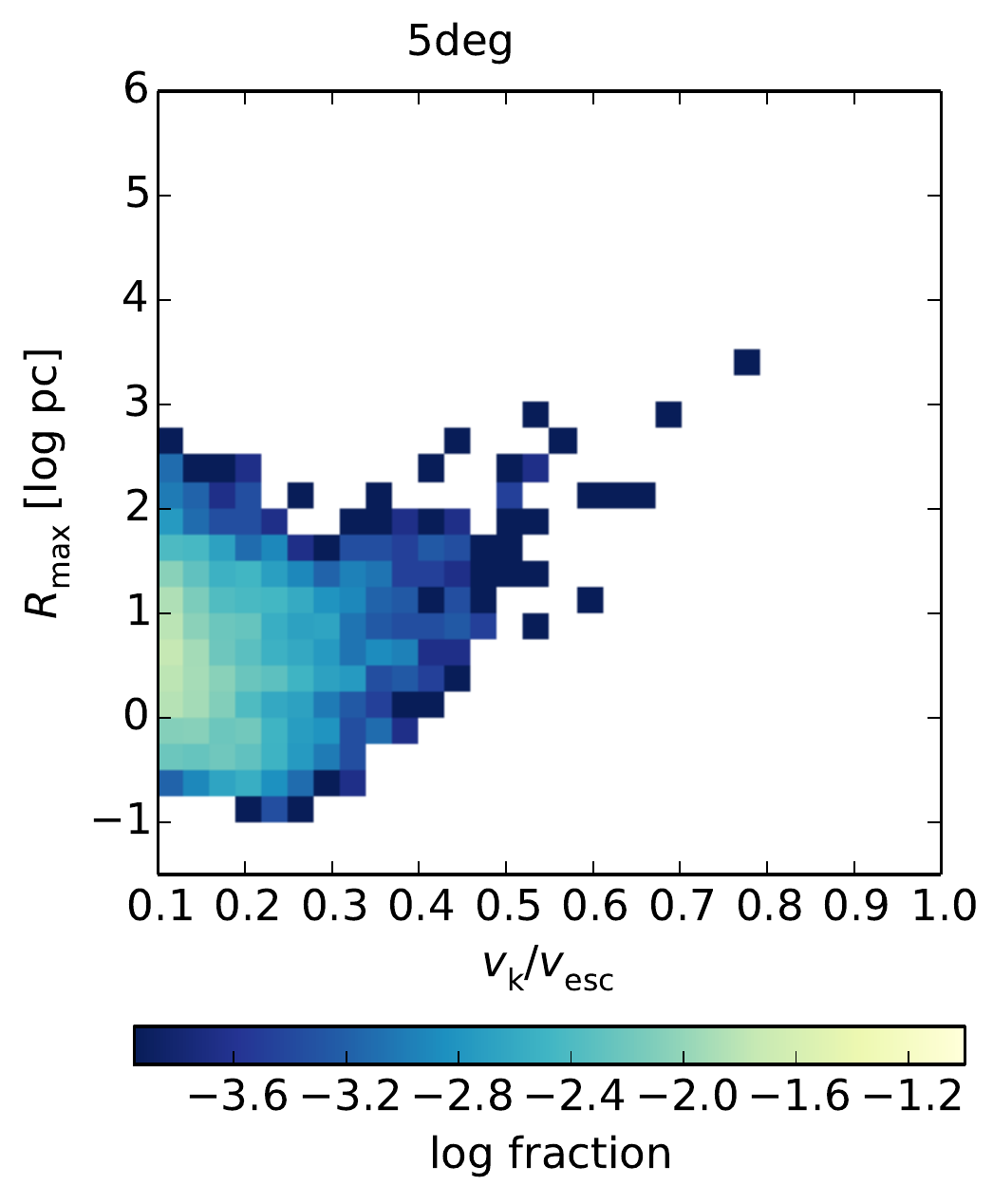}
\caption{The characteristics of bound recoil trajectories (\vk/\vesc\ $< 1$) are shown for three different spin models: \randdry\ ({\em top panels}), \fgasb\ ({\em middle panels}), and \fivedeg\ ({\em bottom panels}). In each case, the {\em left panel} shows the distribution of return times $t_{\rm return}$ for the recoiling BH to settle back to the galactic center, as a function of \vk/\vesc. The {\em right panel} shows the maximum galactocentric distance of the BH orbit, $R_{\rm max}$, versus \vk/\vesc. There is little dependence of either quantity on \vk/\vesc\ for values $\la 0.5$, but marginally bound recoils can wander in the halo for many Gyr. Such events are much rarer in the hybrid model than the random model, and almost never occur in the aligned spin model.  \label{fig:rmaxtreturn}}
\end{center}
\end{figure*}

We note that the merging BH mass ratio distribution in Illustris is qualitatively different from that in the models of VM08, which exhibits a sharp cutoff at $q \ga 0.3$. Because the recoil kick distribution depends sensitively on BH mass ratios, this is an important distinction. While there are uncertainties in the BH merger timescales in Illustris, as discussed above, they are largest for {\em minor} mergers. A more detailed comparison of our results with those of VM08 is given in Section \ref{ssec:prevwork}. We additionally note that recent numerical simulations of BH binaries in circumnuclear gas disks suggest that accretion onto the lower-mass BH may be more rapid, driving the final mass ratios closer to unity than the progenitor distribution \citep{roedig12,farris14}.

\subsection{Recoiling BH dynamics}
\label{ssec:traj}

Figure~\ref{fig:vkvesc} shows the distribution of kick velocities scaled to the host escape speed, for a random spin model (\randdry), a hybrid model (\fgasb), and an aligned model (\fivedeg). For the random spin models, the distribution peaks at \vk/\vesc\ $\sim 0.3$, unlike the hybrid and aligned models, for which the distribution is heavily skewed toward low \vk/\vesc. In the \fivedeg\ model, recoiling BHs almost never escape the galaxy entirely. (No recoils have \vk/\vesc\ $> 1$ in the \fivedeg\ model, but this is limited by mass resolution and by statistics. The largest kicks produced in the \fivedeg\ model could escape the lowest-mass galaxies in our sample, and would easily escape dwarf galaxies unresolved in the simulation.) Escaping BHs are also very rare in the \cold\ spin model, comprising $< 0.05\%$ of all kicks. The escape fractions in the hybrid spin models are 0.4\% and 0.2\% for \fgasa\ and \fgasb, respectively, comparable to the \hot\ spin model (0.6\% escaping). In contrast, the random spin models eject a full 6\% (\randdry) and 12\% (\randhigh) of all merged BHs.

The colored lines in Figure~\ref{fig:vkvesc} show the distribution of \vk/\vesc\ separated by total merging BH mass. Because BH mass correlates with galaxy mass and \vk\ depends only on the BH mass {\em ratio}, large \vk/\vesc\ events are rarer for high-mass BHs. Despite this, in the random {\em and} the hybrid spin models, even the most massive BHs ($> 10^9$ \msun) occasionally escape their hosts. Thus, if superkicks do occur, even massive hosts may be left without a central BH.  

For bound recoiling BHs (\vk/\vesc\ $< 1$), trajectories can be characterized by the maximum apocentric distance from the host ($R_{\rm max}$) and the time spent off-center before returning to the galactic nucleus ($t_{\rm return}$). These quantities are plotted versus \vk/\vesc\ for selected spin models in Figure~\ref{fig:rmaxtreturn}. Naturally, both $R_{\rm max}$ and $t_{\rm return}$  increase with \vk/\vesc, but the scatter is large, owing to the large dynamic range in halo, stellar, and gas model parameters. There is little to no correlation below \vk/\vesc $\sim 0.5$, where recoil trajectories are strongly suppressed by the dense stellar and gas components. These low-velocity recoiling BHs are generally confined to the central kpc of the galaxy, with return time $< 10^7$ yr, and some are displaced by less than a parsec. In contrast, large kicks with \vk/\vesc\ $\ga 0.8$ can displace the BH for $> 1$ Gyr, with $R_{\rm max} \ga 100$ kpc. 

Pre-merger spin alignment greatly suppresses high-velocity recoils. The hybrid \fgasb\ spin model produces a much smaller fraction of recoils with \vk/\vesc\ $> 0.6$ than the random spin model, and the \fivedeg\ model yields no kicks above \vk/\vesc\ $=0.8$. In the latter model, with extremely efficient spin alignment, recoiling BHs are nearly always confined to the central kpc of the galaxy (though again, this does not apply to dwarf galaxies that are below the resolution limit of the simulation).

\subsection{Merging and recoiling AGN properties}
\label{ssec:recoilingagn}

The AGN Eddington ratios ($f_{\rm Edd} \equiv L_{\rm AGN}/L_{\rm Edd}$) in Illustris evolve strongly toward lower values with cosmic time \citep[][Figure 9]{sijack15}. At $z=4$, when BHs have low masses and their hosts are generally gas-rich, the $f_{\rm Edd}$ distribution peaks near unity. At $z=0$, Eddington-limited AGN are very rare, and the median $f_{\rm Edd}$ for all BHs is $3.7\times 10^{-4}$. 

\begin{figure}
\centering
\includegraphics[width=0.495\textwidth,trim=0 12 0 0]{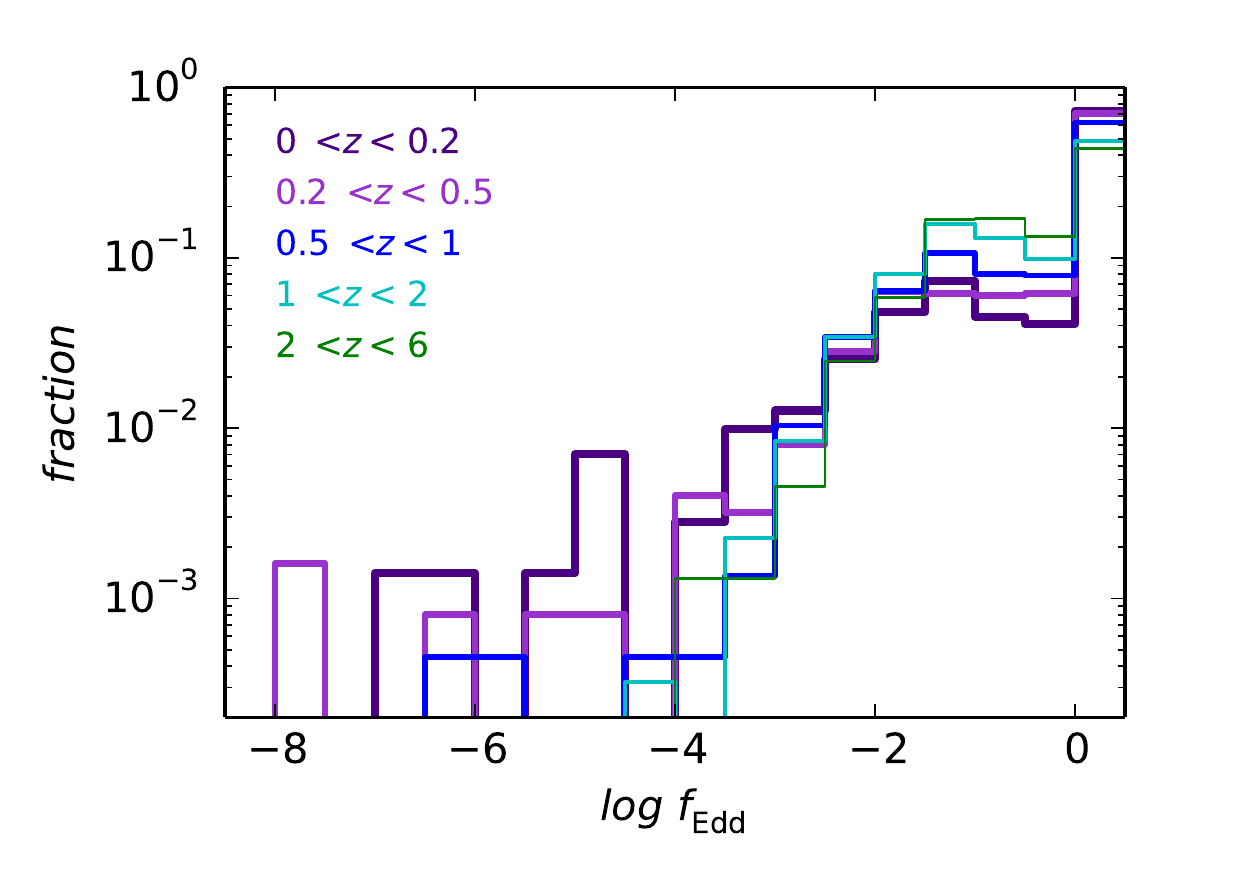}
\includegraphics[width=0.495\textwidth,trim=0 12 0 0]{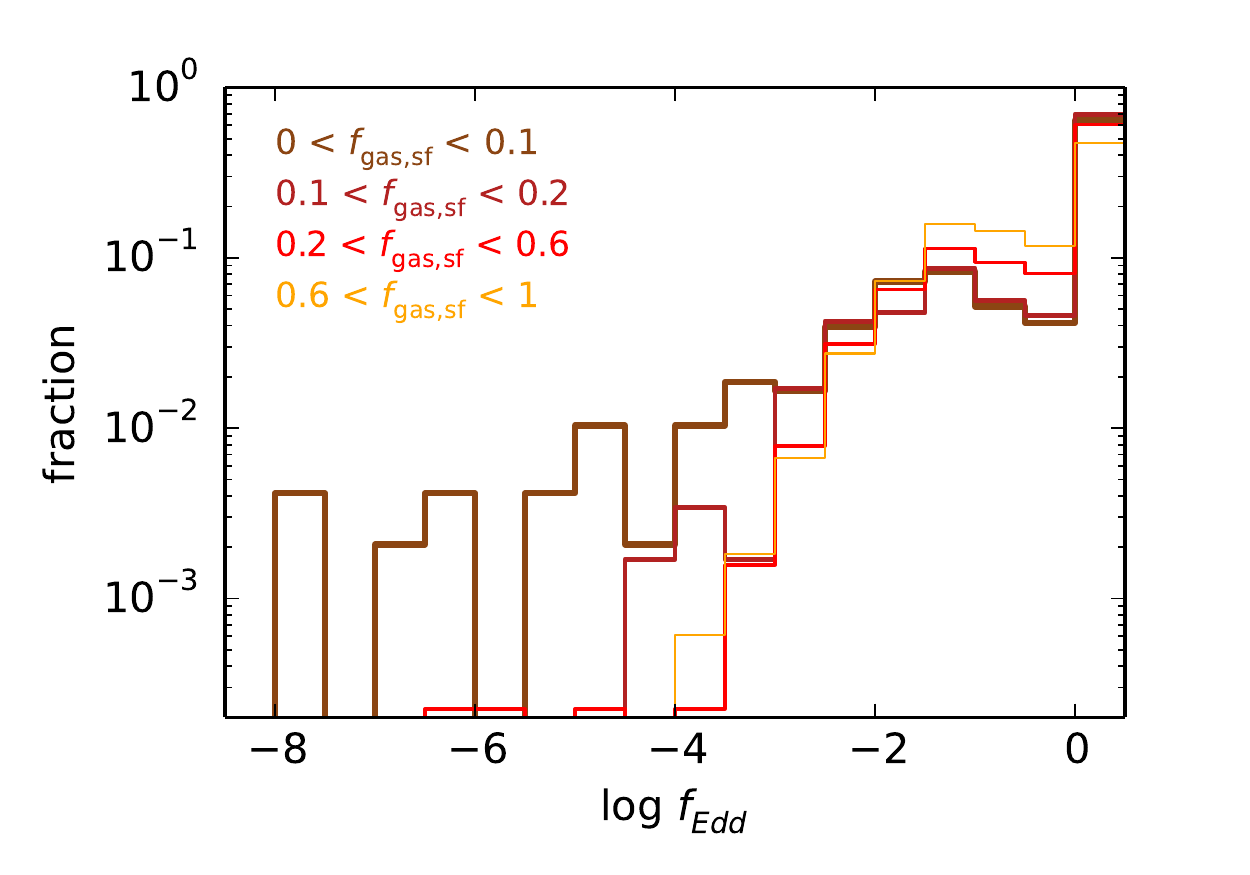}
\caption{Top panel: the distribution of Eddington ratios ($f_{\rm Edd} \equiv L_{\rm AGN}/L_{\rm Edd}$) is shown for merging BHs in Illustris, separated by redshift bin according to the plot legend (thicker lines correspond to lower $z$). Bottom panel: same distribution, but separated by the mass fraction (relative to stars) of cold, star-forming gas in the host (thicker lines correspond to lower $f_{\rm gas,sf}$). Unlike the $f_{\rm Edd}$ distribution for {\em all} BHs \citep[cf.][]{sijack15}, the distribution for merging BHs heavily favors values near unity at all redshifts and for a wide range in gas fractions. \label{fig:fedd}}
\end{figure}

\begin{figure}
\centering
\includegraphics[width=0.495\textwidth,trim=6 12 0 0]{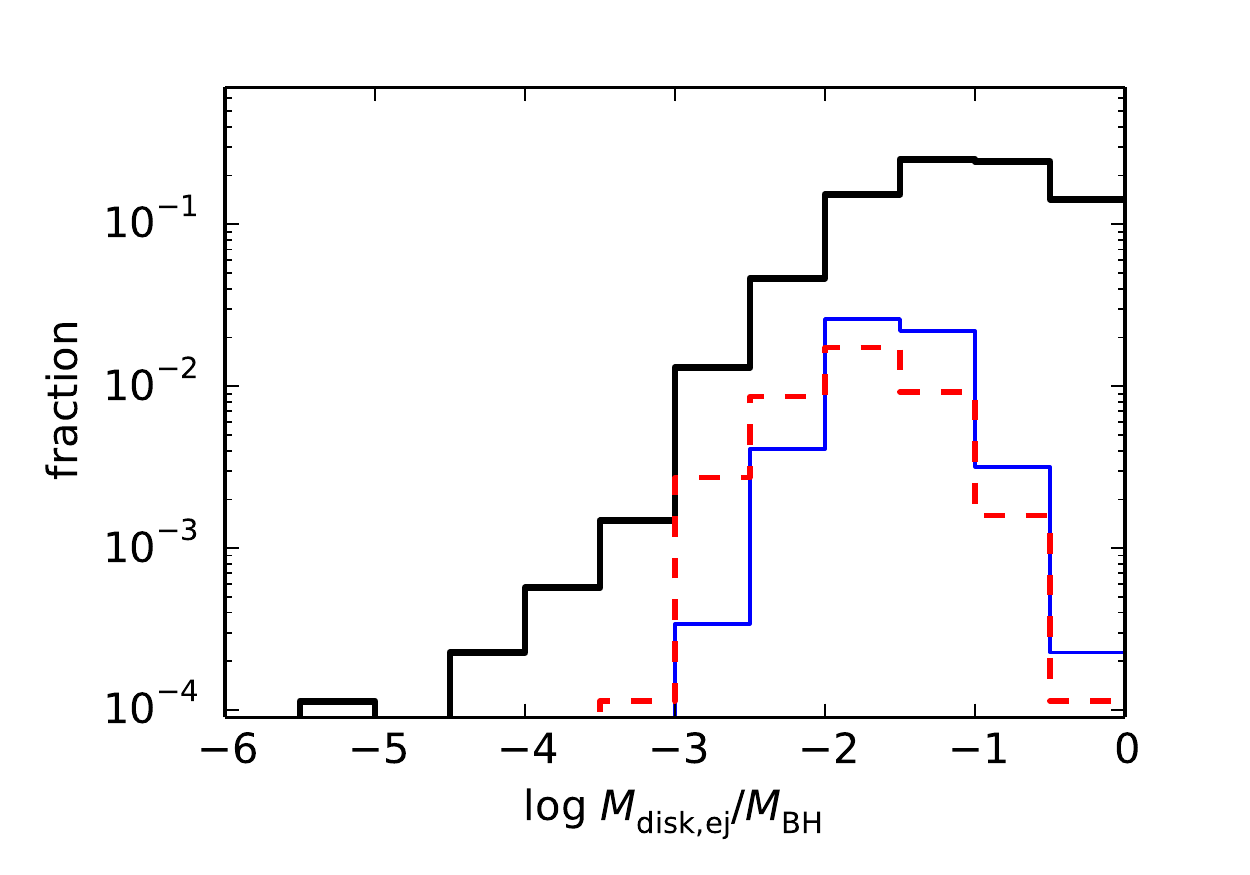}
\caption{The mass distribution of the accretion disk bound to the recoiling BH, scaled to BH mass. The thick black line shows the distribution for all mergers, while the thin lines denote the distributions for only recoil events that produce observable spatially-offset (solid blue line) and velocity-offset (dashed red) AGN. For the offset AGN, the \randdry\ spin model and \hcos\ sensitivity \citep[AB(F814W) $=$ 27.2 mag for point sources,][]{koekem07}, resolution (0.1"), and $\Delta v_{\rm LOS} > 600$ \kms\ are assumed, and a minimum offset lifetime of $10^5$ yr is imposed. Typical disk masses are a few percent of the BH mass. \label{fig:fmdisk}}
\end{figure}

\begin{figure}
\centering
\includegraphics[width=0.236\textwidth,trim=6 0 2 0]{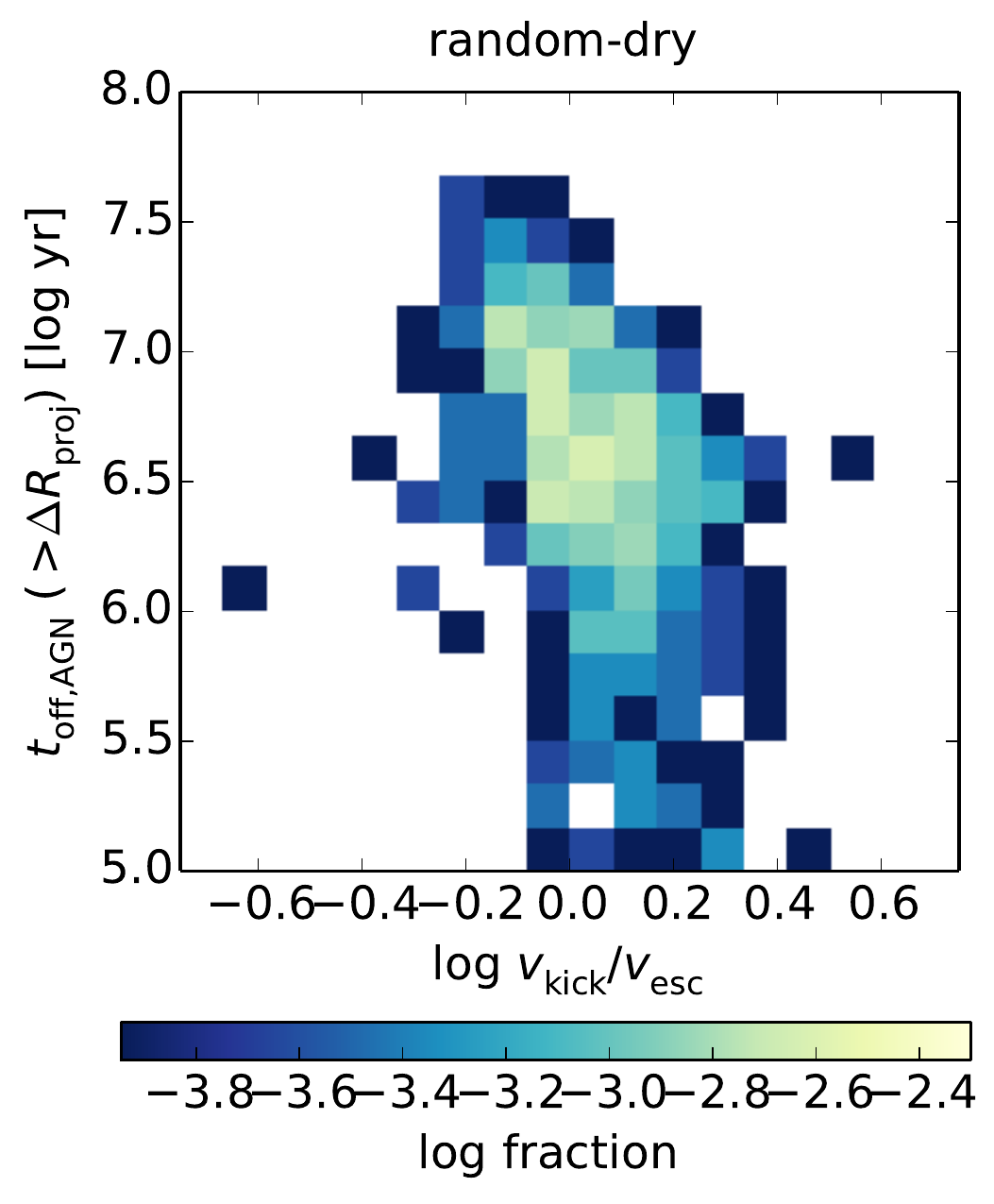}
\includegraphics[width=0.236\textwidth,trim=6 0 2 0]{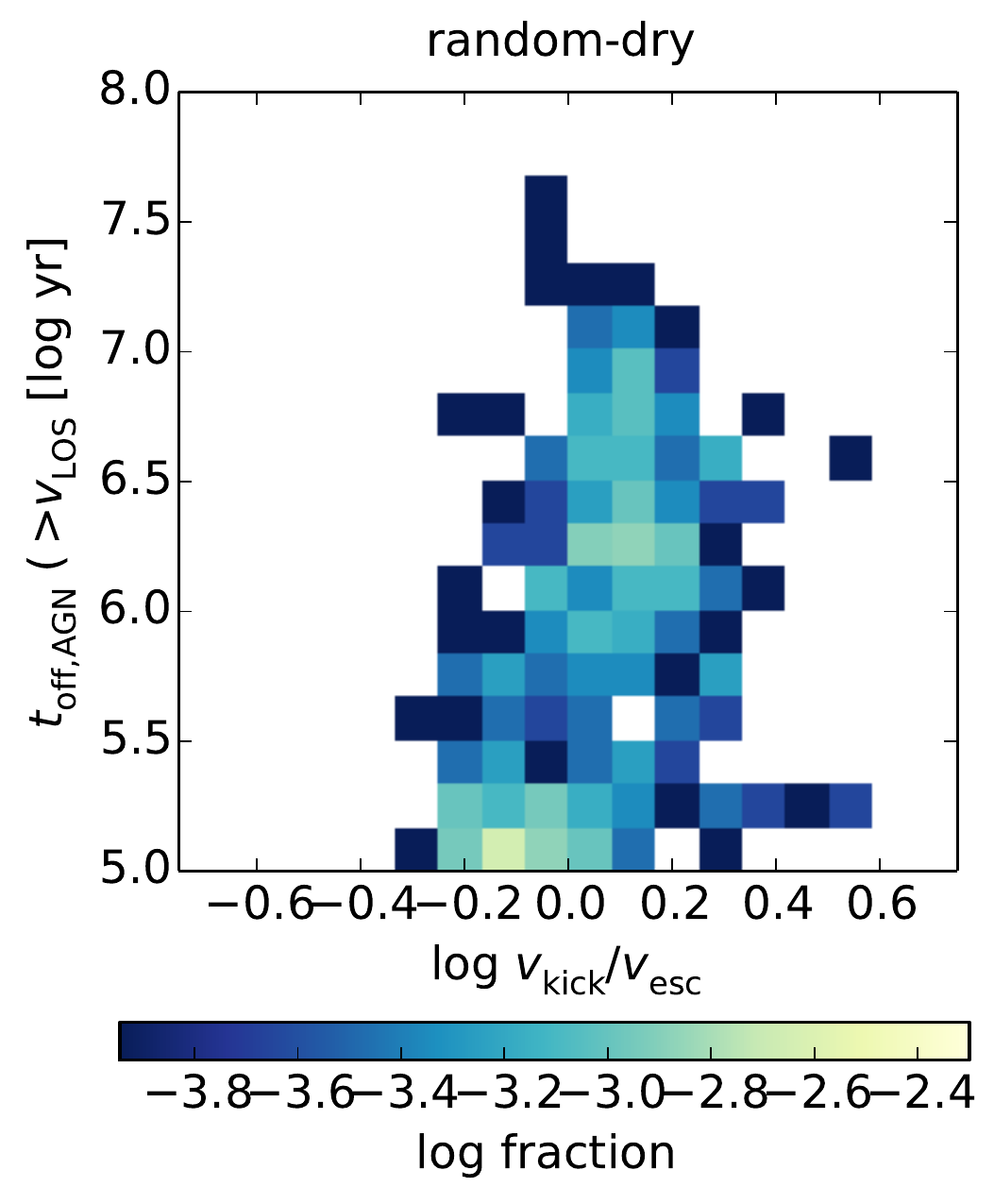}
\includegraphics[width=0.236\textwidth,trim=6 0 2 0]{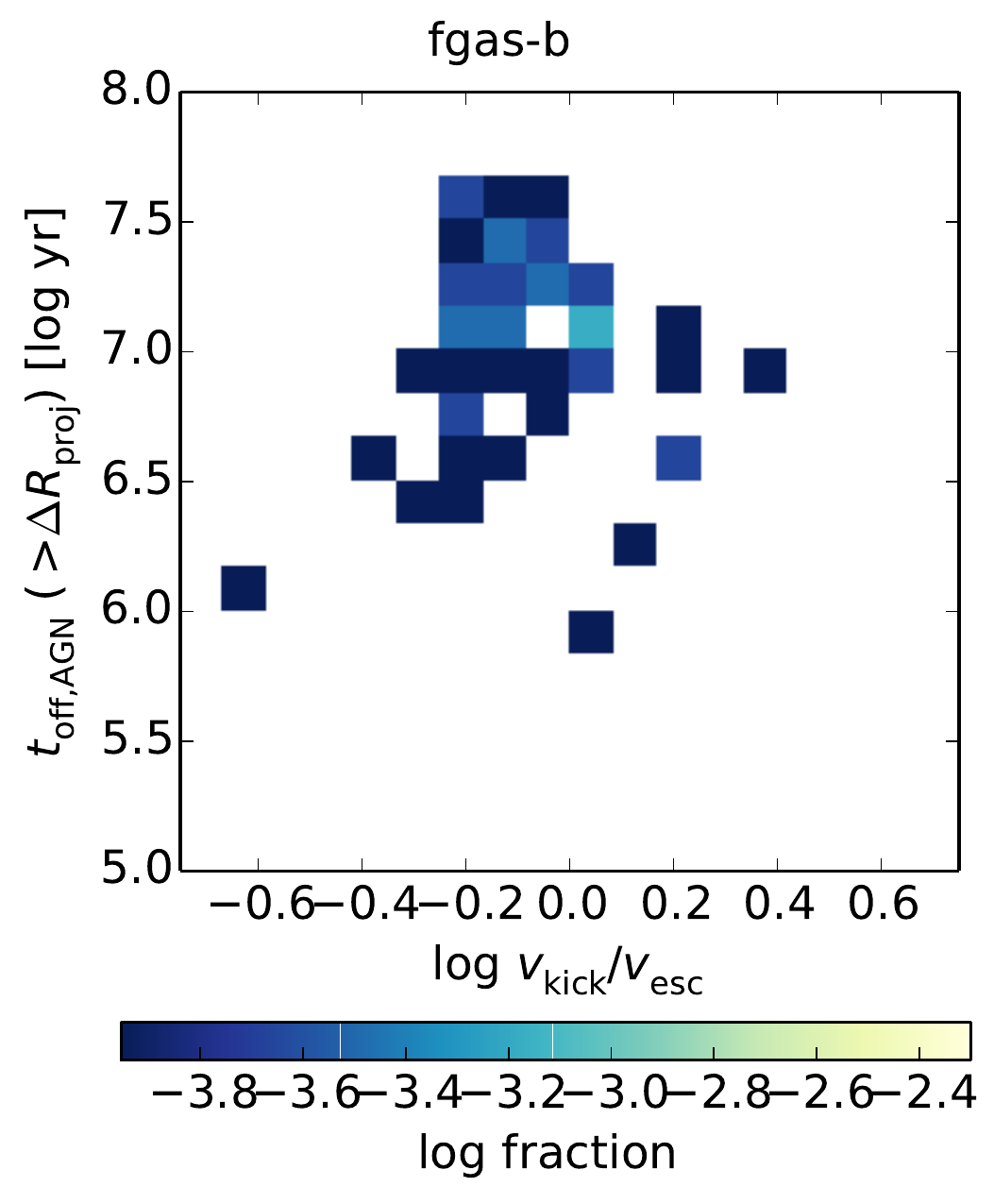}
\includegraphics[width=0.236\textwidth,trim=6 0 2 0]{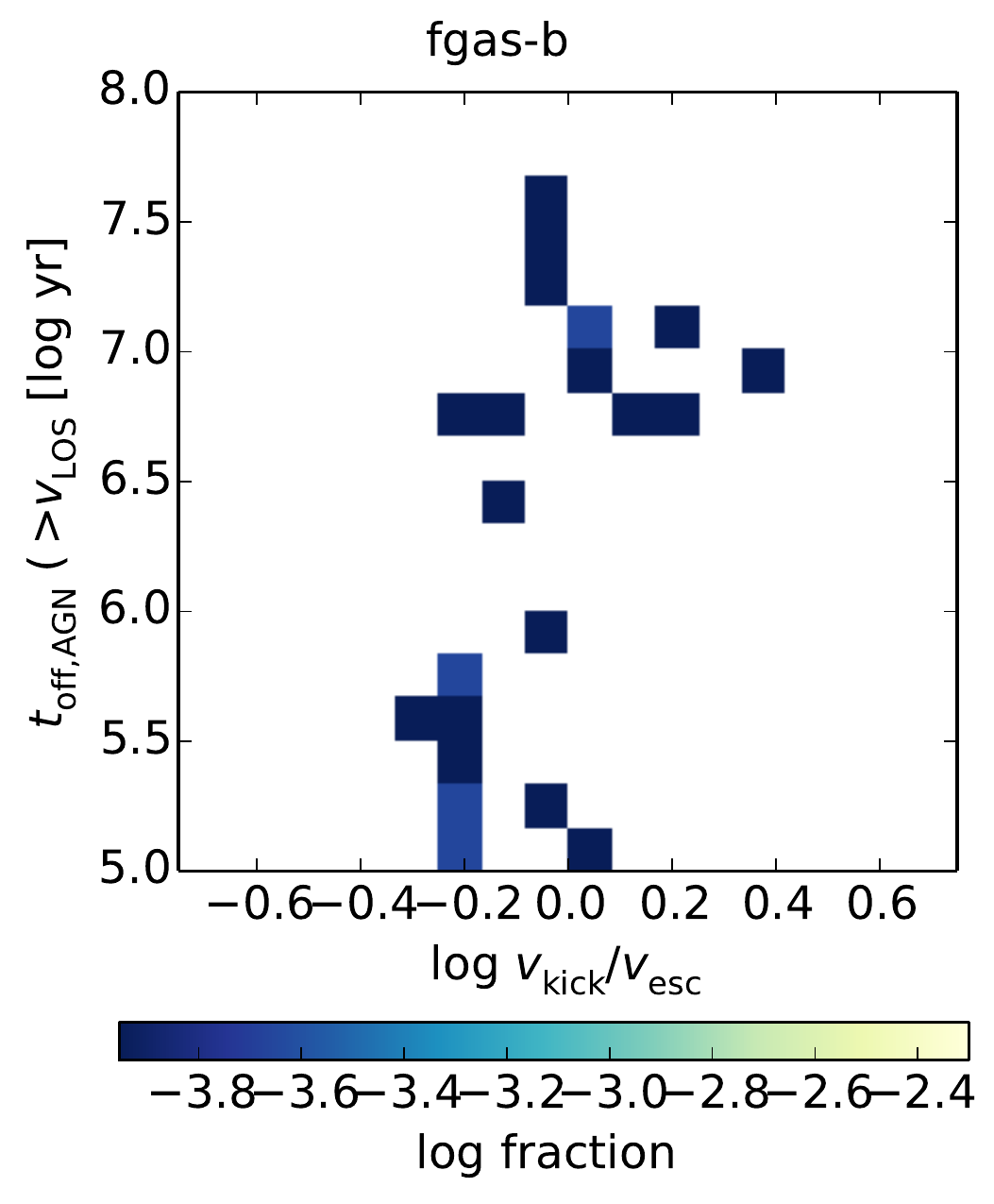}
\caption{The distributions of observable offset AGN lifetime versus \vk/\vesc\ are shown for resolvable projected spatial offsets (left panels) and for LOS velocity offsets (right panels), weighted by the total number of recoil events.  \hcos\ sensitivity and resolution, and a minimum $\Delta v_{\rm LOS} > 600$ \kms, are assumed. The top panels correspond to the \randdry\ spin model, which has 8800 total recoil events, 492 of which yield an observable spatially-offset AGN, and 413 of which produce an observable velocity-offset AGN. The bottom panels correspond to the \fgasb\ model, which has 8989 total events, 54 (30) of which produce observable spatial (velocity) offsets. In all cases, the longest lifetimes occur for kick speeds near the host escape speed. \label{fig:fesc_toff}}
\end{figure}

The Eddington ratios for {\em merging} BHs are quite different. We find that the $f_{\rm Edd}$ distribution is heavily skewed toward unity for merging BHs at all redshifts. Figure \ref{fig:fedd} (top panel) shows the $f_{\rm Edd}$ distribution in several redshift bins, for all merging BHs that meet our minimum-mass criteria. At low redshifts, a tail extends to very low Eddington ratios ($f_{\rm Edd} < 10^{-5}$), but more than half of all merging BHs have accretion rates capped at the Eddington limit at the time of merger. The weak evolution of $f_{\rm Edd}$ with redshift relative to the overall BH population suggests that merger dynamics, rather than cosmic epoch, determine the fueling rate of these AGN. 

The bottom panel of Figure \ref{fig:fedd} shows the $f_{\rm Edd}$ distribution separated by the star-forming gas fraction in the host, illustrating that higher-$f_{\rm gas}$ galaxies have higher Eddington ratios on average, but even mergers with $f_{\rm gas,sf} < 0.1$ have an $f_{\rm Edd}$ distribution skewed toward unity. This indicates that the cold gas in these galaxies is efficiently funneled to the central regions during major mergers. The ability of major mergers to trigger AGN fueling is well-studied in simulations of isolated galaxy mergers \citep[e.g.,][]{wyiloe03,dimatt05,hopkin06a,hopkin08c}, and a clear correlation between merging activity and AGN has been observed in real systems \citep[e.g.,][]{sander88a,urruti08,koss10,elliso11,satyap14,treist12}. Our results demonstrate that this effect is self-consistently produced in the Illustris cosmological simulations, and that a strong correlation exists during the late stages of the galaxy merger. In cases where a significant delay occurs between the BH merger time in the simulation and the actual BH merger on sub-grid scales, the accretion rate may be lower at the time of the merger and recoil, but the longest delays occur for minor mergers that are less likely to produce observable recoils. Once the BH is ejected from the galaxy, however, its accretion rate declines with time, so the recoiling AGN have $L \sim L_{\rm Edd}$ only briefly.

Note that even though the recoiling BH accretion rate in our models declines monotonically after the recoil kick, the total AGN lifetime is generally {\em longer} than if a constant accretion rate (and corresponding AGN lifetime $M_{\rm disk}/\dot M_{\rm BH}$) is assumed. For the minimum observable value of $f_{\rm Edd} = 10^{-2}$ assumed in our models, the AGN lifetime may be nearly 10 times longer than a constant $M_{\rm disk}/\dot M_{\rm BH}$ lifetime.

As described in Section \ref{ssec:model}, recoiling BHs should carry along material that orbits the BH with $v_{\rm orb} \ga $\vk. The amount of mass bound to the recoiling BH, along with the accretion rate at the time of the kick, are used to determine its AGN lifetime. We calculate the mass and radius of this bound material assuming it is composed of an alpha-disk that becomes self-gravitating beyond the radius where the Toomre $Q=1$. The resulting ejected disk masses for recoils with \vk/\vesc\ $> 0.1$ are shown in Figure \ref{fig:fmdisk}, for the \randdry\ spin model. The overall distribution of disk masses peaks at $\sim 0.1 M_{\rm BH}$ and has a tail extending to $< 10^{-5} M_{\rm BH}$. Among observable offset AGN, very few have $M_{\rm disk} \sim M_{\rm BH}$, because these correspond to low-velocity recoils; this is true regardless of the degree of spin alignment. The disk masses for observable velocity-offset AGN are slightly smaller than for spatially-offset AGN, because the minimum velocity offset requirement means that the former have higher kick speeds on average. In the random spin models, most offset AGN have $M_{\rm disk}/M_{\rm BH} =$ 1 - 10\%. In the hybrid and aligned models (with lower kick velocities), typical disk masses are slightly higher, $\sim 3$ - 30\% $M_{\rm BH}$.

Figure~\ref{fig:fesc_toff} shows the relationship between \vk/\vesc\ and the spatially- or velocity-offset AGN lifetime. Here and throughout our analysis, a random viewing angle is assigned to each recoil event, and the offset lifetime is calculated as the total time for which the recoiling AGN is detectable and has a resolvable projected spatial offset or a resolvable line-of-sight (LOS) velocity offset. The \randdry\ and \fgasb\ spin models are shown, assuming \hcos\ sensitivity \citep[AB(F814W) $=$ 27.2 mag for point sources,][]{koekem07}, resolution (0.1"), and $\Delta v_{\rm LOS} > 600$ \kms. 

The spatially-offset AGN lifetimes are longest for recoils at or somewhat below the escape speed (\vk/\vesc\ $\sim$ 0.6 - 1). This is because marginally-bound BHs experience little deceleration and spend most of their time at large apocenters, well separated from the host centroid. Escaping BHs also produce large offsets, of course, but this is countered by the fact that the accretion disk mass carried with the BH decreases $\propto$ \vk$^2$, such that AGN lifetimes are vanishingly small for \vk\ $\gg$ \vesc. Such extreme kicks are also rare. At high redshift, escaping BHs do begin to dominate the population of observable offset AGN, owing to both angular resolution limits and the suppression of low-velocity recoils in gas-rich mergers. The fact that offset AGN lifetimes are maximized for \vk$ \la $\vesc\ has important consequences for the observability of recoils, foremost that spatial resolution should not be a strongly limiting factor in systematic searches for recoiling AGN. 

Similarly, velocity-offset AGN have the longest lifetimes for recoils near the escape speed (Figure \ref{fig:fesc_toff}, right panels). On average, velocity-offset AGN have shorter lifetimes than spatially offset AGN, because many recoil events have \vk\ $> 600$ \kms\ initially but are quickly decelerated in the host potential. This is especially true at high redshift, where gas-rich mergers tend to suppress long-lived recoils. 

These trends are qualitatively similar for models that include pre-merger BH spin alignment, except that there are fewer offset AGN overall, and the high \vk/\vesc\ tail of events is mostly truncated (Figure \ref{fig:fesc_toff}, bottom panels). Recoil events in gas-rich mergers that produce short-lived velocity-offset AGN in the \randdry\ model are mostly suppressed in the hybrid model. 

\begin{figure*}
\centering
\includegraphics[width=0.28\textwidth]{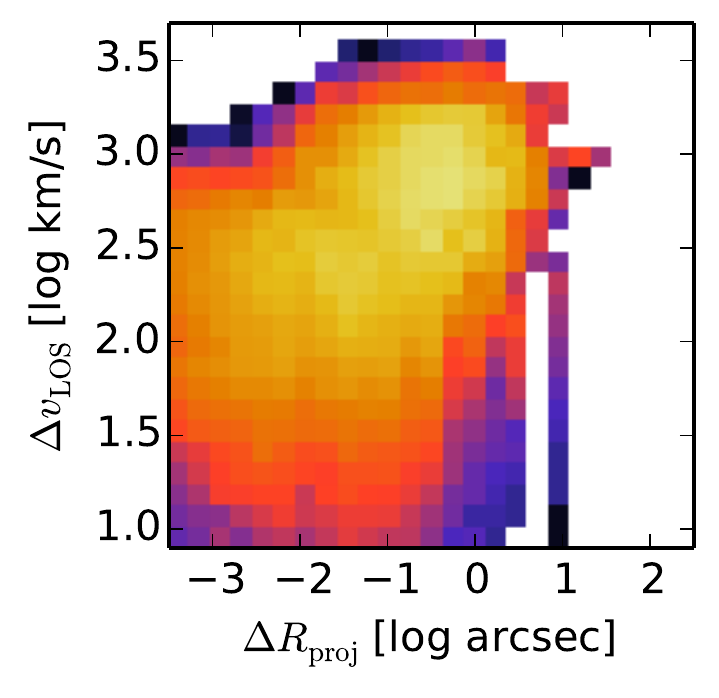}
\includegraphics[width=0.28\textwidth]{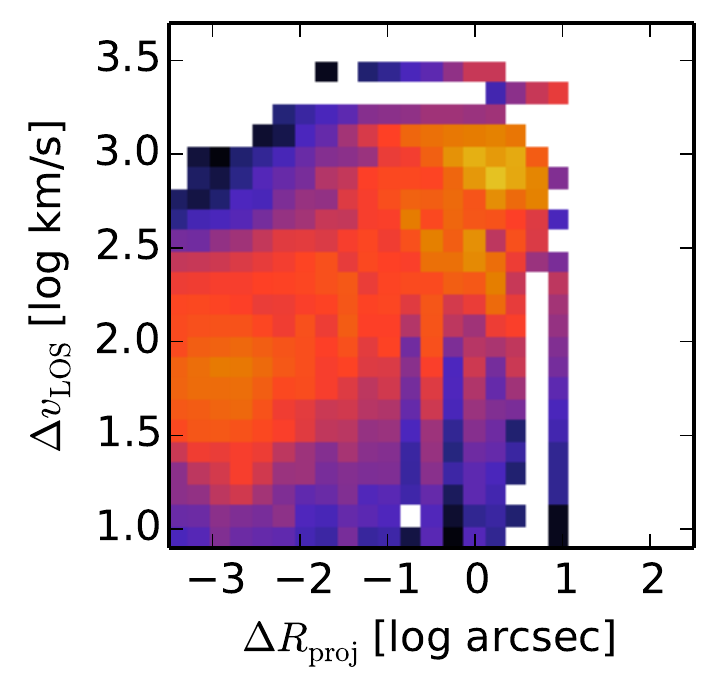}
\includegraphics[width=0.39\textwidth,trim = 6 0 0 6]{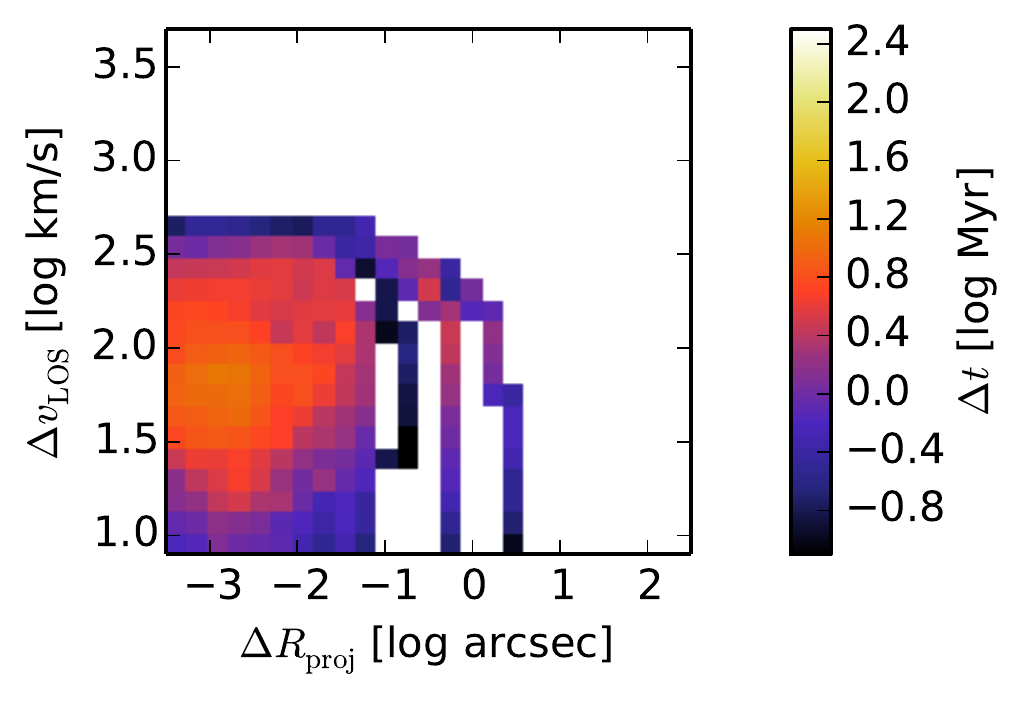}
\includegraphics[width=0.29\textwidth]{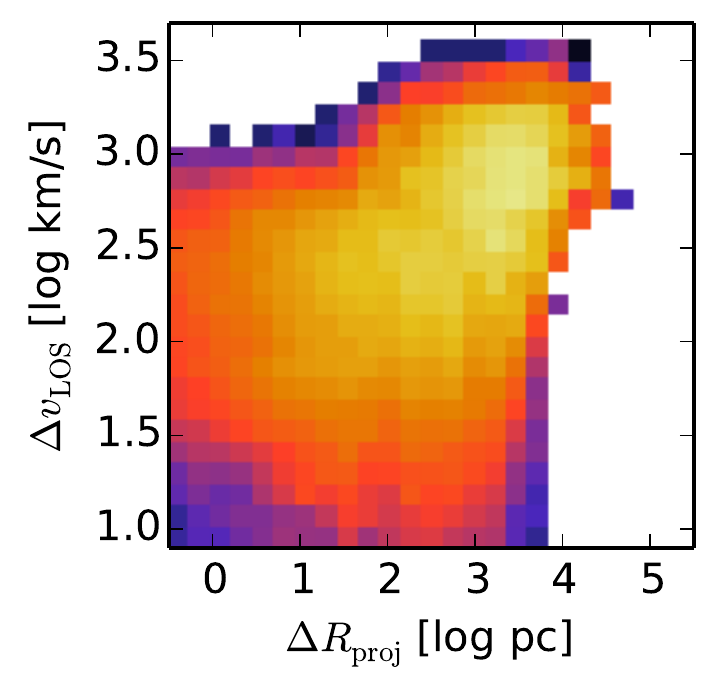}
\includegraphics[width=0.28\textwidth]{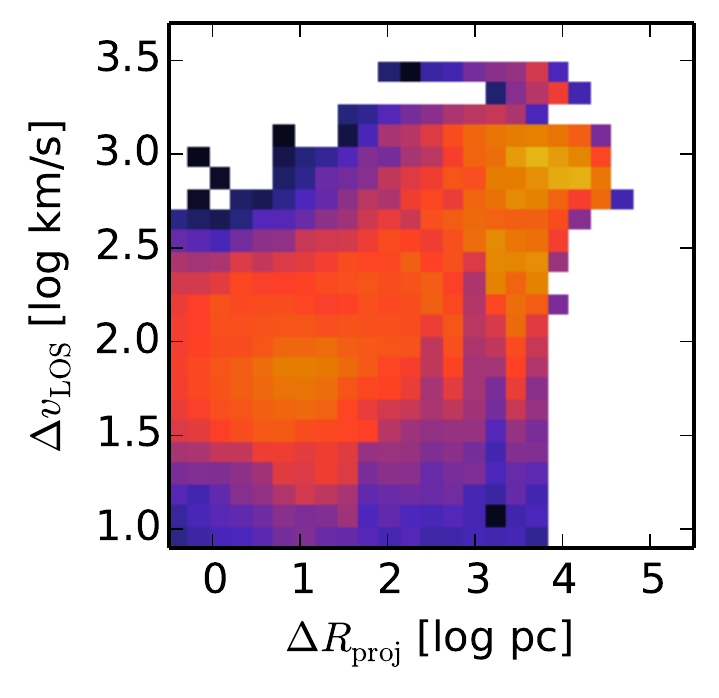}
\includegraphics[width=0.39\textwidth,trim = 6 0 0 6]{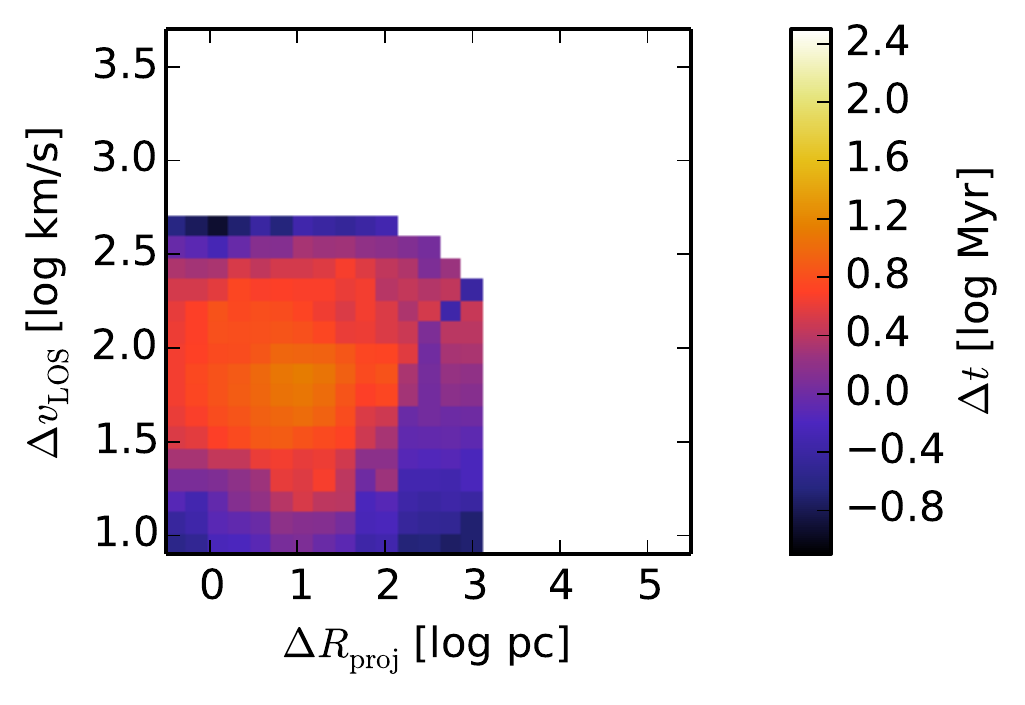}
\caption{The time-weighted distributions of projected spatial offset versus LOS velocity offset are shown for all recoiling AGN. The color scale indicates the amount of time a BH spends at a given ($\Delta R_{\rm proj}, \Delta v_{\rm LOS}$) while active as an AGN and offset from the host nucleus, integrated over all merger events over cosmic time, {\em regardless} of whether the recoil event produces a resolvable offset. The sensitivity of \hcos\ is assumed. The {\em top panels} show the distribution of angular offsets, and the {\em bottom panels} show the physical spatial offsets. The {\em left panels} correspond to the \randdry\ spin model, {\em middle panels} show the \fgasb\ model, and {\em right panels} show the \fivedeg\ model. Large offsets $> 1$ kpc and $> 1000$ \kms\ are {\em favored} in the random spin model but do not occur at all if BH spins are nearly aligned. \label{fig:drdvallAGN}}
\end{figure*}

\subsection{Spatial and velocity offsets}
\label{ssec:offsets}

Figure \ref{fig:drdvallAGN} shows time-weighted distributions of projected spatial ($\Delta R_{\rm proj}$) and velocity ($\Delta v_{\rm LOS}$) offsets for {\em all} recoiling AGN in the simulation (i.e., not only those with resolvable offsets are shown). The panels compare the distributions for the \randdry, \fgasb, and \fivedeg\ spin models. For the \randdry\ model, the distribution is dominated by recoiling AGN with velocities above a few hundred \kms\ and spatial separations $\ga 0.1"$ (physical separations $\ga$ 1 kpc are common). This follows from the result discussed above: recoils with \vk\ $\la$ \vesc\ have the longest offset lifetimes. The largest offsets ($>1000$ \kms\ and $> 10$ kpc) result primarily from BHs that escape their host galaxies, and the angular offset distribution has a tail that extends beyond 10". This demonstrates that {\em seeing-limited surveys could resolve spatially-offset AGN.}

The right panel of Figure \ref{fig:drdvallAGN} demonstrates that if spins are always nearly aligned, spatial offsets $> 0.1"$ are rare, and resolvable velocity offsets ($> 600$ \kms) never occur in the \fivedeg\ spin model. The confirmation of a single recoiling AGN with a resolvable velocity offset would therefore be a strong indication of spin misalignment $\ga 5^{\circ}$. We also see that the $\Delta R_{\rm proj}$, $\Delta v_{\rm LOS}$ distribution is no longer skewed toward the largest offsets in this case, owing to the lack of escaping BHs and short return times in this spin model.

For the \fgasb\ hybrid spin model (middle panels of Figure \ref{fig:drdvallAGN}), the distributions of $\Delta R_{\rm proj}$ and $\Delta v_{\rm LOS}$ offsets are bimodal. Offsets greater than $0.5"$ and 600 \kms\ are produced almost exclusively in gas-poor mergers, for which BH spins are assumed to be random. Small offsets are produced mainly by mergers with $f_{\rm gas,sf}>0.1$, for which BH spins are nearly aligned. This creates distinct signatures in the host galaxy population relative to the non-hybrid spin models, as we show in Section \ref{ssec:hosts}.  

\begin{figure}
\begin{center}
\includegraphics[width=0.495\textwidth,trim=12 0 -12 0]{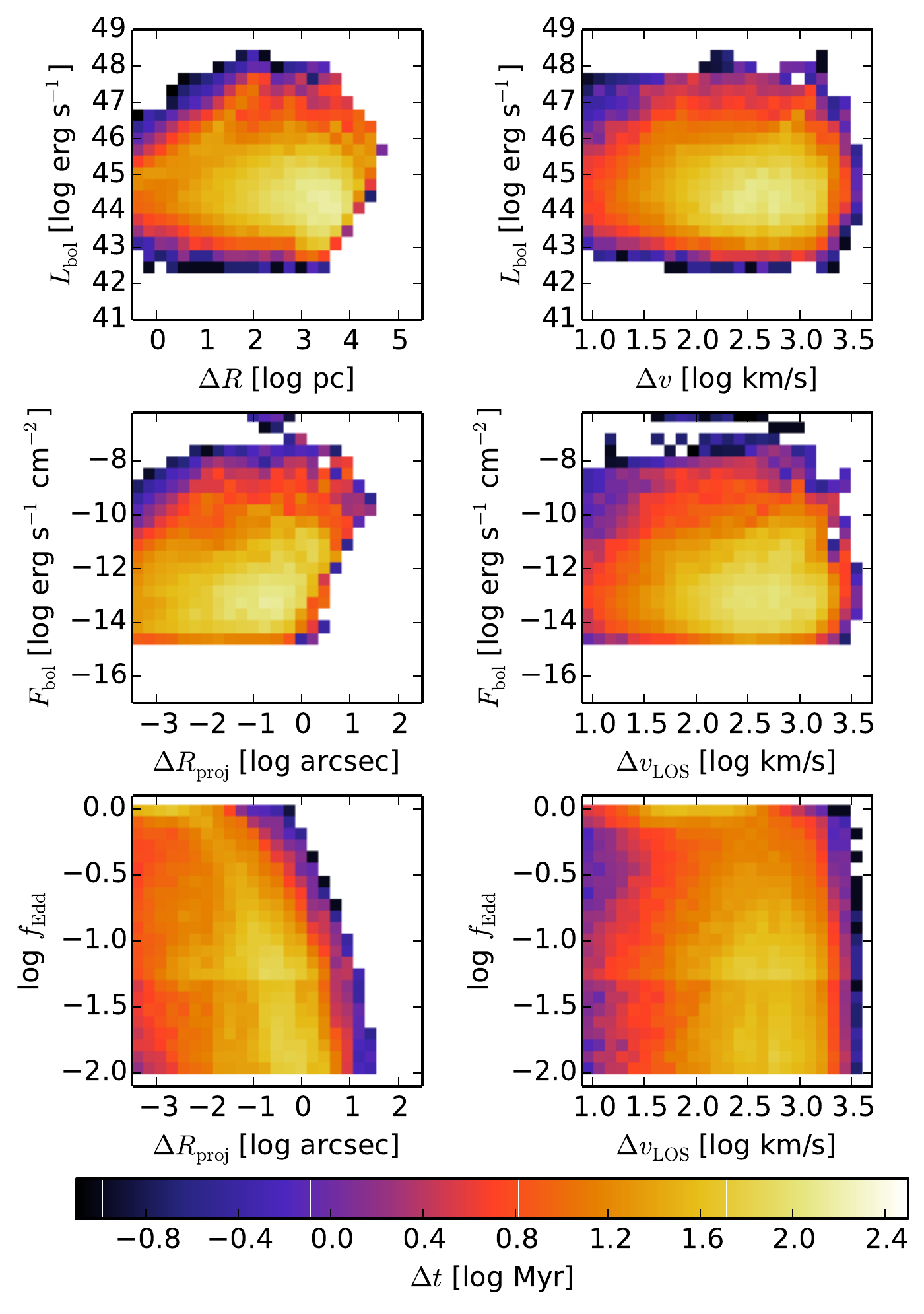}
\caption{Similar to Figure \ref{fig:drdvallAGN}, the time-weighted distributions of projected spatial offsets ({\em left panels}) and LOS velocity offsets ({\em right panels}) are shown for all recoiling AGN in the \randdry\ spin model (and \hcos\ sensitivity), for all timesteps during each recoil event and for all redshifts. Here the offsets are plotted versus the bolometric AGN luminosity ({\em top panels}), bolometric AGN flux ({\em middle panels}) and Eddington ratio $L_{\rm bol}/L_{\rm Edd}$ ({\em bottom panels}). The Eddington ratio is constrained to have a maxmium of 1 and a minimum of $10^{-2}$ in our models. The feature apparent at $f_{\rm Edd}=0.05$ arises from the transition to the radiatively-inefficient accretion regime. Most offset AGN have $f_{\rm Edd} < 0.1$, and for spatial offsets $\ga 1"$, observable offset AGN are rarely flux-limited.\label{fig:drdvallAGNdryrand}}
\end{center}
\end{figure}

\begin{figure}
\begin{center}
\includegraphics[width=0.495\textwidth,trim=12 0 -12 0]{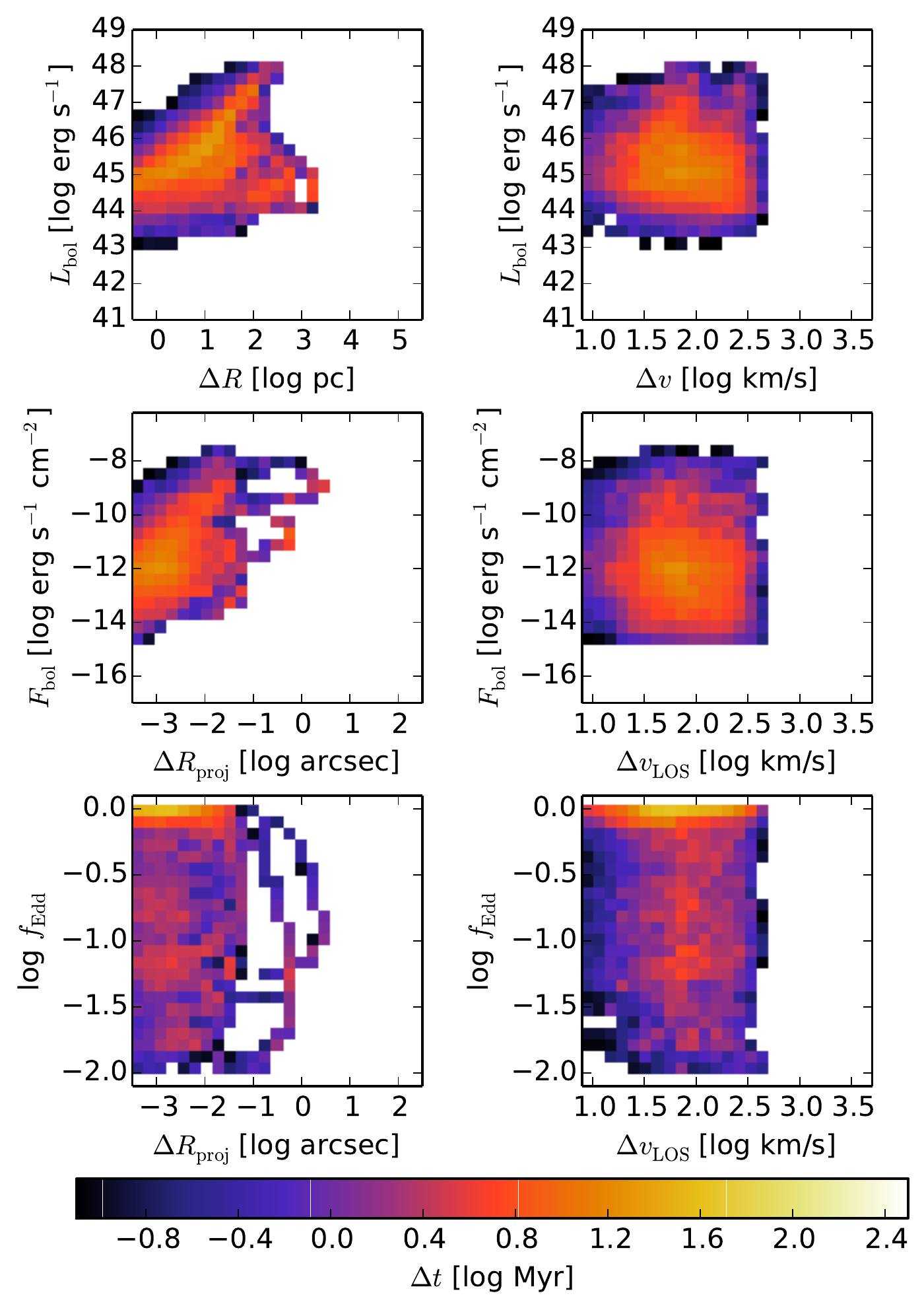}
\caption{The same distributions as in Figure \ref{fig:drdvallAGNdryrand} are shown, but for the \fivedeg\ spin model. The short return times of recoiling BHs in this model prevent the flux limit from being reached in most cases. \label{fig:drdvallAGN5deg}}
\end{center}
\end{figure}

In Figure \ref{fig:drdvallAGNdryrand}  the time-weighted distributions of bolometric AGN flux and Eddington ratio are shown versus $\Delta R_{\rm proj}$ and $\Delta v_{\rm LOS}$, for the \randdry\ spin model. As in Figure \ref{fig:drdvallAGN}, all recoiling AGN are shown, not just those with resolvable offsets. We see that for angular separations $\ga 1"$, recoiling AGN are generally above the flux limit. This reflects the fact that the largest angular separations occur at low redshift, where the minimum AGN luminosity is generally determined by the criterion $f_{\rm Edd} > 10^{-2}$ instead of by the flux limit. Even in the case of velocity-offset AGN, where there is no dependence on angular scale, offset AGN lifetimes at low $z$ are usually limited by the $f_{\rm Edd}$ criterion rather than the flux limit, at least for \hst\ sensitivity. The bottom panels in Figure \ref{fig:drdvallAGNdryrand} confirm that the population of recoiling AGN is dominated by low Eddington ratios ($< 0.1$), particularly for large $\Delta R_{\rm proj}$ and $\Delta v_{\rm LOS}$ offsets. (Lower-velocity recoils typically return to the galactic center before their luminosity can decline substantially.)

The same distributions are shown for the \fivedeg\ spin model in Figure \ref{fig:drdvallAGN5deg}. Owing to the short return times of recoils in this model, the flux limit is rarely reached, and the $f_{\rm Edd}$ distribution is heavily skewed toward high values. 

One important caveat to these findings is that our models include only accretion onto recoiling BHs from the disk carried along with the BH at the time of the kick. As shown in \citet{blecha11}, recoiling BHs on bound orbits may encounter fresh fuel on subsequent passages through the galactic center, substantially extending the AGN lifetime in some cases. This may create a population of offset AGN with smaller spatial offsets, such that the actual distribution is less biased toward large separations than our results suggest. We also assume the recoils always occur in a dense stellar cusp that forms during the galaxy merger \citep[e.g.,][]{mihher94b,hopkin08d,hopkin09b}. If a recoil event instead occurs in a cored elliptical (perhaps following a dry merger, where the BH inspiral time may be long), the BH may have a long phase of small-scale oscillations about the galactic center \citep{guamer08, lena14}. However, any new gas supply encountered by the recoiling BH on small scales would only {\em add} to the offset lifetimes predicted by our models.

Offset AGN that exhibit {\em both} velocity and spatial offsets, such as CID-42 \citep{civano10,civano12b}, will be especially compelling recoil candidates. We find that a {\em majority} of velocity-offset AGN (weighted by offset lifetime) will have projected spatial offsets resolvable with high-resolution imaging (e.g., \hst). This fraction increases with increasing spin alignment, from $\ga 50$\% in the random spin models to $\sim 90$\% in the \cold\ model. Thus, high-resolution imaging may be a fruitful means of follow-up for any velocity-offset recoil candidates. Even with SDSS resolution, $\sim 4$ - $20$\% of velocity-offset AGN may have resolvable spatial offsets; these primarily occur at low redshift ($z < 0.5$). 

In contrast, only $\sim$ 10 - 20\% of spatially-offset AGN should have discernible velocity offsets $\Delta v_{\rm LOS} > 600$\kms\ ($<$ 15\% in the aligned spin models). With a more stringent velocity offset criterion ($\Delta v_{\rm LOS} > 1000$\kms), $< 10$\% of spatially-offset AGN exhibit simultaneous velocity offsets, and no spatial/velocity offset overlap occurs in the aligned spin models. However, we demonstrate below that even this small fraction of spatially-selected offset AGN exhibiting velocity shifts could yield a population of strong recoil candidates.

One difficulty with confirming the nature of objects like CID-42 is that the offset AGN is superimposed on the host galaxy bulge, such that it cannot be easily distinguished from a member of an inspiraling BH pair. CID-42 has no evidence for a second AGN in the host nucleus, but excluding the presence of a {\em quiescent} BH is more difficult. The strongest offset AGN candidates will likely be those that have not only a resolvable spatial offset (and possibly a simultaneous velocity offset), but are also well-separated from the host galaxy's stellar light.

We find that a non-negligible (time-weighted) fraction of spatially-offset AGN resolvable with \hst\ also have separations $\Delta R_{\rm proj} > R_{\rm bulge}$, where $R_{\rm bulge}$ is the radius of the stellar bulge component in our models. Specifically, $\sim 20$ - 30\% of spatially offset AGN have $\Delta R_{\rm proj} > R_{\rm bulge}$ in the random and hybrid spin models, and a similar fraction of AGN with simultaneous spatial and velocity offsets are also found outside the stellar bulge. Separations $> R_{\rm bulge}$ are rare in the aligned spin models. For offset AGN resolvable with SDSS, $\sim$ 40 - 50\% of spatially-offset AGN may have $\Delta R_{\rm proj} > R_{\rm bulge}$ (in the random and hybrid spin models), and $\sim 20$ - 40\% of simultaneously spatially- and velocity-offset AGN should be found at such separations. We caution that the stellar bulges in our galaxy potential models may be more compact than in reality if there is a long delay between the galaxy coalescence and the BH merger and recoil, so $R_{\rm bulge}$ may be underestimated in some cases. Nonetheless, these results indicate that a subsample of offset AGN displaced beyond the host stellar bulge could be found, and these would  be ideal targets for follow-up and confirmation of recoiling BHs.

\subsection{Recoiling AGN observability}
\label{ssec:observe}

\subsubsection{Redshift distribution and general trends}

\begin{figure*}
\centering
\includegraphics[width=0.78\textwidth]{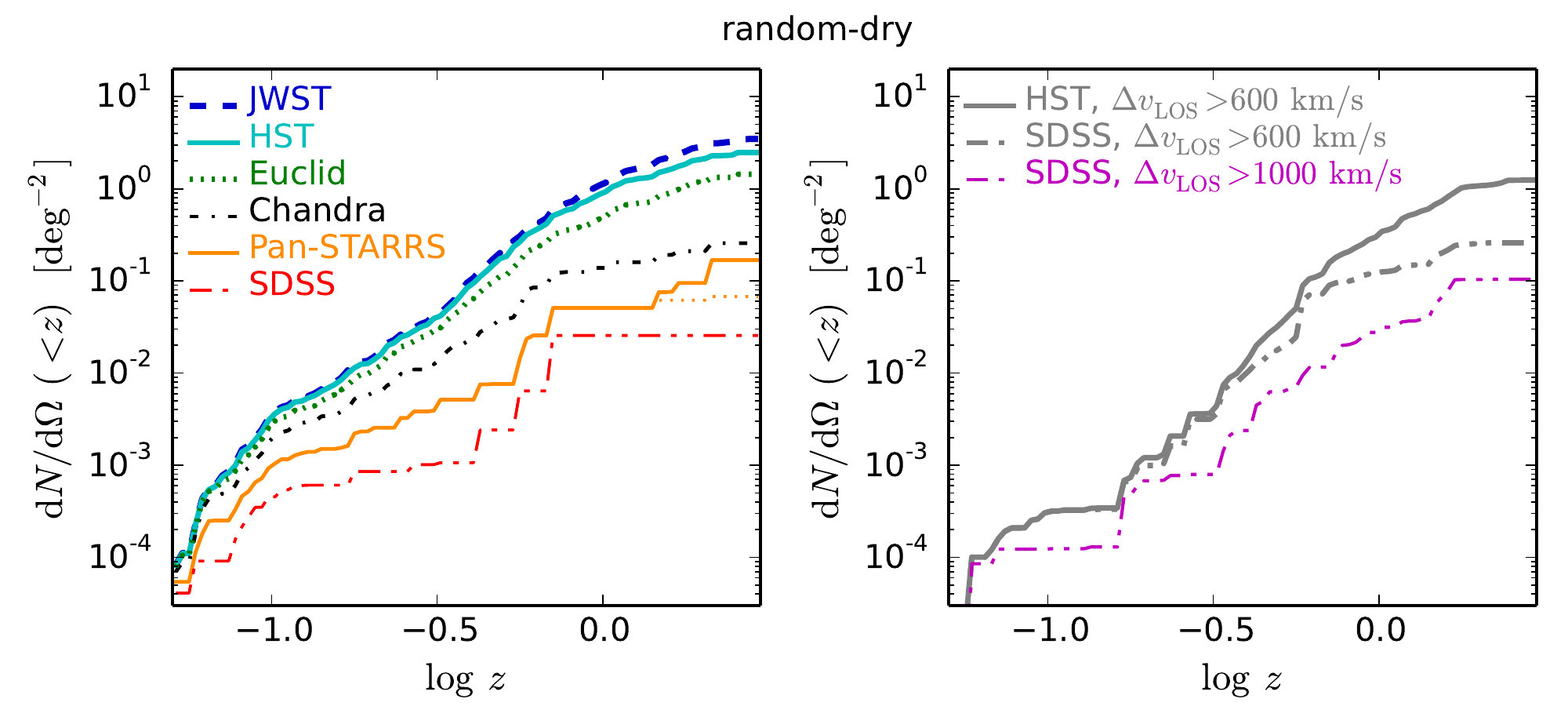}
\includegraphics[width=0.78\textwidth]{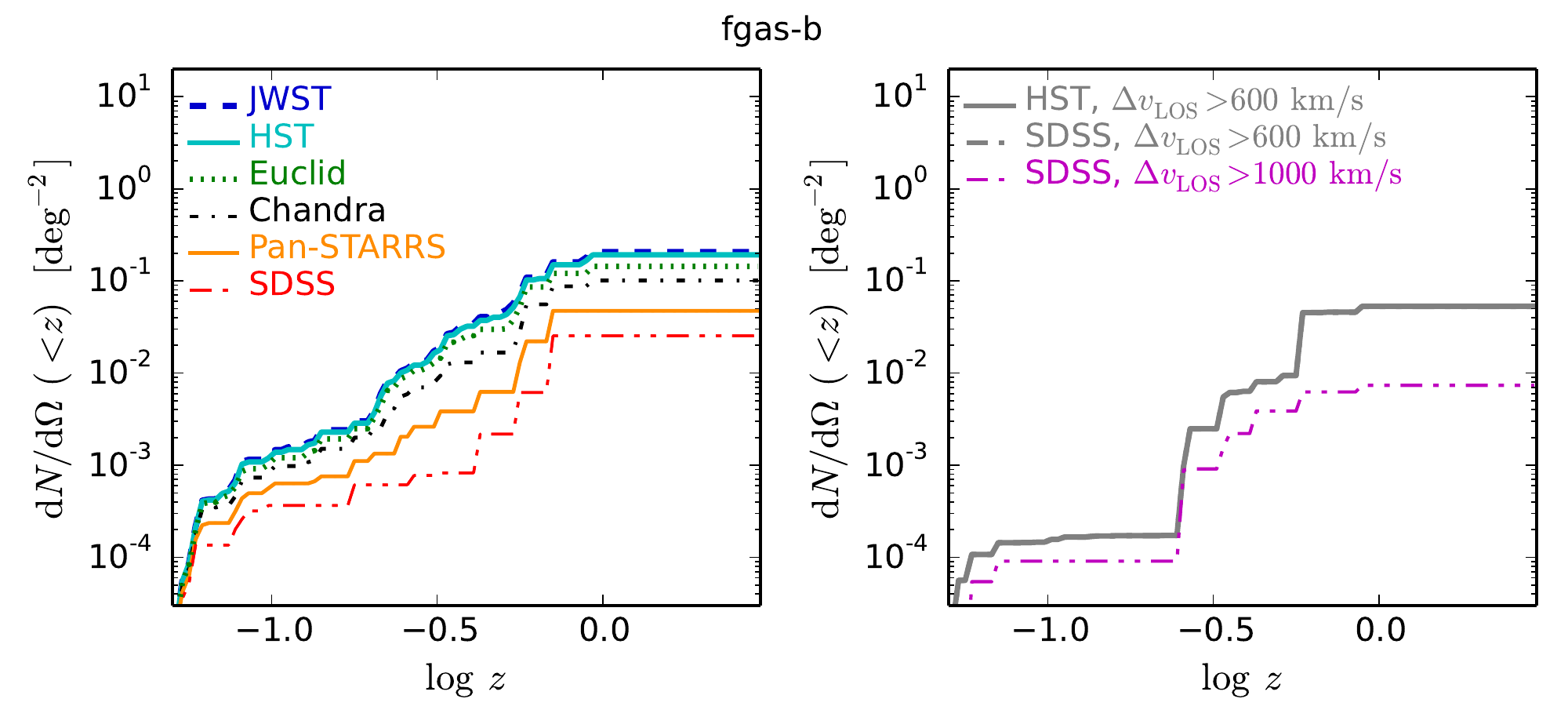}
\includegraphics[width=0.78\textwidth]{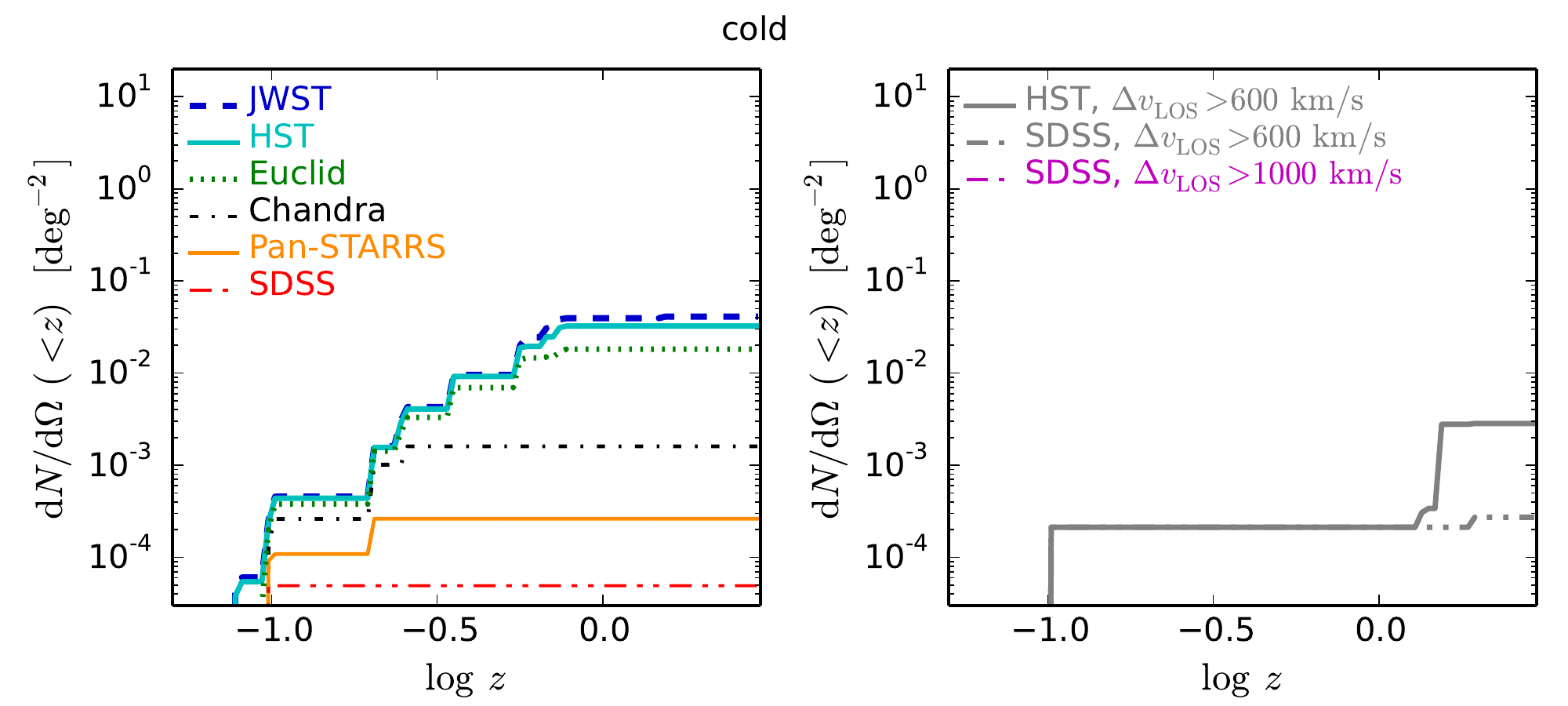}
\caption{The cumulative number of offset AGN observable per deg$^2$ is shown as a function of survey volume (in log $z$), for flux-limited surveys with a given resolution and sensitivity. {\em Top, middle, and bottom panels} show results for the \randdry, \fgasb, and \cold\ spin models, respectively. The {\em left panels} show the number of observable spatially offset AGN. As indicated on the plot legends, the thick solid blue curve corresponds to the sensitivity of \hcos\ in the F814W band, assuming 0.1" resolution, and the dashed cyan curve corresponds to the target \jwst\ sensitivity at 2 $\mu$m, assuming 0.07" resolution. The green dotted line corresponds to the target sensitivity of the Euclid wide survey in the visible band, assuming 0.2" resolution. The black dash-dotted line corresponds to \chandra\ observations at the sensitivity of the C-COSMOS survey at 2-10 keV, for 0.5" resolution, and the thin solid orange line denotes the PS1-MDS result for the $r$-band, assuming a seeing-limited resolution of 1". (The dotted orange line, visible only in the random spin model as distinct from the MDS, corresponds to the PS1-3pi survey). Finally, the dot-dot-dashed red line shows results for the SDSS $r$-band, for a PSF size of 1.4". In each case, the minimum resolvable spatial offset $\Delta R_{\rm proj}$ is assumed to be twice the resolution limit.  The {\em right panels} show predictions for observable velocity offset AGN. Only results for \hst\ (solid line) and SDSS (dot-dot-dashed lines) are shown. Gray lines correspond to a minimum resolvable LOS offset of $\Delta v_{\rm LOS} > 600$ \kms. For SDSS, we also show results for  $\Delta v_{\rm LOS} > 1000$ \kms\ (magenta line). The random models predict up to a few spatially-offset AGN per deg$^2$ detectable with space-based instruments, and up to one velocity-offset AGN per 10 deg$^2$ with $\Delta v_{\rm LOS} > 1000$ \kms\ observable in the SDSS. Even in the hybrid models, where spins are nearly aligned in gas-rich mergers, up to a few spatially-offset AGN per 10 deg$^2$ could be found. \label{fig:source_counts_zbins}}
\end{figure*}

The predicted number of observable offset AGN per square degree is shown is Figure \ref{fig:source_counts_zbins}, as a function of redshift, for various surveys and for three different spin models (\randdry, \fgasb, and \cold). In the \randdry\ spin model, 2 - 3 per deg$^{2}$ spatially offset AGN could be detected in \hcos\ \citep{koekem07} out to $z=3$, and $\sim 1$ per deg$^{2}$ could be found at $z<1$. The predicted source counts decrease with survey sensitivity and spatial resolution; SDSS could see about 1 spatially-offset AGN per 40 deg$^2$, and none should be detectable at $z \ga 0.8$.  

\begin{figure*}
\includegraphics[width=0.82\textwidth]{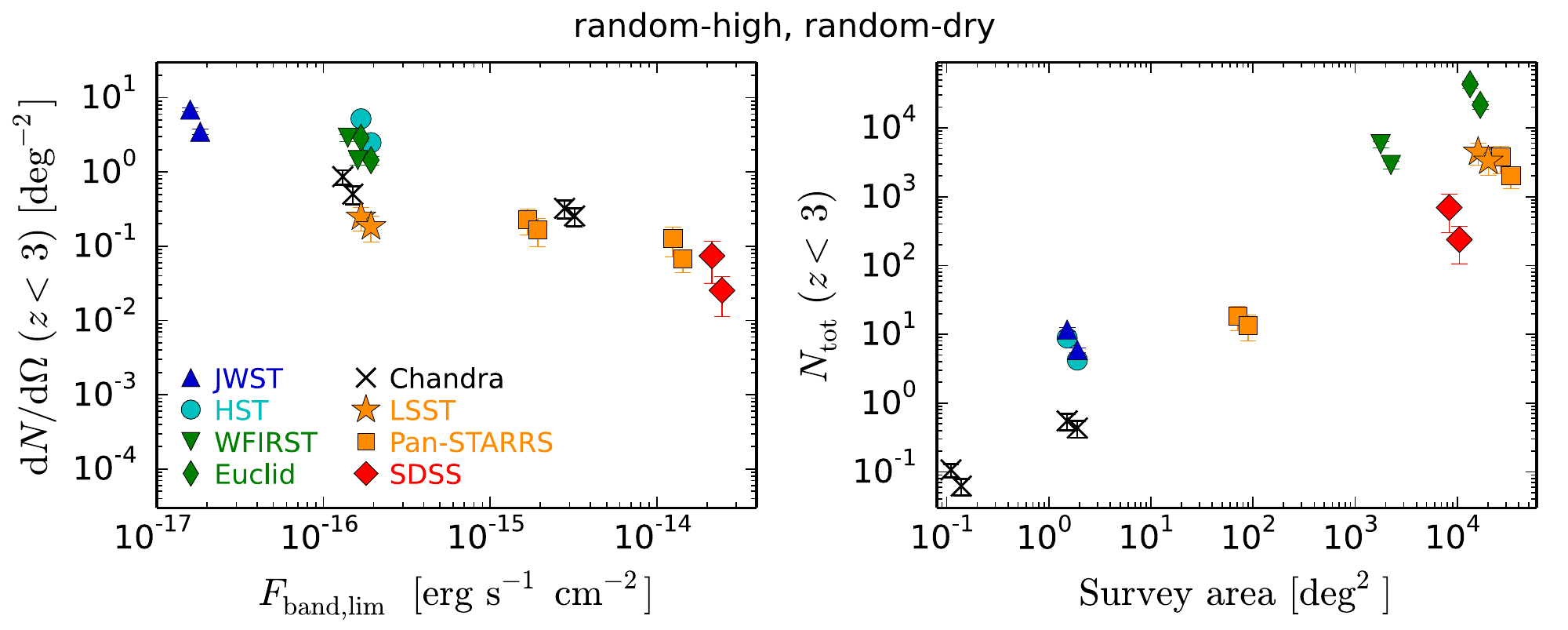}
\includegraphics[width=0.82\textwidth]{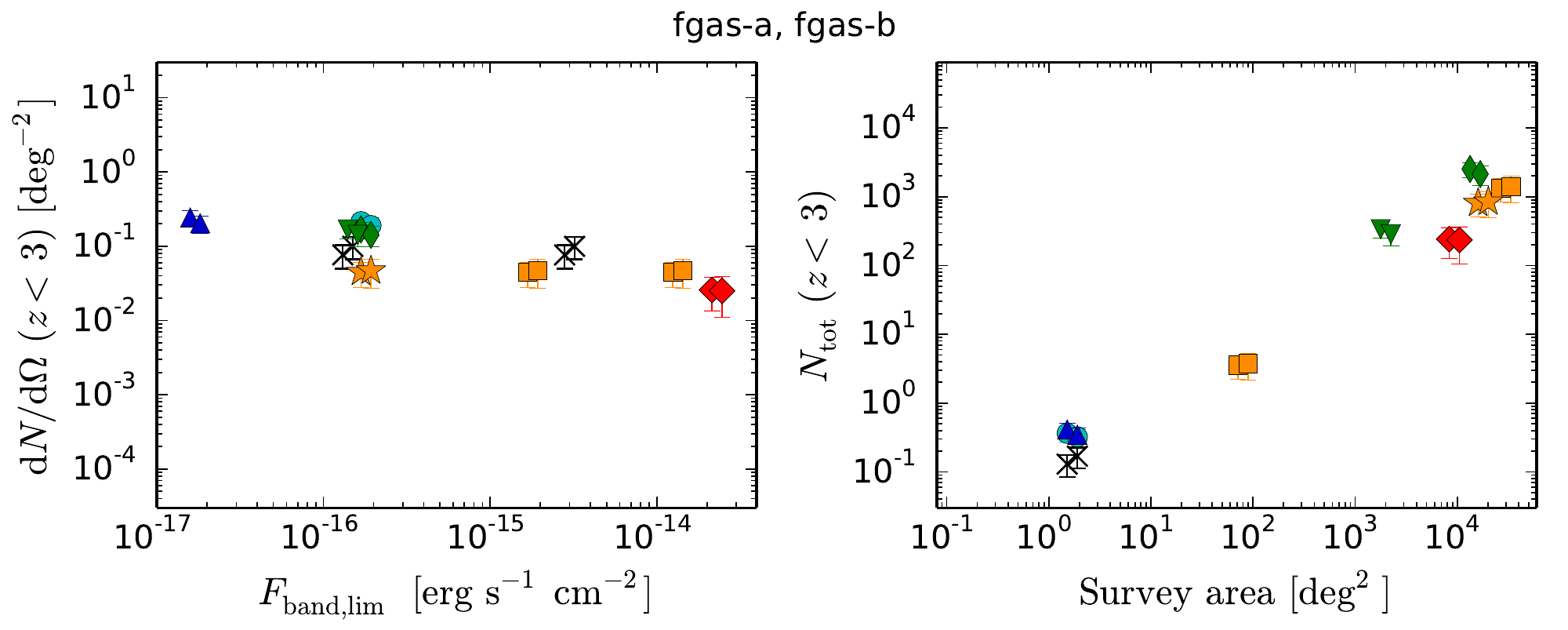}
\includegraphics[width=0.82\textwidth]{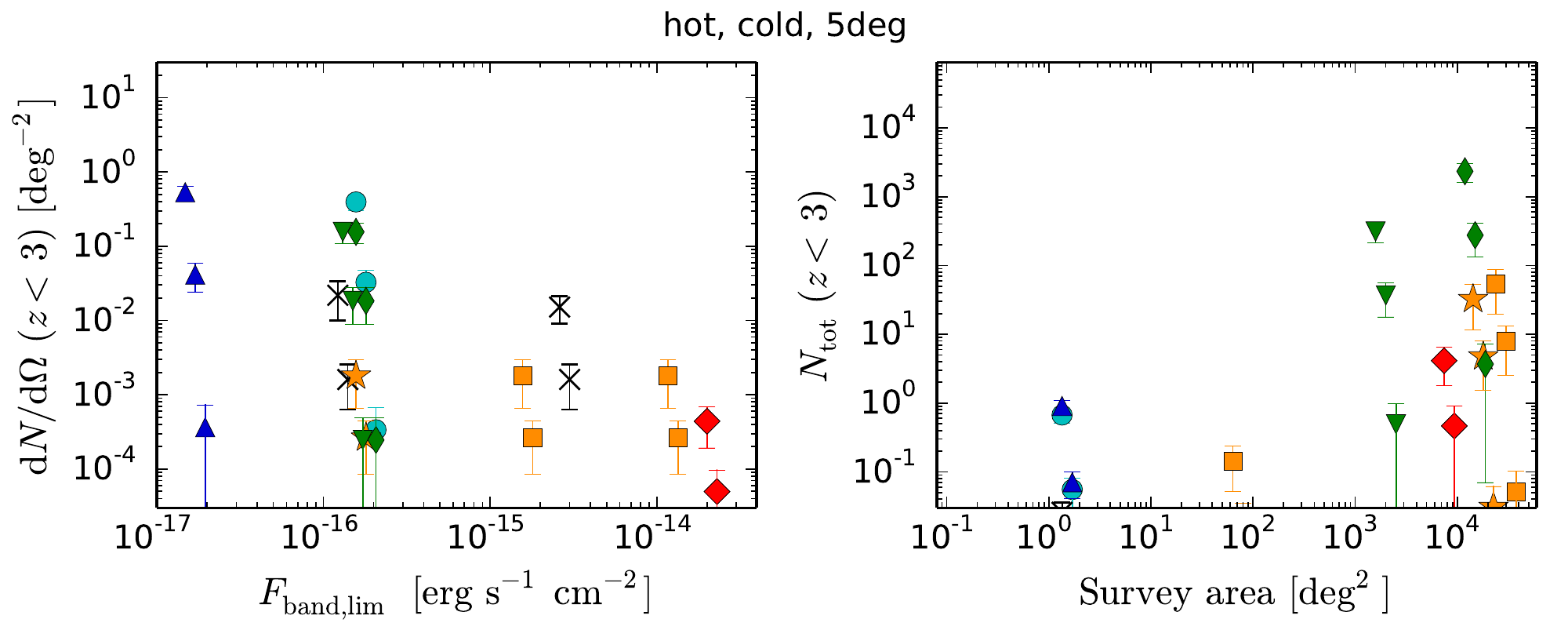}
\caption{The cumulative number of observable spatially offset AGN at $z<3$ is shown for various surveys. The {\em top panels} show the random spin models (\randhigh\ and \randdry, from left to right; a small offset along the x-axis is added for clarity). The {\em middle panels} show the hybrid spin models (\fgasa\ and \fgasb) in the same manner. The {\em bottom panels} show the aligned spin models (\hot, \cold, and \fivedeg). {\em Left column:} For each spin model and observational survey, the cumulative number of observable recoiling AGN with resolvable (projected) spatial offsets {\em per square degree}, out to $z=3$, is shown versus the limiting flux in the relevant band. For the optical and NIR bands, the quantity plotted is $\nu F_{\nu}$ for the effective wavelength of the band, while the X-ray fluxes correspond to $F_{\rm 2-10 keV}$. Points correspond to various observational surveys, as indicated in the plot legend and as follows. Blue triangles: \jwst\ at 2$\mu$m, 0.07" resolution. Cyan circles: \hcos\ in the F814W band, 0.1" resolution. Green inverted triangles: WFIRST high-latitude survey (J band, 0.2" resolution). Green thin diamonds: Euclid wide survey (visible band, 0.2" resolution). Black X's: C-COSMOS and CDF-N (2-10 keV, 0.5" resolution). Orange points: a seeing-limited resolution of 1" is assumed for LSST (Main Survey $r$-band; orange stars) and Pan-STARRS ($r$-band, for PS1-MDS and PS1-3pi; orange squares). Red diamonds: SDSS DR7, assuming a PSF size of 1.4" for the $r$-band. In each case, the minimum resolvable spatial offset $\Delta R_{\rm proj}$ is assumed to be twice the resolution limit. Error bars are derived from time-weighted Poisson statistics for the cumulative number counts. {\em Right column}: the total number of spatially-offset AGN that may be detectable in a given survey is shown, versus survey area. In the same manner as above, the points correspond to \hcos\ and a hypothetical \jwst\ survey of the same area (1.7 deg$^2$), a 15,000 deg$^2$ Euclid wide survey, a 2000 deg$^2$ WFIRST high-latitude survey, the 448 arcmin$^2$ CDF-N and 1.7 deg$^2$ \chandra-COSMOS surveys, the 80 deg$^2$ PS1-MDS, SDSS DR7, the 18,000 deg$^2$ LSST Main Survey, and the 30,000 deg$^2$ PS1-3pi survey. The random spin models predict that $> 10^3$ offset AGN could be found with Pan-STARRS, LSST, and WFIRST, and $> 10^4$ with Euclid. The hybrid models predict $\sim 10^2$ - $10^3$ offset AGN to be detectable in these surveys. In the most extreme aligned model (\fivedeg), however, recoiling AGN may never be detected.  \label{fig:source_counts}}
\end{figure*}

\hst\ could detect {\em velocity}-offset AGN in slightly lower numbers; $\ga 1$ per deg$^{2}$ are predicted out to $z=3$ for a minimum resolvable offset of $\Delta v_{\rm LOS} > 600$ \kms. In general, for angular resolutions better than $\sim 0.5"$, we find that spatially-offset AGN should be found in greater numbers than velocity-offset AGN. The reverse is true for seeing-limited surveys with spectroscopic coverage; in the random spin model, SDSS should find $\sim 10$ times more velocity-offset AGN than spatially-offset AGN for a minimum $\Delta v_{\rm LOS} > 600$ \kms. Even if only offsets $\Delta v > 1000$ \kms\ can be resolved, velocity offsets could still be detected in greater numbers with SDSS.

The predicted number of observable velocity offset AGN (for $\Delta v_{\rm LOS} > 600$ \kms) differs by at most a factor of a few between \hst\ and SDSS, and even less at low redshift, despite the large difference in sensitivity.  As discussed above, the imposed minimum AGN Eddington ratio of $10^{-2}$ generally corresponds to a higher luminosity limit than does the assumed flux limit, for low to moderate redshifts. 
Thus, our results depend only weakly on the absolute flux limit, which becomes important only at higher $z$. 

This trend is even more prominent in the hybrid spin models, where there is almost no dependence on the assumed flux limit. Relative to the random spin models, the hybrid models produce fewer superkicks, and those that do occur are limited to dry mergers, which are found predominantly at low redshifts (virtually none occur at $z>1$). Thus, the AGN lifetime is almost always set by the minimum $f_{\rm Edd}$, and increasing the angular resolution only moderately increases the number of observable spatial offsets for these low-redshift sources ($\sim$ factor of 10-12 higher number counts for a factor of 20 increase in spatial resolution). Note that the SDSS predictions for spatially offset AGN are similar between the random and hybrid spin models; most recoil events in the random spin model that produce a spatial offset observable with SDSS also produce an observable offset in the hybrid spin model. This reflects the fact that SDSS is most sensitive to luminous AGN at low to moderate redshift, which are more likely to occur in dry mergers that can produce superkicks in the hybrid model.  

In the aligned (\cold) spin model, fewer observable recoils are predicted than in the random or hybrid models, as expected, and there is a steeper dependence on angular resolution. Only 1 spatially offset AGN per 30 deg$^2$ should be detectable with \hst, and velocity offset AGN should occur at rate of $< 1$ per 100 deg$^2$.

We note that, despite the prevalence of major mergers among BHs with $M_{\rm BH} < 10^7$ \msun\ (Figure \ref{fig:massratio}), these low-mass BHs do not dominate the number counts of observable offset AGN in any of our models. In fact, for the hybrid spin models, massive BHs ($M_{\rm BH} > 10^8$ \msun) dominate the sample of offset AGN, because these BHs are preferentially hosted in gas-poor galaxies where superkicks can occur. Additionally, for seeing-limited surveys, spatially-offset AGN tend to have high BH masses regardless of spin model. The offset AGN lifetime depends on BH mass, such that low-mass BHs will often exhaust their fuel supply before reaching a resolvable separation from the host galaxy. 

This trend occurs even though massive BHs are generally hosted in massive galaxies, which have high central escape speeds. As a result, only recoiling AGN with \vk/\vesc $\ga 0.8$ produce spatial offsets that are resolvable with seeing-limited surveys. One caveat to this is that our sample of dwarf galaxies is incomplete owing to mass resolution limits. A more complete sample of dwarf galaxies could produce a population of lower-mass AGN with resolvable offsets.

As mentioned above, our accretion disk model generally results in longer recoiling AGN lifetimes than if a constant accretion rate is assumed. This is especially important for spatially-offset AGN, for which offset AGN lifetimes are more often limited by declining luminosity than by a return to the galactic center. (Velocity-offset AGN have shorter lifetimes on average, owing to deceleration by the host galaxy potential.) In most cases, our models predict lower rates of observable spatially-offset AGN by up to a factor of a few if a constant $M_{\rm disk}/\dot M_{\rm BH}$ lifetime is assumed. The difference is largest for the aligned spin models and seeing-limited resolution, where in the constant-$\dot M$ model, few BHs achieve a resolvable separation during their active lifetime. In contrast, velocity offset AGN lifetimes are often limited by deceleration in the host potential, rather than by the luminosity limit, and the predicted numbers of observable velocity offset AGN can be a factor of a few lower {\em or} higher when a constant $\dot M_{\rm BH}$ is assumed. 

\subsubsection{Imaging surveys}

Figure \ref{fig:source_counts} shows the cumulative number of  spatially-offset AGN per square degree at $z<3$ observable in selected current and future surveys, for all spin models considered in this study. We first examine the random spin models: \randhigh\ and \randdry\ (left panels). Up to 7 per deg$^{2}$ spatially offset AGN may be detectable with \jwst. In the hard X-ray band, up to $\sim$ 1 per deg$^{2}$ offset AGN could be detected at the sensitivity of the \chandra\ Deep Field North \citep[CDF-N, ][]{alexan03} survey, and $\sim 0.3$ per deg$^2$ at the limit of the \chandra-COSMOS Legacy \citep[C-COSMOS,][]{civano13} survey. Seeing-limited surveys (assuming 1" effective resolution) may resolve up to 1 spatially offset AGN per 4 deg$^2$, and even SDSS, with an effective PSF size of 1.4", could find up to 1 per 14 deg$^2$.

The bottom panels of Figure \ref{fig:source_counts} show the total predicted number of observable spatially-offset AGN in a given survey, versus survey area. Results for spatial offsets are given for \hcos\ (and a hypothetical \jwst\ survey with the same area), a Euclid wide survey \citep{laurei11}, a 2000 deg$^2$ WFIRST weak-lensing survey \citep{sperge13}, CDF-N, C-COSMOS, PS1-MDS, PS1-3pi, SDSS DR7, and the LSST Main Survey. The random spin models predict that 5 - 9 spatially offset AGN may be found in \hcos, while none are likely to be identified with \chandra\ in the CDF-N or C-COSMOS surveys. In contrast, future wide-area, high-resolution surveys could detect spatially offset AGN in large numbers: WFIRST could detect several thousand such objects, and Euclid could detect tens of thousands.\footnote{The predictions for Euclid assume a point-source sensitivity of 27.3 AB mag in the visible band; the predicted numbers are up to a factor of two lower for an assumed sensitivity of 24 AB mag in the J band.} 

An important prediction of our models is that, because offset AGN spend most of their time at large separations, seeing-limited surveys should be able to resolve spatially-offset AGN. This suggests that offset AGN may be promising targets for ground-based, large-area surveys. For random spins, $\sim$ 15-20 offset AGN may be found in PS1-MDS, and hundreds of spatially offset AGN could be found in SDSS. The predictions are even more optimistic for LSST and PS1-3pi, which could detect up to several thousand spatially-offset AGN. 

Allowing for some amount of pre-merger BH spin alignment reduces the number of observable offset AGN, but does not eliminate the possibility of detecting recoils (Figure \ref{fig:source_counts}, middle and right panels). For the hybrid spin models, less than one offset AGN would be expected in \hcos. However, 2-3 may be detectable in PS1-MDS, and large populations of spatially offset AGN are still detectable in SDSS (up to $\sim 200$ in DR7), LSST (up to $\sim 700$ in the Main Survey), and WFIRST ($>$ 300). The hybrid spin models also predict that the PS1-3pi and Euclid wide surveys could each detect in excess of $10^3$ spatially offset AGN. 

The predictions for the three aligned spin models (right panels of Figures \ref{fig:source_counts}) differ by several orders of magnitude. For high-resolution observations (\hst, \jwst, Euclid, and WFIRST), the \hot\ spin model produces comparable numbers of spatially-offset AGN as the hybrid models. A few tens of offset AGN could be found with LSST and PS1-3pi, and $> 10^3$ could be detected with Euclid. In the \cold\ spin model, a handful might be detectable with LSST and PS1-3pi, but wide-area surveys with space-based instruments (i.e., WFIRST and Euclid) would be required to find a population of offset AGN. In the \fivedeg\ model, the maximum displacements achieved barely exceed the angular resolution of SDSS, and offset AGN are unlikely to ever be detected -- even Euclid is predicted to find at most a few objects over the entire sky.

\subsubsection{Spectroscopic surveys}

Because SDSS also has spectroscopic coverage, we can estimate the number of detectable velocity-offset AGN as well. The random spin models produce $\sim 0.1$ per deg$^{2}$ velocity offset AGN with $\Delta v_{\rm LOS} > 1000$ \kms\ (Figure \ref{fig:source_counts_zbins}, left panel), implying that nearly $10^3$ such objects should exist in the DR7 footprint. 

This result is of particular interest, because to date, the SDSS spectroscopic quasar database is the only data set on which large-scale systematic searches for offset AGN have been carried out. \citet{bonshi07} searched for quasar spectra at $0.1 < z < 0.81$ in which the broad \hbeta\ and Mg II lines have symmetric profiles and consistent velocity shifts relative to the narrow emission lines. They found a null result for $\Delta v > 800$ \kms. Using different selection criteria, \citet{tsalma11} searched for shifted BLs in quasar spectra at $0.1 < z < 1.5$ and identified 32 candidates of interest. \citet{eracle12} searched for BL offsets of $\Delta v > 1000$ \kms\ in \hbeta\ for quasars at $z<0.7$ and found 88 such velocity-offset quasars. However, these BL offsets could also arise from high-velocity outflows, double-peaked emitters, or binary BH motion, and none have been confirmed to date as recoiling BHs. 

For our random spin models, only $\sim$ 10 - 30\% of all velocity-offset AGN occur at $z<0.7$, such that up to several hundred offset AGN are predicted in the relevant redshift range. This is still substantially more than are present in the DR7 quasar catalog, but we stress that these observational results cannot be compared at face value to our models. There are several important distinctions between the offset AGN in our sample and the selection criteria for SDSS quasars \citep{schnei10}, which will cause some offset AGN to be excluded from the latter. The initial selection of SDSS quasar candidates is based on their colors, which means that low-luminosity AGN whose flux is dominated by the host galaxy are excluded. 

The SDSS quasars are also required to have a measurable BL with FWHM $> 1000$ \kms, such that Type II (narrow-line) AGN are excluded as well. However, a BL is also required for detecting velocity-offset AGN, meaning that some fraction of velocity-offset AGN should be undetectable in any case. 

 We have also neglected the effects of obscuration in our models. Obscuring dust in the nuclear region is a very common feature of AGN, required to explain the optical, mid-IR, and X-ray spectral features of many objects, and may often have a toroidal geometry \citep[e.g.,][]{urrpad95,ueda03,stern05}. While recent work has focused primarily on the role of dusty tori in obscuring low-velocity recoil events \citep{raffai16}, we find that dust obscuration is unlikely to be a major factor in the observability of spatially-offset AGN, at least for kick distributions that assume some BH spin misalignment. This is because little of the obscuring material beyond the BL region should remain bound to the BH, and Figure \ref{fig:drdvallAGN} illustrates that offset AGN are most often found at separations larger than the scale of the torus \citep[see also][]{kommer08b}. In contrast, {\em velocity} offsets will occur essentially at the moment of recoil, and thus may be obscured from the observer for the early phase of their lifetime. If obscuring material extends beyond the dusty AGN torus to galactic scales, owing perhaps to a coincident nuclear starburst, some recoiling AGN could be obscured for a significant portion of their offset lifetimes. Recall, however, that a majority of velocity-offset AGN should also have spatial offsets detectable with \hst, and up to 20\% may have spatial offsets detectable with SDSS. Thus, obscuration may have at most a moderate effect on the observability of velocity-offset AGN.

The SDSS DR7 quasars are additionally required to be fainter than $i \approx 15$ mag (to avoid saturation of the spectrograph) and more luminous than $M_i = -22$. These magnitude limits, at least, have only a modest effect on the offset AGN population at $z<0.7$, if random spins are assumed. If we approximate these cuts for our sample (using the same optical bolometric corrections from \citet{hopkin07a} that are used throughout the text), the predicted number of velocity-offset AGN ($\Delta v > 1000$) is reduced by about 30\% for the \randdry\ model, and by only 5\% in the \randhigh\ model. 

An obvious alternative explanation for the observed rarity of large BL offsets in SDSS is that pre-merger spin alignment plays at least some role in reducing the number of superkicks with \vk $> 1000$ \kms. The hybrid spin models predict that $\sim$ 70 (\fgasb\ model) to 220 (\fgasa) velocity-offset AGN with $\Delta v > 1000$ \kms\ may exist in the SDSS footprint, most of which are found at $z<0.7$. With the estimated magnitude limits described above, these numbers are reduced to $\sim 30$ and 200, respectively. Thus, even if spins are preferentially aligned only in gas-rich mergers, our results imply that a population of velocity-offset AGN may be missed by the SDSS DR7 selection criteria.

In the \hot\ aligned-spin model, up to 10 velocity-offset AGN (with $\Delta v_{\rm LOS} > 1000$ \kms) could be found with SDSS, with only $\sim 2$ at $z<0.7$. None are predicted to be observable in the \cold\ or \fivedeg\ models. Thus, follow-up studies of candidate recoils in SDSS could place indirect constraints on pre-merger BH spins.

\subsection{Host galaxy properties}
\label{ssec:hosts}

\begin{figure}
\centering
\includegraphics[width=0.495\textwidth,trim=12 0 0 0]{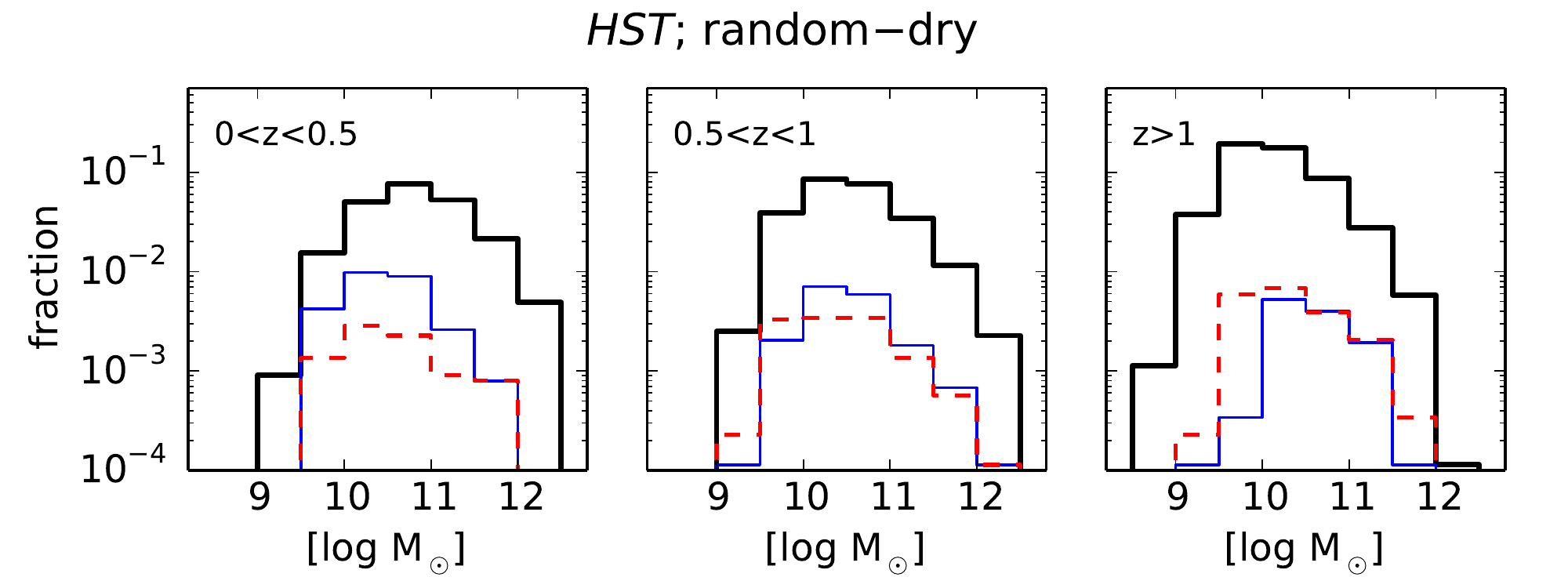}
\includegraphics[width=0.495\textwidth,trim=12 0 0 0]{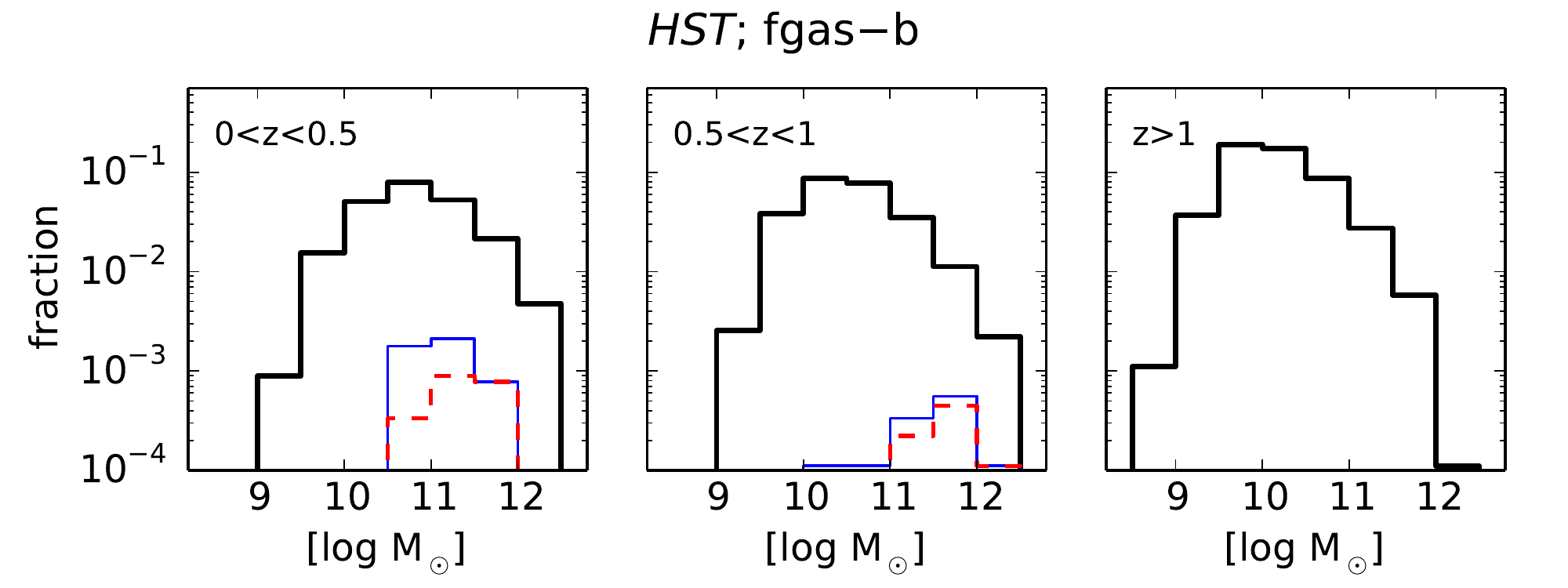}
\includegraphics[width=0.495\textwidth,trim=12 0 0 0]{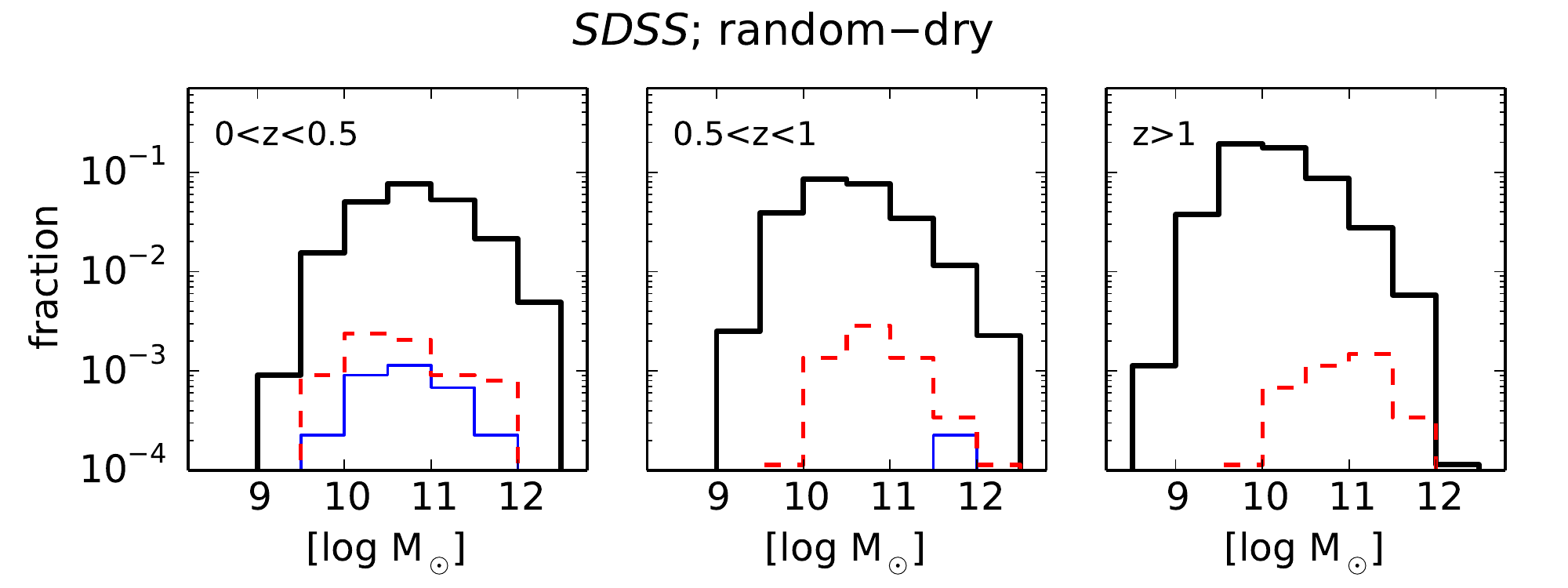}
\caption{{\em Top row:} the distributions of host stellar masses are shown for all merging BHs in Illustris (thick black line), in three redshift bins. The host stellar mass is defined as the combined mass of the progenitor galaxies in the simulation. The thin solid blue line denotes the subset of galaxy merger remnants that host an observable spatially offset AGN, for the \randdry\ spin model and assuming \hcos\ sensitivity and resolution. The thin red dashed line similarly denotes the hosts of observable velocity-offset AGN ($\Delta v_{\rm LOS} > 600$ \kms). In both cases, only offset AGN with a minimum offset lifetime $> 10^5$ yr are shown. {\em Middle row:} same as top row, but for the \fgasb\ spin model. {\em Bottom row:} same, as top row, but assuming SDSS sensitivity and resolution. The host mass distribution is broad but peaks at log ($M_*$/\msun) $=$ 10 - 11 for the random and aligned spin models. In the hybrid spin models, higher-mass hosts are preferred despite their higher escape speeds, owing to their low gas content, and observable recoils are very rare at $z>1$. \label{fig:mstar_hist}}
\end{figure}

To characterize the host galaxies of offset AGN, we first examine their stellar mass distribution and its evolution with redshift (Figure \ref{fig:mstar_hist}). The stellar mass distribution for all BH merger hosts peaks at $\sim 10^{10}$ \msun\ at the highest redshifts, and evolves via hierarchical growth toward larger masses at lower $z$. The distribution is limited at both the low- {\em and} high-mass end by the resolution of the simulation; specifically, the minimum BH mass and stellar/halo particle cuts imposed on each progenitor galaxy exclude many dwarf-dwarf mergers and satellite-halo mergers. 

For the random spin models, and assuming \hcos\ sensitivity and resolution, the offset AGN hosts have a broad mass distribution that peaks at log $(M_*$/\msun) $\sim$ 10 - 11 (Figure \ref{fig:mstar_hist}, top panel). At $z<1$, offset AGN favor lower-mass hosts relative to the overall distribution. Nominally, low-mass galaxies should more easily produce offset AGN owing to their low central escape speeds, and indeed, for the random spin models, nearly 30\% of BH mergers occurring in galaxies with log $(M_*/$\msun) $ < 10$ at low redshifts give rise to an observable spatially-offset AGN.

At $z > 1$, BH merger hosts have lower $M_*$ on average, but the spatially-offset AGN hosts do not. At most a few percent of low-mass galaxy merger remnants (log $(M_*/$\msun) $< 10$) produce observable spatial offsets. In part, this is because low-mass galaxies generally host smaller BHs with lower AGN luminosites. These low-mass bulges are also increasingly gas-rich at high $z$, making it difficult for BHs to escape the nucleus. 

\begin{figure}
\centering
\includegraphics[width=0.5\textwidth,trim=14 0 -18 0]{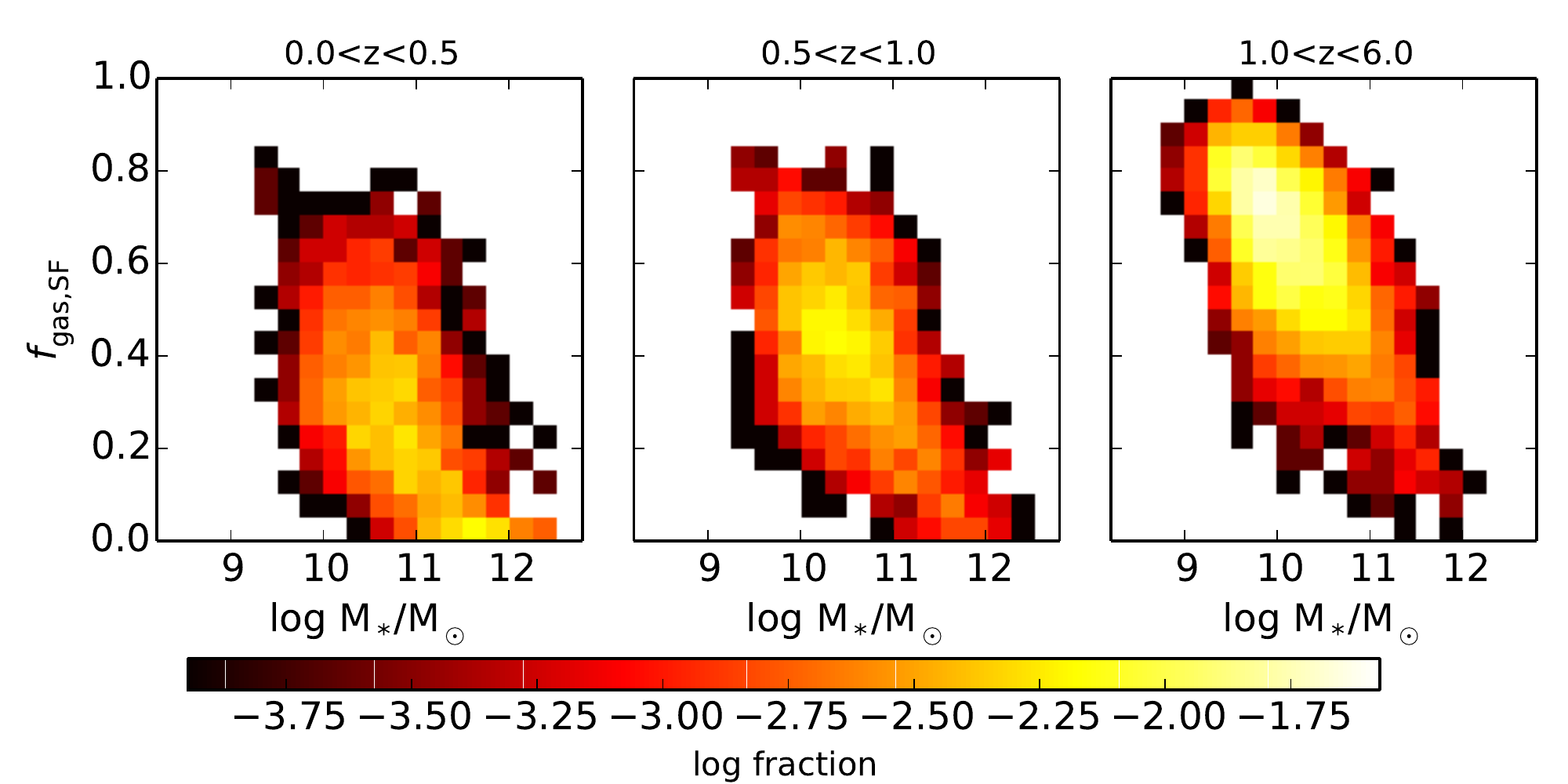}
\includegraphics[width=0.5\textwidth,trim=14 0 -18 0]{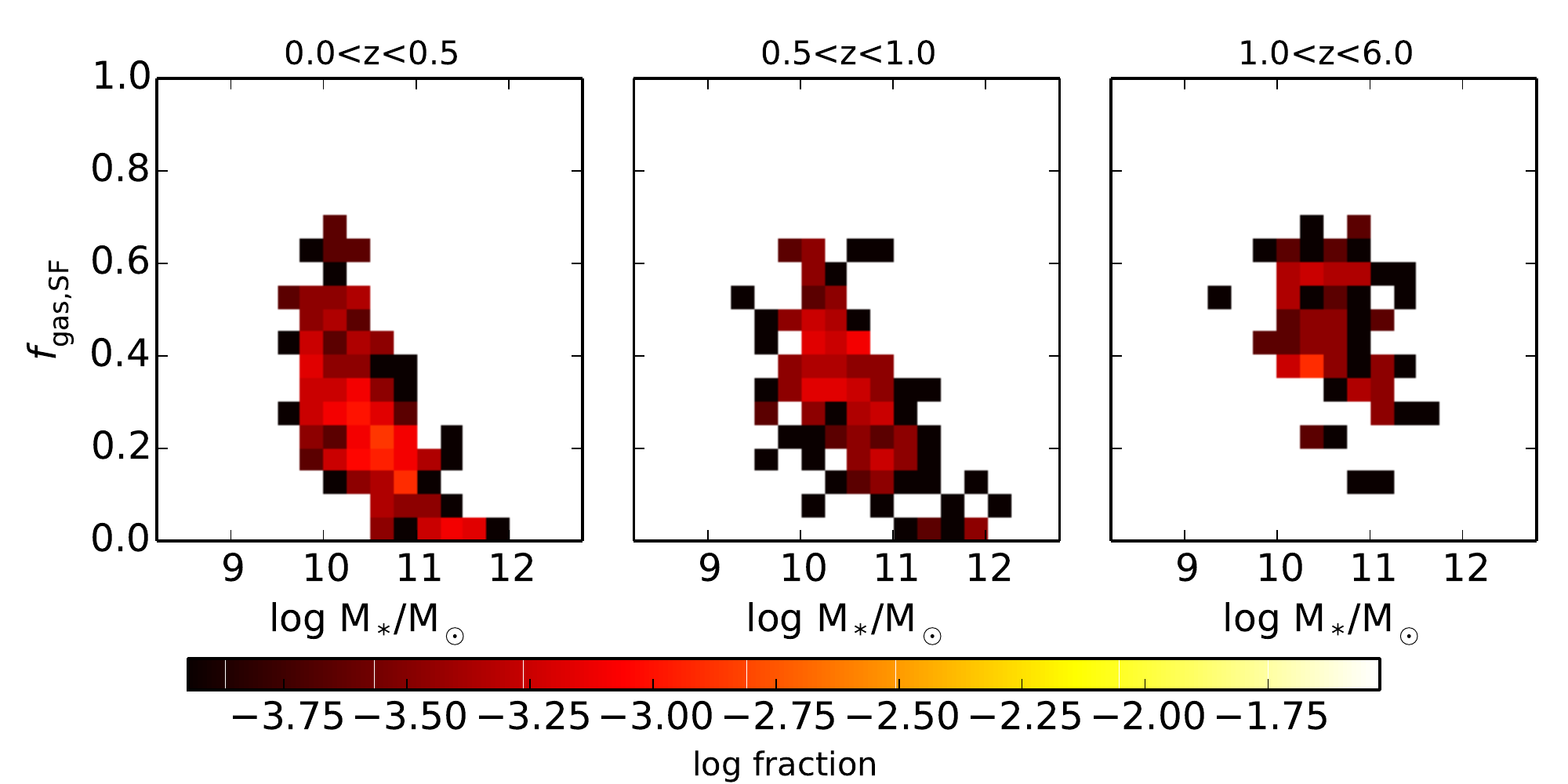}
\includegraphics[width=0.5\textwidth,trim=14 12 -18 0]{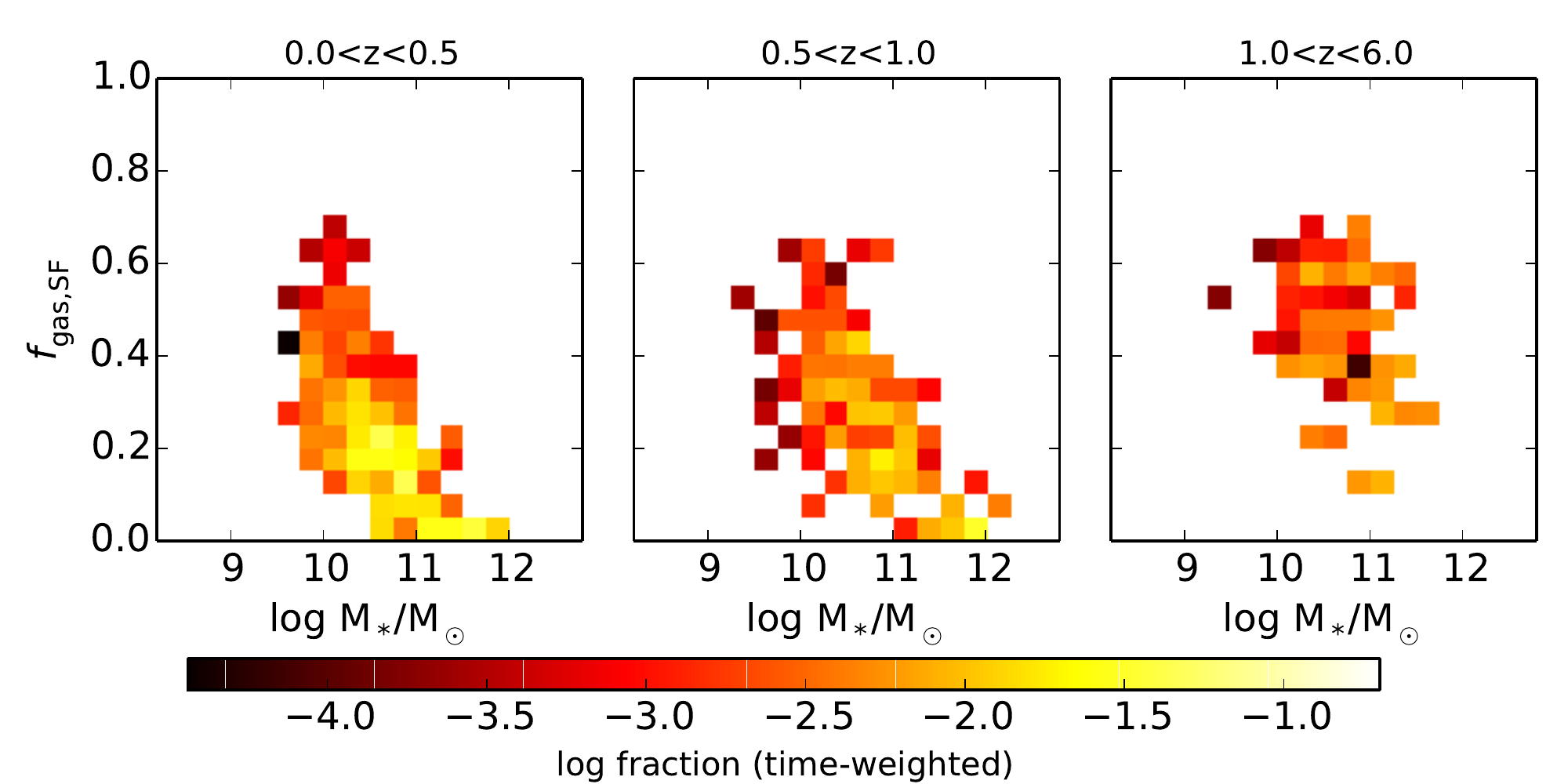}
\caption{The distributions of stellar mass ($M_*$) versus the fraction of cold, star-forming gas ($f_{\rm gas,sf}$, relative to $M_*$) are plotted for different redshift bins. The top row shows the distributions for all merging BHs in our Illustris sample (8800 mergers), while the middle and bottom rows show only the 492 mergers that produce observable spatially-offset recoiling AGN in the \randdry\ model, assuming \hcos\ resolution and sensitivity. In the middle row, the distributions are weighted by number, and in the bottom row, they are weighted by time. The overall population of merger hosts evolves from low masses and high gas content to higher-mass, quenched galaxies. In contrast, the hosts of observable offset AGN favor lower gas fractions at all redshifts.\label{fig:mstar_fgas}}
\end{figure}

Figure \ref{fig:mstar_fgas} shows the distribution of $M_*$ versus the star forming gas fraction ($f_{\rm gas,sf}$) for all BH merger hosts and for spatially-offset AGN hosts. The merging galaxy population shows a strong ``quenching" trend evolving from low-mass, gas-rich galaxies at high $z$ to higher-mass, gas-poor galaxies at $z\sim 0$. Comparing with the hosts of spatially offset AGN, we see that they follow the same general trend, but that galaxies with low- to moderate gas fractions are favored; extremely gas-rich systems rarely produce offset AGN. This is even clearer in the bottom panel of Figure \ref{fig:mstar_fgas}, which shows the same distribution weighted by offset AGN lifetime. The longest-lived offset AGN occur preferentially in galaxies with low $f_{\rm gas,sf}$.

The $M_* - f_{\rm gas,sf}$ distributions for velocity-offset AGN hosts are similar to those for spatially offset AGN. Velocity-offset AGN occur more often at $z>1$ in low-mass, gas-rich galaxies than do spatial offsets, but these generally have short lifetimes, $< 1$ Myr. The suppression of recoils in high-$z$, gas-rich galaxies is another reason that our results depend only weakly on the assumed flux limit, because it is primarily at high redshifts where the minimum observable flux is more constraining than the minimum $f_{\rm Edd} = 10^{-2}$.

In the aligned spin models, offset AGN are much rarer overall, and almost none occur in the \fivedeg\ model. Recoil velocities are generally too small to produce observable offsets in either gas-rich galaxies or massive, quenched galaxies. As a result, offset AGN in the aligned spin models favor similar host stellar masses as the random models (log $(M_*/$\msun) $\sim$ 10 - 11). No spatially-offset AGN are found at $z>1$ in the \cold\ spin model, or at $z>2$ in the \hot\ model. 

When the degree of BH spin alignment depends on the host gas fraction, as in the hybrid spin models, recoils are preferentially suppressed in gas-rich hosts by spin alignment in addition to the aforementioned dynamical effects. As a result, offset AGN are preferentially found in higher-mass hosts, which have lower gas content on average. This is consistent with the above observation that massive BHs dominate the offset AGN population in the hybrid models. In the \fgasb\ model (Figure \ref{fig:mstar_hist}, middle panels), the extreme alignment of the \fivedeg\ spin model for gas-rich hosts eliminates virtually all observable offsets in hosts with log $(M_*/$\msun) $< 10.5$. At $z>1$, there are no merger remnants with $f_{\rm gas,sf}<0.1$, and no observable offset AGN. The effect is less severe in the \fgasa\ model, where spins are less highly aligned, but higher-mass, low-$f_{\rm gas}$ galaxies are still favored as offset AGN hosts. 

To summarize, we find that observable offset AGN favor host masses of log ($M_*$/\msun) $=$ 10 - 11 in the random spin models and that higher-mass hosts are favored in the hybrid spin models. Offset AGN are more likely to be found in hosts with low to moderate cold gas fractions, especially in the hybrid spin models. These trends indicate that the host properties of offset AGN could provide indirect information about BH spin alignment. However, because the distributions of $M_*$ and $f_{\rm gas,sf}$ are broad, a statistical sample of offset AGN would be required to distinguish between spin models. Alternatively, finding even one example of an offset BH in a very massive galaxy would strongly argue for spin misalignment in at least some cases.

Distinguishing between BH spin models based on host stellar masses would also likely require space-based observations. The bottom panels of Figure \ref{fig:mstar_hist} show the $M_*$ distribution for all BH merger hosts and for offset AGN hosts in the \randdry\ model, as in the top panels, but for SDSS resolution and sensitivity. Relative to the population detectable with \hcos, there are fewer offset AGN overall, and the preference for low-mass hosts largely disappears. Because lower-mass BHs (hosted in lower-mass galaxies) have shorter recoiling AGN lifetimes, they are more likely to exhaust their fuel supply before reaching an offset resolvable with ground-based instruments. Also, many low-mass BHs are too faint to be detected in SDSS at $z \ga 0.5$. 

We again note that the BH merger sample is incomplete in the dwarf mass regime (log $(M_*/$\msun) $\la 10$) owing to mass resolution limits. However, the BH occupation fraction in such galaxies is not empirically constrained in any case. At the massive end, the resolution limits on minor mergers are unimportant, because these events do not produce observable recoils.

It is important to bear in mind that these galaxy merger remnant models are constructed using the {\em progenitor} galaxy properties in the simulation snapshot prior to each BH merger; as such, they do not account for evolution that may occur between the snapshot and the time of merger. In particular, the gas fractions shown in Figure \ref{fig:mstar_fgas} can be considered ``initial" values, in that some of this gas will form stars or be expelled during the galaxy merger, prior to the BH merger and recoil.  

\subsection{Dependence on model parameters}
\label{ssec:paramstudy}

\subsubsection{Accretion disk parameters}
Accretion onto recoiling BHs is assumed to occur via a thin, radiatively-efficient ``alpha"-disk in the inner regions. At low Eddington ratios, theory predicts that accretion will become radiatively inefficient, and our accretion disk model will become unphysical. In order to avoid artificially long offset AGN lifetimes, a minimum Eddington ratio for the recoiling AGN must be imposed. This is especially important at low redshift, where AGN observations are generally not flux-limited. In our fiducial models, we adopt a minimum $f_{\rm Edd} = 10^{-2}$ for recoiling AGN, motivated by the distribution of Eddington ratios in observed quasars \citep[e.g.,][]{shekel12}. This is a conservative choice; massive BHs in particular may be detectable at lower Eddington ratios. Here we explore the consequences of adopting a lower minimum value of $f_{\rm Edd} = 10^{-3}$. 

\begin{figure}
\begin{center}
\includegraphics[width=0.4\textwidth]{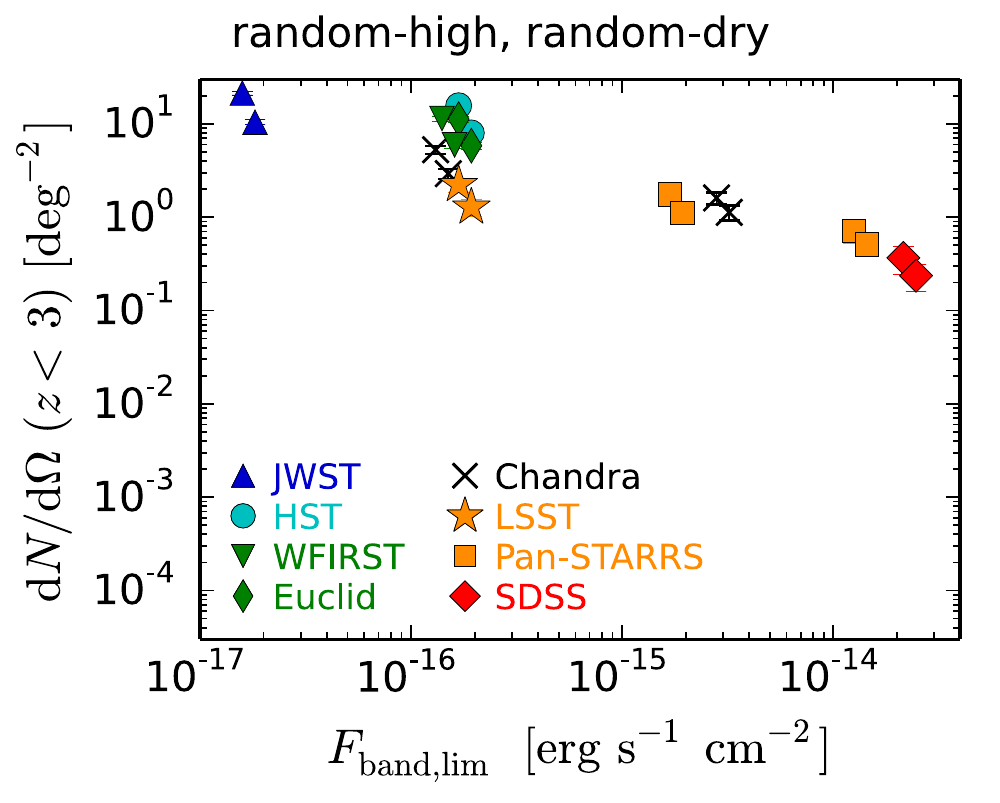}
\includegraphics[width=0.4\textwidth]{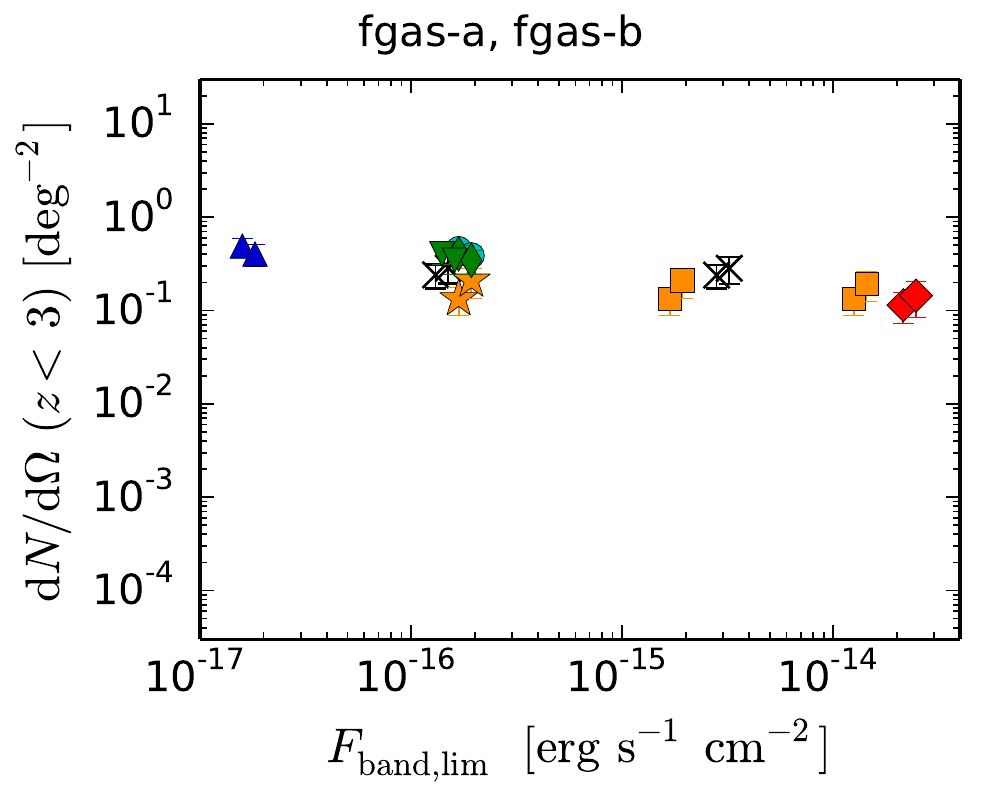}
\includegraphics[width=0.4\textwidth]{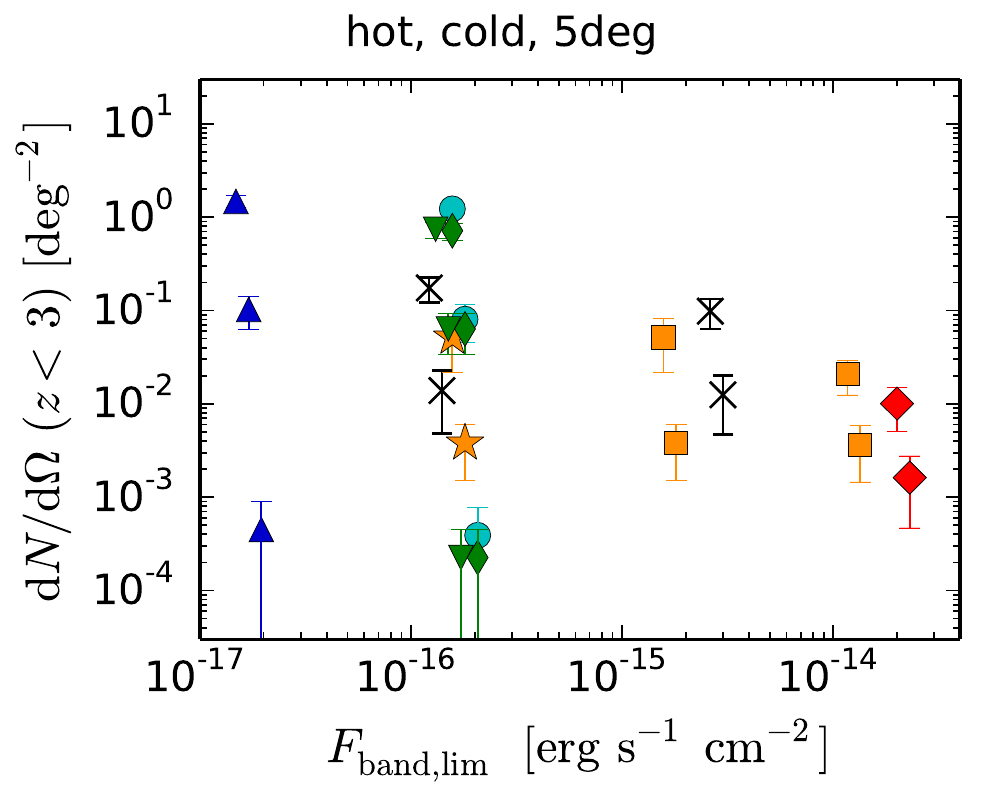}
\caption{Same as the left column of Figure \ref{fig:source_counts}, but for an $\alpha$ viscosity parameter of 0.03. The predicted rates of observable spatially-offset AGN are generally a factor of a few higher than in the fiducial model, but in the aligned spin models the predictions for seeing-limited observations increase by more than an order of magnitude. Similar enhancements relative to the fiducial model are seen for the $f_{\rm Edd,min} = 10^{-3}$ and radiatively-efficient models. \label{fig:source_counts_alpha}}
\end{center}
\end{figure}

The statistics of recoil events and properties of their hosts do not change; only the observable lifetime of each recoil event depends on the minimum Eddington ratio. Spatially-offset AGN lifetimes can be $\ga$ 100 Myr in the $f_{\rm Edd,min} = 10^{-3}$ model, while the maximum lifetimes for $f_{\rm Edd,min} = 10^{-2}$ are $\sim 40$ Myr. Similarly, velocity-offset AGN have a maximum lifetime of $\sim 60$ Myr, versus a maximum of $\sim 30$ Myr for $f_{\rm Edd,min} = 10^{-2}$. (Velocity-offset AGN lifetimes are more often limited by deceleration in the host potential than by the AGN luminosity.)

In most cases, setting $f_{\rm Edd,min} = 10^{-3}$ produces a factor of a few more spatially offset AGN per square degree than in the fiducial model. Because the AGN reach larger separations before falling below the minimum $f_{\rm Edd}$, the biggest increase in number of observable recoils occurs for seeing-limited observations, where spatial resolution is most often a limiting factor. Particularly in the aligned spin models, where large spatial offsets are rare, the $f_{\rm Edd,min} = 10^{-3}$ model produces at least 10 times more spatially-offset AGN per deg$^2$. Specifically, the \hot\ spin model predicts 1-2 spatially-offset AGN in \hcos\ and $> 10^3$ in the LSST Main Survey, while even the \cold\ spin model predicts nearly 100 spatially-offset AGN to be observable with LSST. Velocity-offset AGN number counts for $f_{\rm Edd,min} = 10^{-3}$ are more similar to the fiducial model, increasing by less than a factor of two. This is because at low redshift, the offset AGN lifetimes are still more often limited by $f_{\rm Edd,min}$ than by the absolute flux limit, even for $f_{\rm Edd,min} = 10^{-3}$. The results are more sensitive to the assumed flux limit than in the fiducial models, but only moderately so.

We also consider how our results depend on the choice of $f_{\rm riaf} = 0.05$, the critical Eddington ratio at which accretion transitions to a radiatively-inefficient regime. If this regime is ignored entirely, such that a constant radiative efficiency of 0.1 is used at all times, the maximum spatially-offset AGN lifetime is $\sim 90$ Myr.  Similar to the $f_{\rm Edd,min} = 10^{-3}$ model, these longer lifetimes increase the number of predicted observable recoils by up to a factor of a few for most models, and by more than an order of magnitude for seeing-limited observations in the aligned spin models. 

Note that assumption of a time-dependent accretion rate is especially important in these models, owing to their longer AGN lifetimes. For both the $f_{\rm Edd,min} = 10^{-3}$ and radiatively-efficient models, assuming a constant accretion rate with AGN lifetime $M_{\rm disk}/\dot M_{\rm BH}$ reduces the predicted number of spatially-offset AGN by a factor of a few to a few tens in most cases. Velocity offset AGN lifetimes are also generally shorter by a factor of a few if a constant accretion rate is assumed.

Finally, the $\alpha$ viscosity parameter is also important in determining the accretion disk lifetime. Our fiducial model assumes $\alpha = 0.3$, motivated by models of self-gravitating accretion disks \citep[e.g.,][]{goodma03}. The nature of the viscous accretion flow is uncertain, however, and $\alpha$ is commonly assumed to be lower. Accordingly, we have also tested a model with $\alpha = 0.03$. Figure \ref{fig:source_counts_alpha} shows the resulting predictions for the number of observable spatially-offset AGN per deg$^2$, in the same manner as Figure \ref{fig:source_counts}. The lower value of $\alpha$ yields longer offset AGN lifetimes and higher average luminosities. The predicted number of observable recoils is higher by a factor of a few to $\ga 10$ than in the fiducial model, a slightly larger enhancement than seen in the $f_{\rm Edd,min} = 10^{-3}$ or radiatively-efficient models. Again, predictions for seeing limited observations in the aligned spin models are most sensitive to the change in accretion disk parameters.

Overall, we find that our results depend on the accretion model for recoiling AGN, but because the fiducial model is designed to be conservative, variations in the accretion model parameters tend to {\em increase} the observability of recoils. Moreover, the aligned spin models are the most sensitive to changes in the accretion rate; for the random and hybrid models, the alternate accretion models tested here generally increase the predicted number of observable offset AGN by factors of a few at most. 

\subsubsection{Circumnuclear gas disk}

Our fiducial galaxy models include a compact circumnuclear gas disk with a mass equal to the total mass in cold, star-forming gas in the progenitor galaxies. This is well-motivated by the correlation of galaxy mergers with both nuclear starbursts and AGN fueling, which is observed empirically and in simulations. Simulations have also shown that the steep central potentials created during gas-rich mergers can greatly suppress recoil trajectories \citep{blecha11,sijack11}. As discussed in Section \ref{ssec:hosts}, the gas component included in our galaxy models does indeed reduce offset AGN lifetimes in gas-rich galaxies, such that recoils are more likely to be detected in merger remnants with relatively low gas fractions. 

However, the simplified nature of these potential models cannot fully capture the complexity of a galaxy merger remnant. In particular, while recoiling BH orbits will generally be centrophilic, irregularities in the potential will prevent them from returning exactly to the galaxy center, thus delaying their orbital decay. We have also neglected the consumption of cold gas by star formation in the interval between the simulation snapshot time (when the progenitor galaxy properties are obtained) and the BH merger time, as well as during the recoil event. The amount of cold gas present in the merger remnant model may therefore be systematically overestimated. In order to understand how these assumptions regarding the gas disk model influence our results, we consider the case in which the gas disk is neglected entirely in the galaxy potential model. (The host gas fraction is still used to determine pre-merger BH spin alignment in the hybrid spin models.)

Naturally, galactic escape speeds are lower in the absence of a gas disk component. For the random spin models, the escape fractions are significant: 20\% and 30\% of recoil events for the \randdry\ and \randhigh\ models, respectively. Moreover, even when scaling the kick speed to the escape speed, return times (and maximum displacements) are larger for a given \vk/\vesc\ in the model without gas. Recoils with \vk/\vesc\ $\sim 0.7$ can displace the BH from the galactic center for up to 1 Gyr, a factor of a few longer than in the fiducial model (Figure \ref{fig:rmaxtreturn}). Recoils with low to moderate \vk/\vesc\ ($\la 0.6$) are most strongly affected by the shape of the inner density profile.  When a dense gas component is included, some BHs with \vk/\vesc $< 0.6$ have return times $< 10^5$ yr (Figure \ref{fig:rmaxtreturn}), but in the model without a gas disk, such events are rare even in the aligned spin models. 

\begin{figure}
\includegraphics[width=0.495\textwidth,trim=8 18 -12 0]{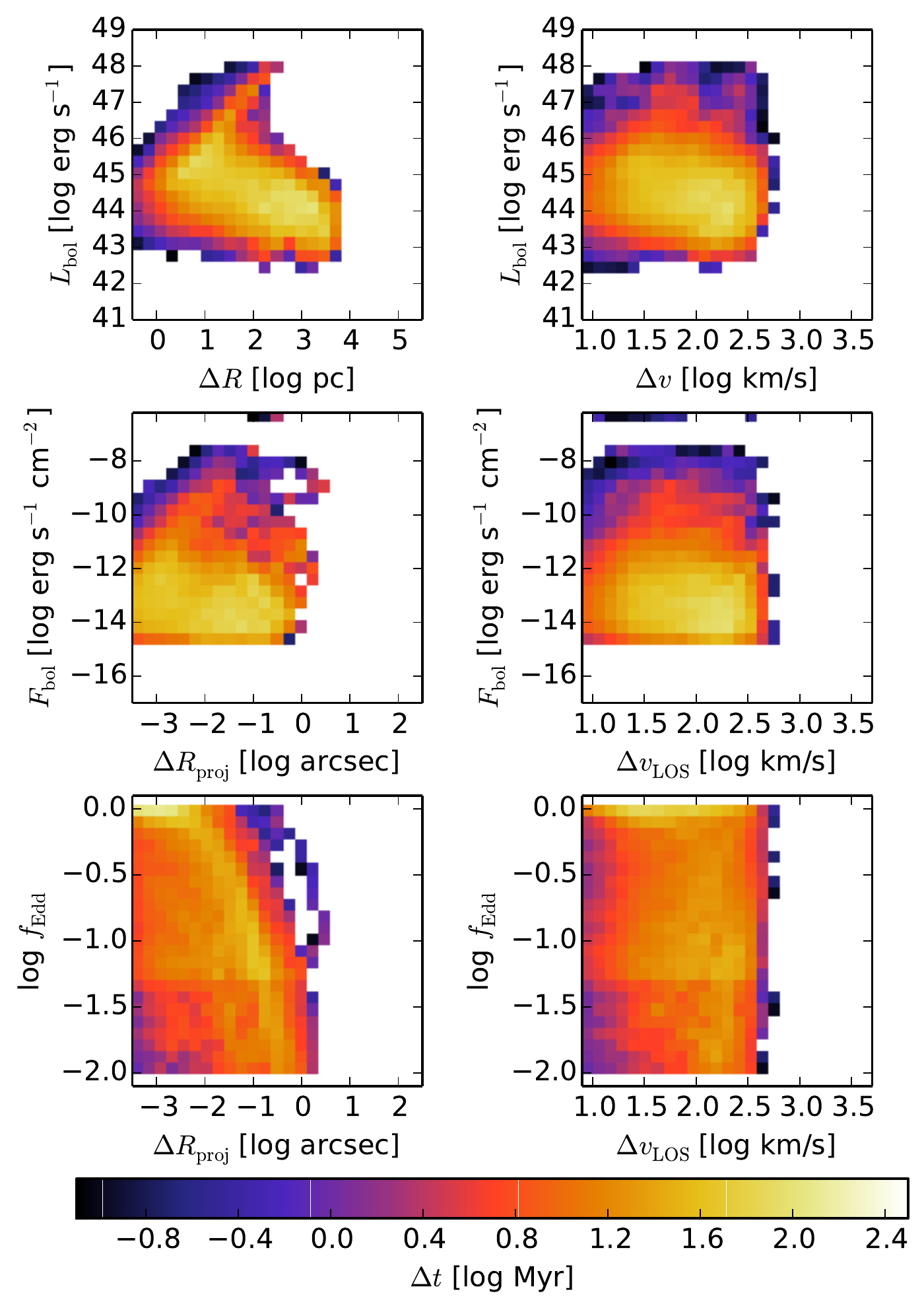}
\caption{Same as Figure \ref{fig:drdvallAGN5deg}, but for a model in which no gas disk is included in the merger remnant potential. The \fivedeg\ spin model is shown. Recoil events with \vk/\vesc\ $\la 0.6$, which dominate the offset AGN population in this model, are most affected by the absence of a gas disk component. \label{fig:nogasdrdvallAGN5deg}}
\end{figure}

As a result, offset AGN lifetimes are longer in the model with no gas disk, and a larger fraction of recoil events produce resolvable spatial offsets. Figure \ref{fig:nogasdrdvallAGN5deg} shows the same distributions of offset AGN properties as in Figure \ref{fig:drdvallAGN5deg}, for the \fivedeg\ spin model and \hst\ resolution and sensitivity, but for the model with no gas disk. The longer return times for moderate-velocity recoils create a population of flux-limited offset AGN at $\Delta R_{\rm proj} \la 0.5"$ that is largely absent in the fiducial model.  These offset AGN would be resolvable with \hst, \jwst, Euclid, and WFIRST; the no-disk model predicts that up to 14 spatially-offset AGN per deg$^2$ could be found in \hcos, and 7 per deg$^2$ with Euclid and WFIRST. Even the cold spin model predicts that 1-2 spatially-offset AGN could be detected in \hcos. For coarser angular resolutions ($> 0.5"$), though, the predicted numbers of spatially-offset AGN are closer to the fiducial model predictions, because these offsets mostly result from larger kicks that are less strongly affected by the inner potential. 

The offset AGN that are observable in the no-disk model but not the fiducial model are almost exclusively hosted in gas-rich mergers ($f_{\rm gas,sf} > 0.3$), and most occur at high redshift ($z>1$). Because gas-rich galaxies have lower masses, on average, low-mass hosts are more heavily favored in the no-disk model. Also, in contrast to the fiducial model, there is no strong preference in the no-disk model for offset AGN to be hosted in gas-poor galaxies.

\subsubsection{Gas-rich merger definition}

We distinguish between gas-rich and gas-poor galaxy mergers to determine which BH binaries are likely to have an ample reservoir of cold gas capable of aligning the BH spins. Gas-poor mergers are also assumed to require a longer timescale for the BHs to merge, though this has little effect on our results relative to the mass-ratio dependence of our merger-timescale prescription. We have adopted a conservative definition of gas-rich mergers as those with a star-forming gas fraction $f_{\rm gas,sf} > 0.1$. According to this definition, more than 85\% of mergers at $z<1$, and nearly all of those at $z>1$, are gas-rich. In the hybrid spin models, a less stringent definition of gas-rich mergers would result in fewer aligned spins and a greater number of superkicks. 

We test this assertion by considering a model with a critical gas fraction of 0.2, for which $\sim 75\%$ of galaxies at $z<1$ are defined as gas-rich. For the random and aligned spin models, where the critical gas fraction enters only into the merger delay timescale, the results do not change significantly. The predicted numbers of observable recoils differ by a few percent from the fiducial model. For the hybrid spin models, the model with critical $f_{\rm gas,sf} = 0.2$ predicts a factor of 2 - 3 more observable recoiling AGN (for spatial and velocity offsets) than the fiducial model. The preference for high-mass host galaxies also largely disappears, because greater numbers of low-mass galaxies are defined as gas-poor, and thus can produce high-velocity kicks. 

In reality, BH spin alignment should depend on the mass (and angular momentum) of gas in a circumnuclear disk around the BHs, rather than the total gas mass in the galaxy. However, because it is unclear what fraction of the cold gas will condense into a coherent disk during the merger, we have opted to tie spin alignment to the global character of the merger (gas-rich or gas-poor). For completeness, we have also tested a model in which the total star-forming gas mass is compared to the total BH mass, with gas-rich galaxies defined as those with $M_{\rm gas,sf}/M_{\rm BH} > 10$ (such that $M_{\rm disk}$ \ga $M_{\rm BH}$ if $\ga 10\%$ of the total cold gas mass is condensed into a circumnuclear disk). In this case, $> 95\%$ of mergers are gas rich at $z<1$, indicating that this is an even more conservative definition of gas-rich mergers than our fiducial model. 

Again, for random and aligned spins, there is negligible difference between this model and the fiducial case. For hybrid spin models, despite the small fraction of mergers that do not have aligned BH spins, only a factor of $\sim 2$ fewer offset AGN are predicted with this definition of gas-rich mergers. We also note that, while offset AGN tend to have high BH masses in the hybrid spin models regardless, this effect is even stronger when spin alignment is tied to the BH mass.

\subsubsection{BH merger timescales}

As discussed in Section \ref{ssec:sims}, a delay is added to the time of each BH merger in our model, because BH binary evolution cannot be modeled on sub-grid scales, and because minor-merger timescales in particular may be underestimated by the BH prescriptions in Illustris. If we examine the results without this delay timescale added, we find relatively minor differences; the delay timescale lowers the predicted numbers of offset AGN by less than a factor of two in most cases. In part, this owes to the small fraction of {\em major} mergers for which the BH binary is still unmerged at $z=0$ according to this prescription (2.5\% of mergers with log $q > -1.5$). Also, the unmerged binaries that are removed from our sample are balanced to some extent by higher-redshift binaries that have long delay timescales, such that the corresponding recoil events occur at lower redshift, where they are more easily observable. 

As shown in Figure \ref{fig:mrgrate}, the delay timescale shifts the peak of BH merger activity to slightly lower redshift, from $z \sim 2$ to $z \sim 1.5$. Although this does not produce a large effect on the observability of recoils, these longer merging timescales may have a more significant impact on GW signals from BH mergers. The inspiral timescales of BH binaries on subgrid scales, and the resulting GW signal, will be explored in detail in Kelley et al. (in preparation).

\section{Discussion}
\label{sec:discuss}
 
\subsection{Detection and confirmation of recoiling AGN}
\label{ssec:confirmation}

Our models predict the numbers of spatially or kinematically offset AGN that are a) resolvable and b) above the sensitivity limit for a given observational survey. Once a candidate recoiling AGN is identified via a spatial or kinematic offset, however, extensive multi-wavelength follow-up observations may be required to rule out alternate scenarios. Alternative explanations for existing recoil candidates include dual or binary BHs in which only one BH is active \citep{komoss08,bogdan09, dotti09, comerf09b,batche10, robins10, tsalma11,eracle12, lena14}, extreme double-peaked emitters \citep{shield09b, tsalma11, steinh12, eracle12}, chance superposition \citep{shield09a,heckma09}, or unusual supernovae \citep[SNe,][]{koss14}. Interestingly, \citet{decarl14} find that neither the recoil scenario nor other proposed explanations for the velocity-offset quasar J0927+2943 are consistent with recent observations. 

Type IIn SNe may resemble AGN for months or years after the explosion; they commonly have both broad and narrow emission lines and slow decay times owing to interactions of shocked ejecta with circumstellar material. However, as their spectral features evolve and they eventually fade, such events should be distinguishable from AGN via continued monitoring and multiwavelength observations. 

Extreme double-peaked emitters in which only one peak is visible may also resemble recoiling AGN, in that they can produce systemic velocity shifts in broad emission lines \citep{chehal89,eracle95}. Different BLs often have very different velocity shifts in such cases, while to first order, recoiling AGN should exhibit a consistent shift in all broad lines (owing to the bulk motion of the AGN). Similarly, the orbital motion of sub-parsec binary BHs in which only one BH is active may give rise to velocity-shifted BLs. In this case, the velocity shift will evolve on the orbital timescale, while recoiling AGN should exhibit a constant velocity shift.

One of the more challenging alternative scenarios to rule out for candidate recoils is that of a well-separated ($\sim$ kpc scale) BH {\em pair} in which only one BH is actively accreting. This is the case for CID-42, which has two distinct nuclei separated by 2.5 kpc, one of which is a confirmed AGN with a BL offset of 1300 \kms\ \citep{civano10,civano12b,novak15}. Note that the magnitude of this spatial and velocity offset is quite consistent with the predictions of our models, if spins are not highly aligned (Figure \ref{fig:drdvallAGN}). The offset BL and lack of evidence for an AGN in the second nucleus support the recoil scenario, and indirect arguments suggest that the host galaxy properties are more consistent with a post-merger phase, but it is difficult to rule out the possibility that a {\em quiescent} BH is harbored in the second nucleus \citep{blecha13a}. 

An additional challenge in confirming spatially-offset AGN is that recoil events occur, by definition, in recently-merged galaxies. The host may therefore be highly disturbed, such that identifying the galaxy centroid is nontrivial. In cases where the BH binary inspiral time may be long, for example in gas-poor mergers, the host may have a more relaxed morphology by the time the BH merger and recoil occurs.  However, we have demonstrated that if spins are not highly aligned, offsets larger than a kpc should be common. In such cases, it is not necessary to pinpoint the host centroid to great accuracy in order to determine that the AGN is offset. This is crucial for any programme to search for offset AGN in ground-based, all-sky surveys, where galactic morphological features often will not be resolved in great detail. Similarly, because observable recoiling AGN also tend to have large velocity offsets ($\ga$ 500 - 1000 \kms), any ambiguity in the rest-frame redshift of the disturbed host (owing to residual motion of the NL gas, for example) is unlikely to preclude the detection of such offsets. 

As noted earlier, our models do not account for the possibility that a recoiling BH may encounter fresh fuel as it settles back to the galactic center, which would produce recoiling AGN with much smaller spatial offsets. This could be especially relevant for recoils in dry merger remnants, where the BH may undergo long-lived, small-amplitude oscillations in a core potential \citep{guamer08, lena14}. Regardless, our results suggest that if BH spins are not highly aligned, a population of spatially-offset recoiling AGN may be resolvable with seeing-limited observations.

The most unambiguous recoiling AGN candidates will likely be those with both spatial and velocity offsets, as in CID-42, or with a projected spatial offset large enough that the AGN is well-separated from the stellar light of the host galaxy. Our results indicate that a majority of velocity-offset AGN should also have spatial offsets resolvable with \hst, and 5 - 20\% of spatially-offset AGN resolvable with ground-based observations should also have LOS velocity offsets $> 600$ \kms. Additionally, up to 50\% of spatially-offset AGN resolvable with the ground-based imaging (and up to 25\% of those resolvable with \hst) should appear at radii beyond the host stellar bulge. Thus, the possibility exists of finding hundreds of such objects with Pan-STARRS and LSST, if spins are not strongly aligned. 

Recoiling AGN with large spatial offsets also tend to have low Eddington ratios, as their accretion rate declines over time. We assume that accretion transitions to a radiatively-inefficient regime below $f_{\rm Edd} = 0.05$;  this regime may often be associated with radio emission in AGN. Thus, high radio-to-optical flux ratios in offset AGN, or even small-scale radio jets, could be another good indication of a past recoil event. 

We note that the hybrid and aligned  spin models predict few recoils to be detectable at $z>1$. For the random spin models, this is not the case; velocity-offset AGN in particular may be found in equal or greater numbers at $z>1$ as at lower redshift (Figure \ref{fig:source_counts_zbins}). This raises the interesting possibility of detecting recoiling AGN near the cosmic peak of quasar and galaxy merging activity, and such detections would place additional indirect constraints on pre-merger BH spins. Evidence of past BH mergers at high redshift would also place stronger limits on BH merging timescales than could be obtained in the local Universe. However, if such events are observable, they would be more challenging to confirm with follow-up observations than their low-redshift counterparts.

Our models provide only analytic approximations to the properties and structure of the host galaxies of recoiling AGN, which should highly dynamic during major mergers. The total stellar mass and star-forming gas content are readily obtained from the progenitors, however, and we can identify trends in these properties of offset AGN hosts. Specifically, although the host stellar mass distribution is broad, stellar masses of log ($M_*$/\msun) $\sim$ 10 - 11 are favored in the random and aligned spin models. In the hybrid spin models (for which spin alignment occurs only in gas-rich mergers), offset AGN occur preferentially in higher-mass hosts, log ($M_*$/\msun) $\ga 11$. This is a testable prediction, but a statistical sample of offset AGN would be required to differentiate between these distinct but overlapping distributions. 

We also find that merger remnants with low to moderate star-forming gas fractions are preferred as offset AGN hosts, as recoil trajectories tend to be quickly suppressed in gas-rich mergers. This is especially true of the hybrid spin models, where almost no observable recoils occur in gas-rich merger remnants. We suggest that a systematic search for offset AGN might focus initially on bulge-dominated or disturbed galaxies with low to moderate star formation rates.

Again, we caution that our sample is incomplete at the low-mass end owing to resolution limits, such that offset AGN in low-mass or dwarf hosts are largely excluded from our analysis. This is an important caveat; although the merger rate and BH occupation fraction of dwarf galaxies are poorly constrained, the low escape speeds of low-mass galaxies should result in resolvable offsets in many if not most recoil events, even in galaxies that are gas-rich. The recoil candidate SDSS 1133 is one possible example of this \citep{koss14}. Low-mass merger remnants may therefore host a population of spatially-offset AGN in addition to those predicted by our models. 

As a final caveat, our results assume that recoiling AGN generally resemble ``normal," stationary AGN, in that they retain an accretion disk and broad line region and have an SED that is well-modeled with standard AGN templates. This seems a reasonable first-order assumption, given that the region producing most of the AGN emission will typically have orbital speeds $v_{\rm orb} \gg$ \vk\ and will feel the recoil event as a negligible perturbation. Nonetheless, the possibility that recoiling AGN appear qualitatively different from stationary AGN cannot be discounted, and until recoil events can be unambiguously confirmed, this remains an uncertainty for any search for such objects. 

The detection of even one recoiling AGN with a large velocity offset ($\ga 1000$ \kms) or spatial offset ($\ga 10$ kpc), or alternatively the discovery of a massive galaxy with no central BH, would strongly indicate that binary BH spins are not efficiently aligned in all cases. Our results suggest that up to hundreds or thousands of spatially offset AGN may be detectable with Pan-STARRS and LSST. While these numbers are dependent on model assumptions, as discussed in Section \ref{ssec:paramstudy}, we have attempted to be conservative in constructing the fiducial models. If such a population of recoiling AGN is identified, then further distinctions between BH spin and recoil velocity distributions may be possible. If no convincing recoil candidates are found in large-area surveys, the simplest explanation may be that spin alignment is effective for most binary BHs, or that spin magnitudes are low. Constraints on the efficiency of spin alignment could also provide indirect information about the nature of the accretion flow, specifically whether it is likely to be coherent on timescales required to align the BH spins, or if it is more often characterized by chaotic, randomly-oriented streams.

\subsection{Comparison with previous work}
\label{ssec:prevwork}
 
Our predicted rates of observable offset AGN are substantially higher than those of VM08, the only previous study of recoiling AGN in a cosmological framework. Here, we identify some key differences that contribute to this discrepancy. 

VM08 used semi-analytic merger tree models to determine the BH merger rate and host properties, similarly integrating recoil trajectories for each event in a galaxy potential model. Our predictions for the number of observable spatially-offset AGN per deg$^2$ in the \randhigh\ spin model (Figure \ref{fig:source_counts}) can be compared with their results for \hst, \jwst, and \chandra. For our fiducial models, these numbers are comparable to the predictions of  the VM08 ``halo-only" models, and they are a factor of $> 10$ - 100 above their ``halo$+$bulge" predictions. Given that our fiducial model includes not only a halo and compact stellar bulge but also a dense gas disk, the recoiling BH return times are much shorter than in a halo-only model. Thus, our model predictions are in fact significantly higher than those of VM08. 

Neglecting the gas component in our models increases the discrepancy with the VM08 results for a halo$+$bulge potential, up to a factor of several hundred for \hst\ and \jwst. If we include only a DM halo potential, $\sim$ 15 - 25 times more spatially offset AGN are predicted than in our fiducial model. 

The recoil kick velocity distribution is sensitive to the BH mass ratio distribution, but the kick formula of \citet{baker08} used in VM08 depends even more sensitively on $q$ than is predicted in more recent work \citep[e.g.,][]{vanmet10,lousto12}. Specifically, the \citet{baker08} formula scales with $\eta^3$, where $\eta \equiv q/(1+q)^2$ is the symmetric mass ratio, while the more recent formulae include an $\eta^2$ term. As a result, the \citet{baker08} formula produces fewer high-velocity recoils, and the steeper dependence on $q$ amplifies discrepancies in the underlying distributions. Using this formula in our models lowers our predictions by a factor of $\sim 2$.

The ejected accretion disk model of \citet{loeb07} used in VM08 has a steep dependence on BH mass ($\propto M_{\rm BH}^{2.2}$). Thus, the disk must be capped at $M_{\rm disk} = M_{\rm BH}$ in most cases and depends sensitively on the underlying BH mass distribution. Our fiducial models instead assume the disk model of \citet{bleloe08}, which accounts for disk self-gravity and yields typical disk masses of a few percent of the BH mass. Furthermore, the accretion rate in our models is assumed to decline over time as the disk diffuses outward. This increases the predicted number of spatially-offset AGN by a factor of a few relative to a constant-$\dot M$ model.  

If we attempt to compare as closely as possible with the results of VM08, using the same potential models, recoil kick formula, and ejected accretion disk model, we still predict spatially-offset AGN to be observable at rates that are tens to hundreds of times higher than in any of their models. The discrepancy is larger when comparing to the halo$+$bulge models than the halo-only case, presumably because the bulge mass is directly scaled to the BH mass and thus is sensitive to differences in the underlying distribution. Owing to the significant differences in numerical techniques and prescriptions for BH seeding, growth, and feedback between the semi-analytic models and cosmological simulations, it is not surprising that the resulting BH merger and recoil properties are qualitatively different. The offset AGN population has a nontrivial dependence on the details of each method, but nonetheless, a few important differences can be pinpointed.

First, as noted in Sec~\ref{ssec:mrgrate}, the BH mass ratio distribution in Illustris is qualitatively different from that shown in VM08, the latter having a sharp deficit of mergers with $q>0.3$. It is suggested that this deficit owes to the minimum mass threshold ($M_{\rm tot} > 10^5$ \msun) imposed, indicating that the BH mass distribution differs qualitatively from that in Illustris (at least for merging BHs), with lower-mass BHs favored in the latter.  VM08 also show for comparison a $q$ distribution derived from Monte Carlo sampling of their $z=0$ BH mass function, which is in much better qualitative agreement with the Illustris result. We can also compare our results with the semi-analytic model results of \citet{baraus12}; they find fewer $q > 0.3$ BH mergers than in Illustris as well, but the discrepancy is much smaller. If we consider only recoil events with $q<0.3$ in our models, the rates of observable recoils are reduced by a factor of $\sim$ 5 - 6.

While it is possible that $q\sim1$ mergers are over-represented in Illustris, owing to the uncertainties in merger timescales discussed in Section \ref{ssec:sims}, the inspiral times are most likely to be underestimated for {\em minor} mergers. We have compensated for this  by imposing a minimum BH mass of $10^6$ \msun\ and a merger delay timescale that scales with $q$. 

If the merging BH mass distribution is also different in VM08 than in Illustris -- specifically if lower masses are favored in the former, this would contribute to the discrepant results as well. The mass distribution is not explicitly given in VM08, but the reasonable agreement of the integrated merger rate for $M_{\rm BH} > 10^{6-7}$ \msun\ between Illustris and the other semi-analytic models \citep{sesana04,baraus12} suggests that the merging BH mass distribution in Illustris does not differ too greatly from those studies, at least at the high-mass end and integrated over cosmic time.

We again note that any differences in the mass and mass ratio distributions between Illustris and the models of VM08 are amplified by the sensitive dependence of the \citet{baker08} kick formula and the \citet{loeb07} disk model to these parameters. However, these factors alone do not appear to account for all of the difference between our comparison model and the results of VM08.

The cumulative BH merger rate for $M_{\rm BH} > 10^6$ \msun\ agrees well between Illustris and semi-analytic models, and the rate peaks at similar redshift \citep{sesana04,baraus12}. However, as noted in Section \ref{ssec:mrgrate}, there are fewer high-redshift mergers and more low-redshift mergers in Illustris, owing to the different BH seed formation, growth, and merger prescriptions. Differences in the redshift distribution of mergers are likely a significant contributor to the higher rates of observable recoils predicted in this work. Empirical constraints on the BH merger rate are insufficient to distinguish between these models, but the agreement of the Illustris galaxy merger rate with observations is encouraging \citep{rodgom15}, as is the agreement with the observed BH mass function and quasar luminosity function \citep{sijack15}.

\section{Summary}
\label{sec:summary}

We have developed models for AGN that are spatially or kinematically offset from their host galaxies owing to GW recoil kicks following a major merger. Using data from the Illustris cosmological simulations, we determine the characteristics of recoiling AGN and their host galaxies over cosmic time. Because recoil events may be preferentially suppressed in gas-rich mergers, the use of hydrodynamic simulations is critical for determining which recoiling AGN may be observable. Each BH merger in our sample is assigned: a) a recoil kick velocity based on the BH mass ratio and assumed spin distribution, b) an accretion disk that remains bound to the recoiling BH, and c) a merged host galaxy model based on the progenitor galaxy properties. The recoil trajectory is integrated in this galaxy potential, and the AGN luminosity is calculated at each timestep. 

This study is the first to consider hydrodynamic effects on recoiling BHs in a cosmological framework, and the first to examine the effect of pre-merger BH spin alignment on the observability of recoils. Mergers between BHs with aligned or nearly-aligned spins produce dramatically lower recoil velocities than do randomly-oriented spins, and misaligned spins are required to produce superkicks that can eject BHs from galaxies entirely. BH spin alignment is predicted to occur via interaction with a circumbinary gas disk, though the degree of alignment is not well constrained and likely depends on the properties of the gas disk. We therefore consider three categories of BH spin models: those with randomly-oriented spins, those in which BHs always undergo some amount of alignment, and ``hybrid" models in which spin alignment occurs only in gas-rich mergers. 

Our main results can be summarized as follows.

\begin{itemize}

\item Recoiling AGN do {\em not} spend most of their time at small spatial and velocity offsets, unless pre-merger BH spins are highly aligned. The longest-lived offset AGN are produced by recoils near the host escape speed. This occurs because marginally-bound and escaping BHs experience the least deceleration as they leave the host nucleus, and because marginally-bound recoiling BHs spend most of their time at large apocenters. Recoil events with \vk $\gg$ \vesc\ are intrinsically rare and have short AGN lifetimes because little gas remains bound to the recoiling BH. \\

\item Spatially-offset AGN are most commonly found at projected separations $> 0.1"$ from the host galaxy, with a tail extending beyond 10" (if BH spins are not strongly aligned). This indicates that {\em even seeing-limited observations can resolve spatially-offset AGN}, making them promising targets for current and future wide-area surveys such as SDSS, Pan-STARRS, LSST, Euclid, and WFIRST. These spatial offsets often correspond to physical projected separations $> 1$ kpc, such that they may be detectable even if the host galaxy is disturbed and its centroid ill-defined.\\

\item If BH spins are randomly oriented, our models predict that nearly 10 spatially-offset AGN may be detectable in \hcos. The Pan-STARRS1 Medium Deep Survey could find $\sim 15-20$ offset AGN, and hundreds could be found in SDSS DR7. The all-sky surveys LSST and PS1-3pi could detect up to several thousand spatially-offset AGN, as could a wide-area WFIRST survey. An all-sky survey with Euclid could detect up to several tens of thousands of such objects. \\

\item The random-spin models also predict hundreds of velocity-offset AGN with $\Delta v_{\rm LOS} > 1000 $ \kms\ in the SDSS DR7 footprint. At face value, this is in tension with the findings of \citet{bonshi07}, \citet{tsalma11}, and \citet{eracle12} that large BL offsets are rare among SDSS quasars. Many of the offset AGN in our sample may be low-luminosity or obscured AGN excluded from the SDSS quasar catalog, but this may also be a preliminary indication that pre-merger spin alignment is effective in at least some mergers, or that BH spins are low. \\

\item A population of recoiling AGN may be detected in current and future surveys {\em even if some pre-merger BH spin alignment occurs}. If spin alignment is {\em always} very efficient (within $\sim$ 5 $^{\circ}$ of the orbital angular momentum), recoils are unlikely to ever be detected. But if only moderate spin alignment occurs, up to a few tens of spatially-offset AGN could be detected with LSST or Pan-STARRS, and if spins are only aligned in gas-rich mergers, hundreds of recoiling AGN may be observable in such surveys. In the latter case, over 2000 recoiling AGN could be detected with Euclid. \\

\item For comparable assumptions regarding the recoil velocity distribution and host galaxy properties, we find a much higher incidence of observable recoiling AGN than a previous study (VM08). We argue that this discrepancy owes primarily to  systematic differences in the underlying redshift, mass and mass ratio distributions of merging BHs in Illustris versus the semi-analytic models, as well as the use of more physically-motivated models for recoiling AGN accretion disks and an updated recoil kick formula in this work. \\ 

\item A {\em majority} of recoiling AGN with resolvable LOS velocity offsets may have simultaneous spatial offsets resolvable with \hst, and $\sim 4-20$\% of velocity-offset AGN detectable with SDSS may also have spatial offsets resolvable with SDSS. This indicates that high-resolution imaging could be key for confirming any velocity-offset recoil candidates. \\

\item 
A smaller but non-negligible fraction of spatially-offset AGN ($< 10$\%) should simultaneously have LOS velocity offsets $\Delta v > 1000$ \kms. 
We also find that up to 50\% of spatially-offset AGN resolvable with seeing-limited observations (and up to 25\% of those resolvable with \hst) should be at projected separations larger than the stellar bulge radius, such that they could be distinguished from inspiraling dual BHs. AGN exhibiting simultaneous spatial and velocity offsets, or that are well separated from the host galaxy stellar light, or both, would be among the strongest candidate recoils.\\

\item The preferred stellar masses of offset AGN hosts differ depending on the degree of pre-merger BH spin alignment. If the spins are always drawn from the same distribution (random or aligned), the host stellar mass distribution peaks around log ($M_*$/\msun) $=$ 10 - 11. For the hybrid spin models, where the degree of BH spin alignment depends on the host gas content, {\em high-mass} hosts are preferred despite their larger escape speeds, because they have lower gas fractions on average and are more likely to yield superkicks from misaligned BH spins. These models also yield very few observable recoils at $z>1$, owing to the gas-rich nature of high-redshift galaxies.\\

\item In all cases, offset AGN are most likely to be found in hosts with low to moderate gas fractions, owing to the dynamical suppression of recoil trajectories by dense circumnuclear gas in gas-rich systems. The longest-lived offset AGN occur preferentially in galaxies with low gas fractions.

\end{itemize}

Our quantitative predictions for the observability of recoiling AGN have some dependence on model assumptions, including the host galaxy potential model and accretion disk parameters. As discussed in Section \ref{ssec:paramstudy}, variation with these parameters is often substantially larger than the statistical errors shown in Figure \ref{fig:source_counts}. Because the fiducial model is conservative by design, reasonable variations in model parameters tend to {\em increase} the predicted number of observable offset AGN. However, the results also depend on the underlying distributions of BH mass, accretion rate, and merger rate in Illustris. Owing to uncertainties in the sub-grid BH prescriptions, we cannot exclude the possibility of systematic bias in the merging BH population, but the success of Illustris in reproducing key observables of the BH population \citep{sijack15}, and the reasonable level of convergence of our results (Appendix \ref{ssec:converge}) are encouraging. 

These results strongly motivate a systematic search for offset AGN in current and future large-area surveys. Discoveries of GW recoils would confirm this prediction of strong-field gravity, and they would provide robust evidence of recent BH mergers, thereby constraining the timescales for BH binary inspiral. We demonstrate that observations of recoiling AGN could also place indirect constraints on the efficiency of pre-merger spin alignment, which in turn could provide new evidence regarding the nature of the BH accretion flow. 

\section*{Acknowledgements}

We would like to thank Cole Miller, Suvi Gezari, and Julie Comerford for useful discussions and comments on the manuscript, and we thank Enrico Barausse, Marta Volonteri, and Alberto Sesana for helpful discussions and comparisons with their model data. LB acknowledges support provided by NASA through Einstein Fellowship grant PF2-130093. LH acknowledges support from NASA grant  NNX12AC67G and NSF grant AST-1312095. VS acknowledges support through the European Research Council through ERC-StG grant EXAGAL-308037. Simulations were run on the Harvard Odyssey and CfA/ITC clusters, the Ranger and Stampede supercomputers at the Texas Advanced Computing Center as part of XSEDE, the Kraken supercomputer at Oak Ridge National Laboratory as part of XSEDE, the CURIE supercomputer at CEA/France as part of PRACE project RA0844 and the SuperMUC computer at the Leibniz Computing Centre, Germany, as part of project pr85je. 

\bibliography{refs_illustris_recoil_v3}

\appendix
\section{Resolution Convergence}
\label{ssec:converge}

\begin{figure}
\includegraphics[width=0.495\textwidth]{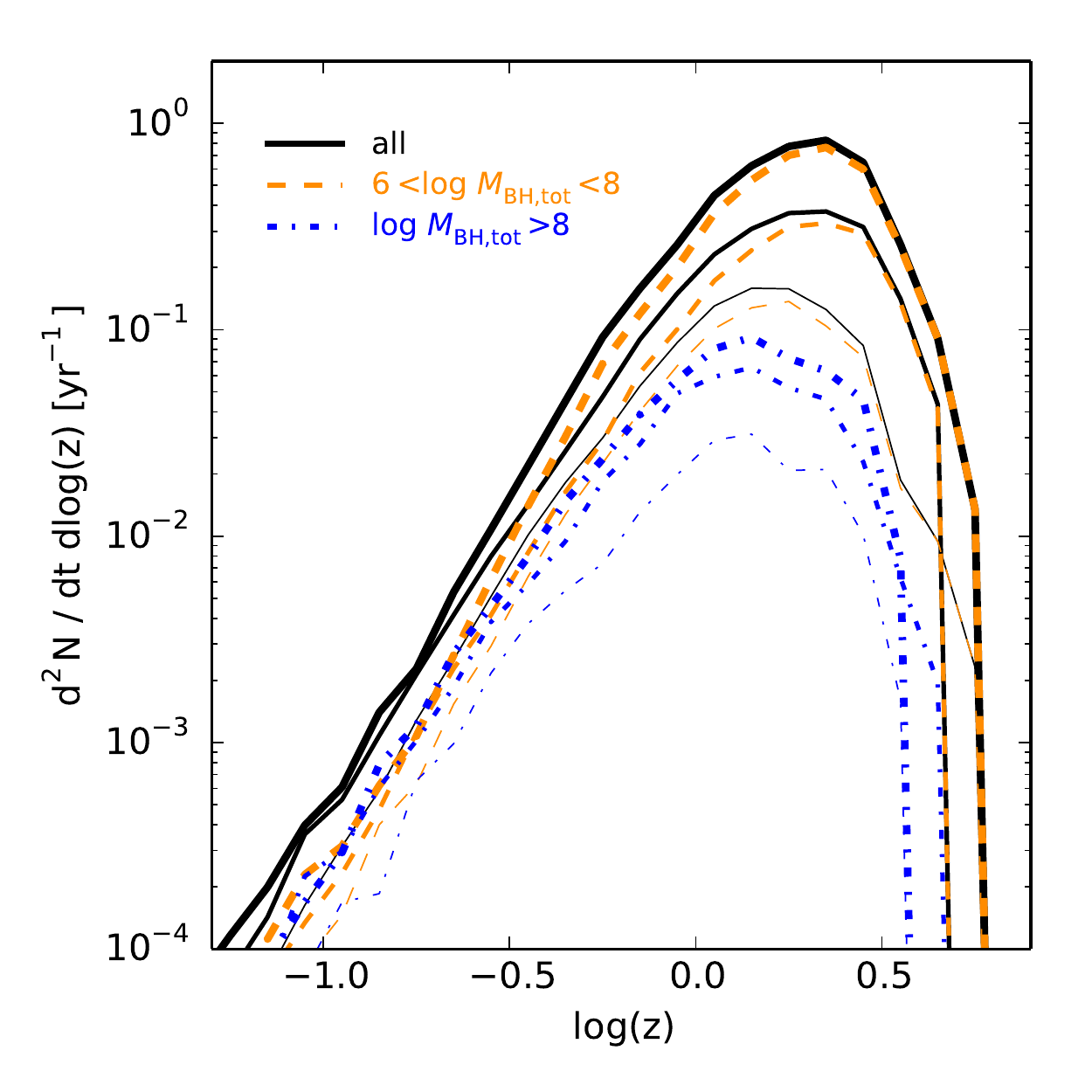}
\includegraphics[width=0.495\textwidth]{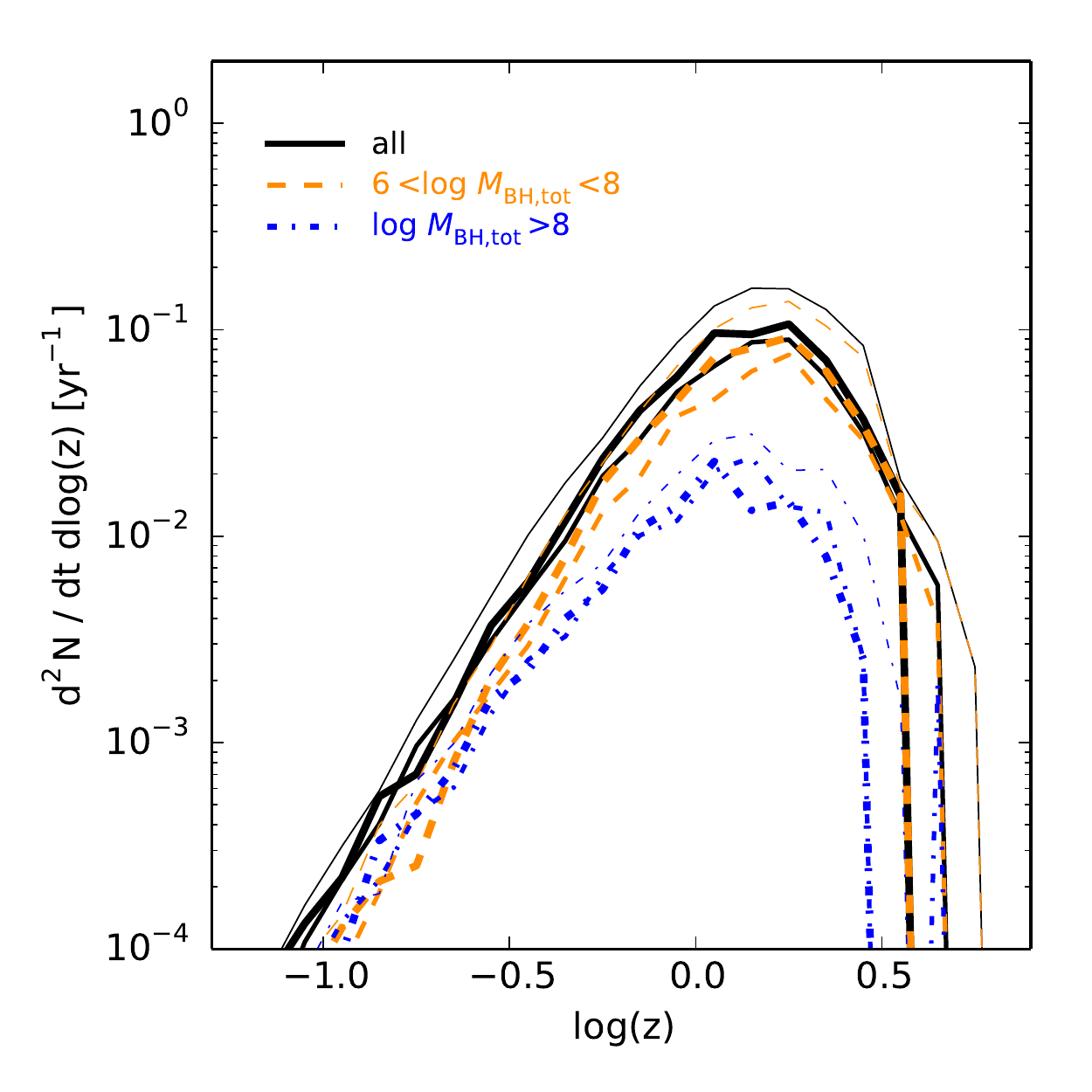}
\caption{{\em Top panel:} the BH merger rate in Illustris is shown in the same manner as in Figure \ref{fig:mrgrate}, but here the resolution convergence between Illustris-1, 2, and 3 is shown. Only mergers for which each progenitor galaxy has at least 300 DM and 80 stellar particles are included. As before, the solid black lines denote the total merger rate for our BH sample with a delay timescale added to each merger, as described in the text. The orange dashed and blue dot-dashed lines show the merger rate separated by total BH mass: $6 < $ log $M_{\rm BH} < 8$ and log $M_{\rm BH} > 8$, respectively. In all cases, thicker lines denote higher resolution simulations. {\em Bottom panel:} same as top panel, but with a constant minimum {\em mass} imposed on progenitor galaxies, rather than a minimum particle number. A minimum stellar (halo) mass of $6.4\times 10^9$\msun\ ($1.2\times 10^{11}$ \msun) is assumed, corresponding to 300 DM (80 stellar) particles in Illustris-3, such that this curve is identical between the two panels. The convergence improves substantially in this case. \label{fig:converge_mrgrate}}
\end{figure}

The results presented in Section \ref{sec:results} are based on the highest-resolution Illustris simulation (Illustris-1). Here we discuss the level of convergence between the three Illustris simulations (Illustris-1, Illustris-2, and Illustris-3), which differ in mass resolution. The BH merger rate for the three simulations is shown in Figure \ref{fig:converge_mrgrate} (top panel), for all BHs and divided by total BH mass bins. There is a trend toward convergence for high-mass mergers ($M_{\rm tot} > 10^8$  \msun) and at low redshift ($z\la 0.3$), but there is a clear dependence of merger rate on mass resolution, with higher merger rates in the higher-resolution simulations.

As described in Section \ref{ssec:model}, we consider only mergers for which each progenitor has $M_{\rm BH} > 10^6$ \msun\ and is hosted in a galaxy with at least 80 stellar and 300 DM particles. Thus, it is natural to expect a lower merger rate in the lower-resolution simulations, where low-mass galaxies are excluded from our analysis. The bottom panel of Figure \ref{fig:converge_mrgrate} compares the merger rate for the three Illustris simulations with a consistent {\em mass} cut applied to the BH host galaxies in each case, rather than a  minimum particle number. Here the Illustris-3 curves are the same as in the top panel, corresponding to a minimum stellar (halo) mass of $6.4\times 10^9$ \msun\ ($1.2\times 10^{11}$ \msun), and the Illustris-1 and 2 curves assume the same minimum mass. The convergence improves substantially in this case. This demonstrates that our fiducial sample from Illustris-1 has a higher BH merger rate primarily because more low-mass galaxies can be resolved.   

\begin{figure*}

\includegraphics[width=0.41\textwidth]{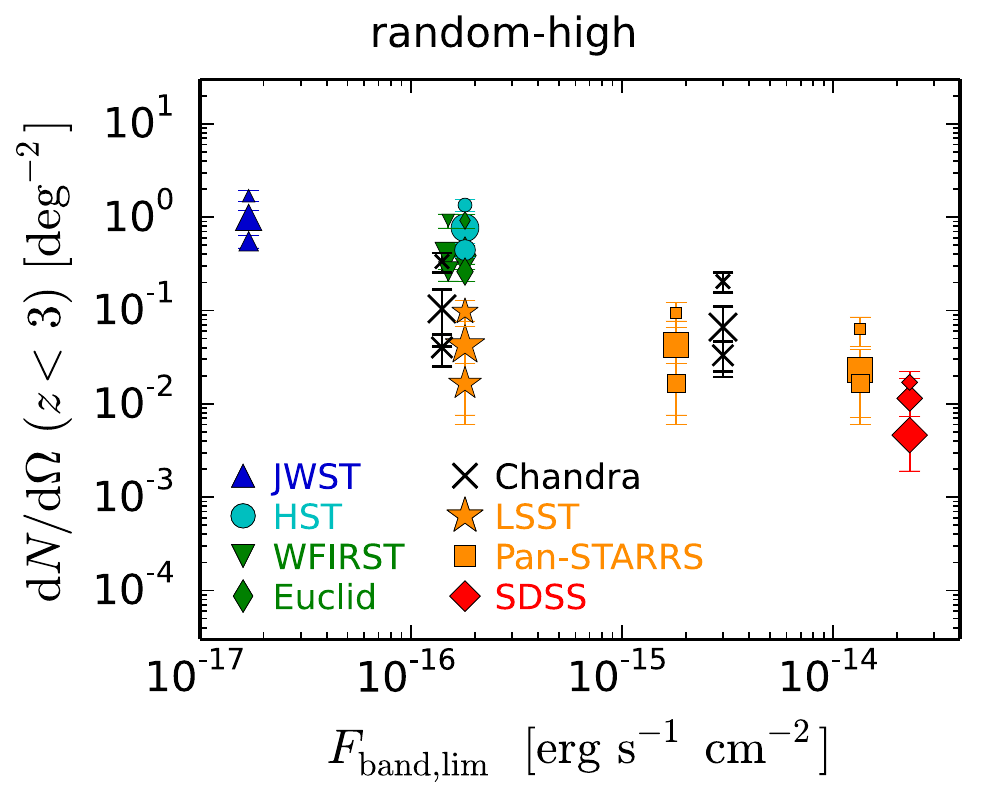}
\includegraphics[width=0.41\textwidth]{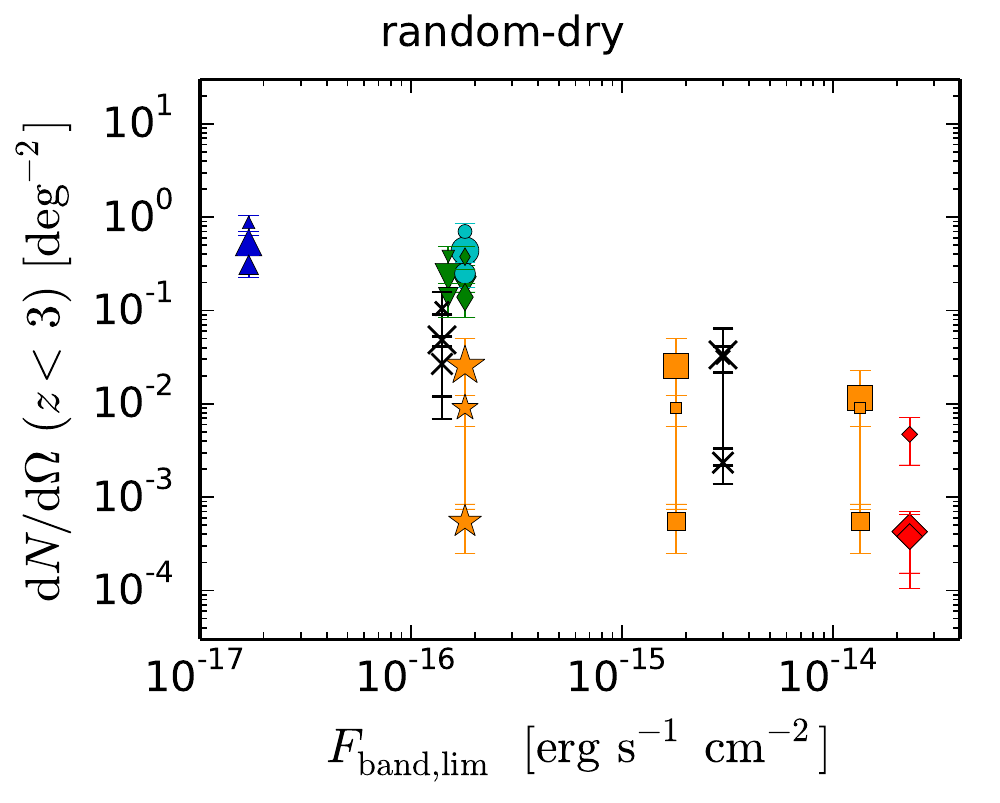}
\includegraphics[width=0.41\textwidth]{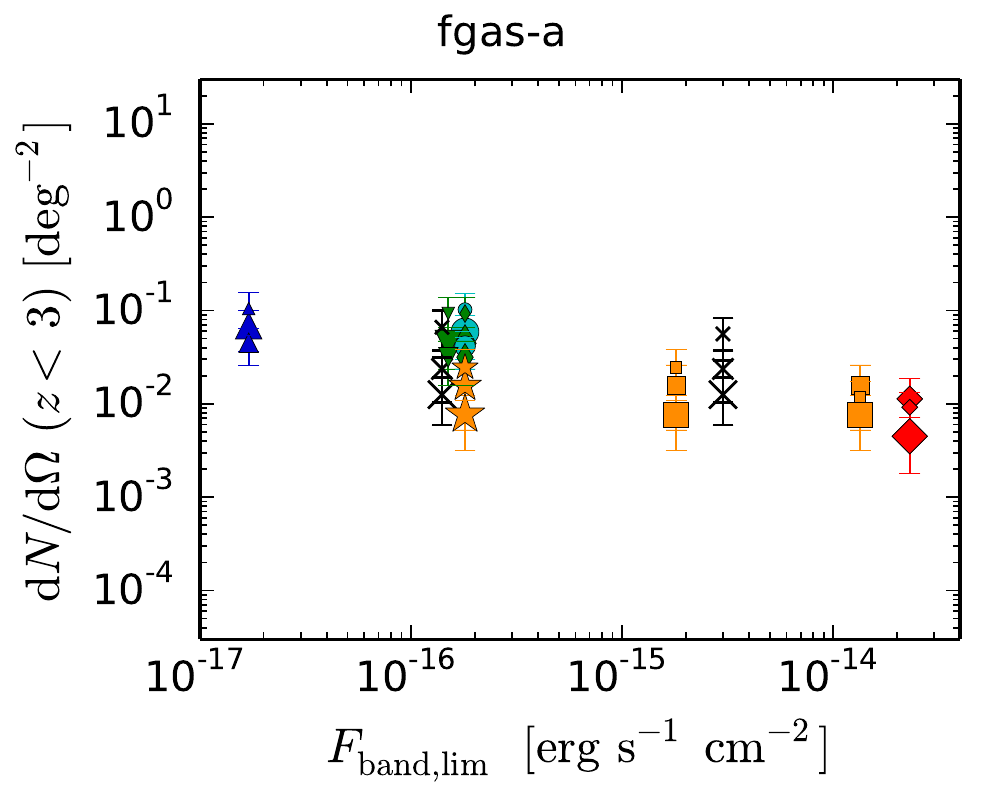}
\includegraphics[width=0.41\textwidth]{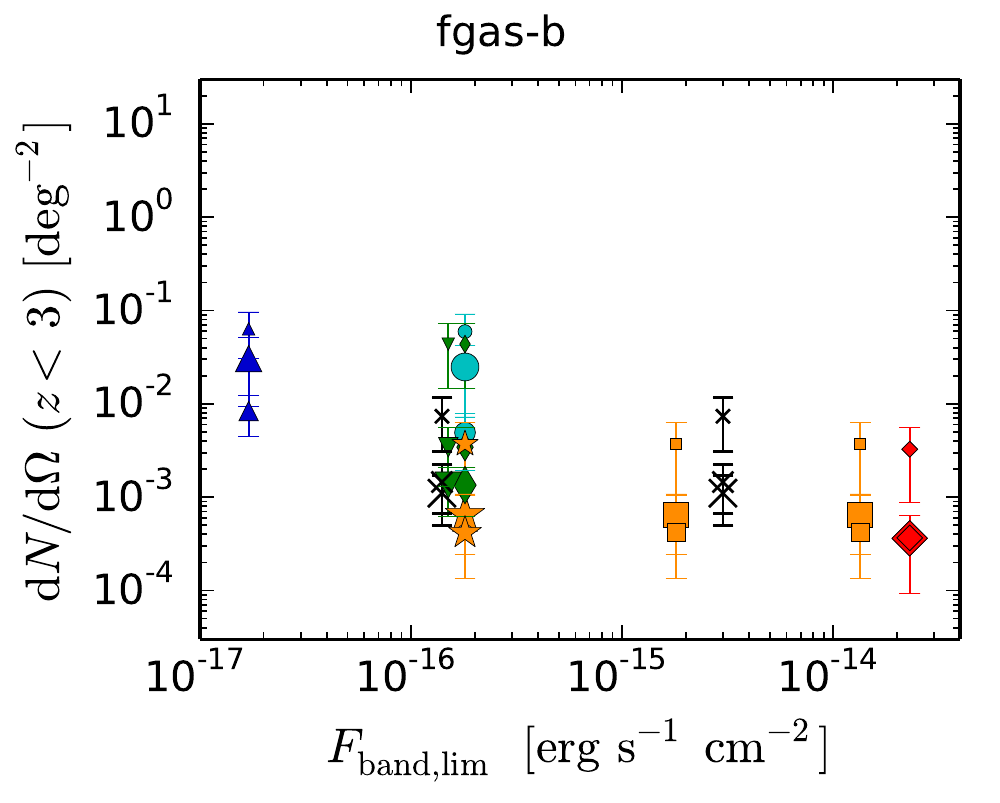}
\includegraphics[width=0.41\textwidth]{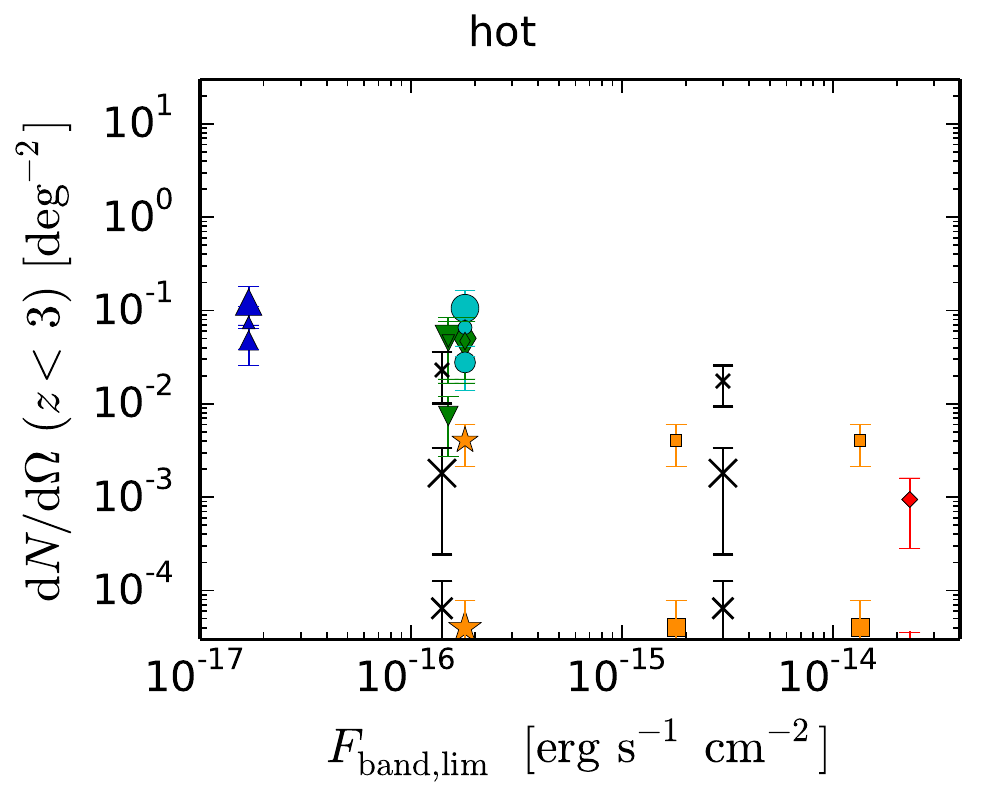}
\includegraphics[width=0.41\textwidth]{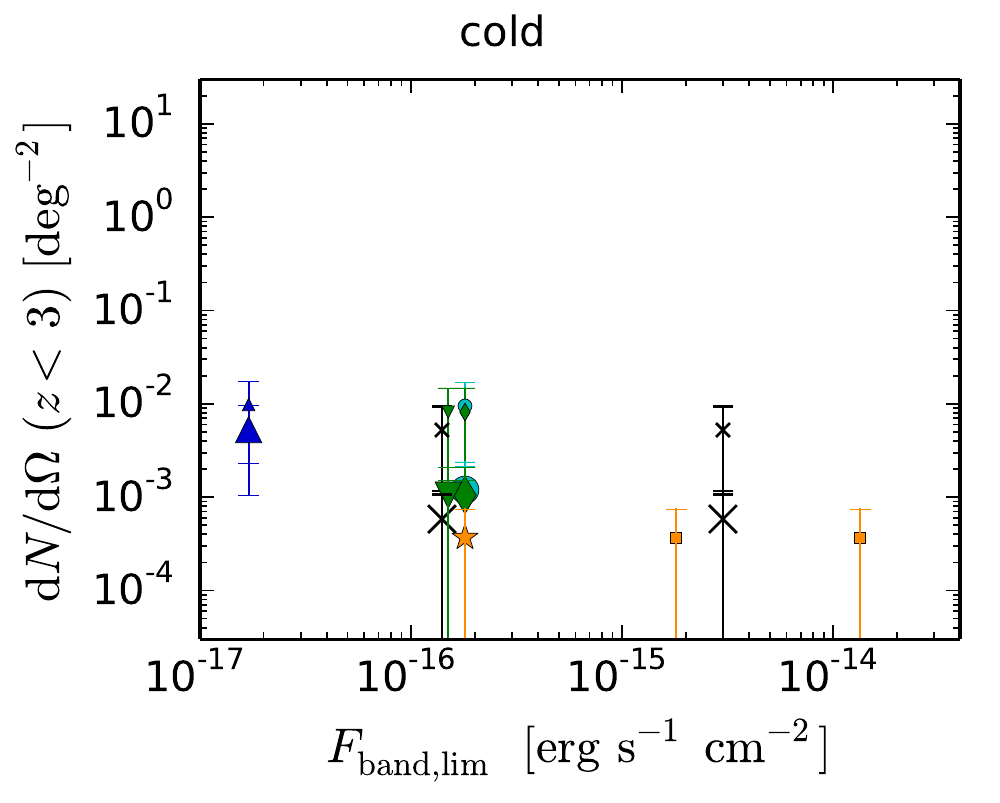}
\caption{As in the top panels of Figure \ref{fig:source_counts}, each panel shows the cumulative number of observable spatially-offset AGN per deg$^2$, out to $z=3$, for various surveys as a function of limiting flux in the corresponding band. Different panels correspond to different spin models, as indicated on the plot labels. The \fivedeg\ spin model is not shown, as the number counts are nearly all below the plot range. The points are in the same style as Figure \ref{fig:source_counts}, except that here results are compared between Illustris-1, 2, and 3. Point size denotes the simulation resolution; the largest points correspond to Illustris-1. As in the bottom panel of Figure \ref{fig:converge_mrgrate}, a constant minimum DM and stellar mass are assumed for the progenitor halos of merging BHs in each simulation, rather than a constant minimum particle number. As a result, the predictions here for Illustris-1 are lower than in Figure \ref{fig:source_counts}, where lower-mass halos are included. \label{fig:compare_source_counts}}
\end{figure*}

Figure \ref{fig:compare_source_counts} illustrates the level of resolution convergence of the predicted number of observable spatially-offset AGN in our fiducial models. Again, a constant minimum progenitor halo mass is used for the comparison. In general, the convergence is quite reasonable; in most cases the differences between the simulations are within the Poisson errors. Convergence is poorest for the aligned spin models (\hot\ and \cold) in seeing limited surveys, where recoils are unlikely to be observed in any case. The higher values for Illustris-3 in these models reflect its higher BH merger rate when a consistent minimum mass cut is applied in all simulations (Figure \ref{fig:converge_mrgrate}, bottom panel).

\end{document}